\def\CHfour{{\fam0 CH_4}}
\def\CHthree{{\fam0 CH_3}}
\def\CHthreeNHtwo{{\fam0 CH_3NH_2}}
\def\CHtwoNHtwo{{\fam0 CH_2NH_2}}
\def\CHtwoNH{{\fam0 CH_2NH}}
\def\HtwoCN{{\fam0 H_2CN}}
\def\CN{{\fam0 CN}}
\def\HCN{{\fam0 HCN}}
\def\CHthreeCN{{\fam0 CH_3CN}}

\def\CHthreeOH{{\fam0 CH_3OH}}
\def\CHthreeO{{\fam0 CH_3O}}
\def\CHthreeCHO{{\fam0 CH_3CHO}}

\def\CHtwoOH{{\fam0 CH_2OH}}
\def\CH{{\fam0 CH}}
\def\COtwo{{\fam0 CO_2}}
\def\CO{{\fam0 CO}}

\def\CthreeHtwo{{\fam0 C_3H_2}}

\def\CtwoHtwo{{\fam0 C_2H_2}}
\def\CtwoHthree{{\fam0 C_2H_3}}
\def\CtwoHfour{{\fam0 C_2H_4}}
\def\CtwoHfive{{\fam0 C_2H_5}}
\def\CtwoHsix{{\fam0 C_2H_6}}
\def\C{{\fam0 C}}
\def\oneCHtwo{{\fam0 ^{1}CH_2}}
\def\threeCHtwo{{\fam0 ^{3}CH_2}}

\def\hd189{{\fam0 HD 189733b}}
\def\hd209{{\fam0 HD 209458b}}
\def\HCthreeN{{\fam0 HC_3N}}
\def\HtwoCO{{\fam0 H_2CO}}
\def\HCO{{\fam0 HCO}}
\def\HNCO{{\fam0 HNCO}}

\def\HtwoCCO{{\fam0 H_2CCO}}
\def\HtwoO{{\fam0 H_2O}}
\def\Htwo{{\fam0 H_2}}
\def\H{{\fam0 H}}

\def\M{{\fam0 M}}
\def\NO{{\fam0 NO}}
\def\NH{{\fam0 NH}}
\def\NHtwo{{\fam0 NH_2}}
\def\NHthree{{\fam0 NH_3}}
\def\Net{{\fam0 Net:}}
\def\N{{\fam0 N}}
\def\Ntwo{{\fam0 N_2}}

\def\NNH{{\fam0 NNH}}
\def\NtwoHtwo{{\fam0 N_2H_2}}

\def\NtwoHthree{{\fam0 N_2H_3}}
\def\NtwoHfour{{\fam0 N_2H_4}}
\def\OH{{\fam0 OH}}
\def\OoneD{{\fam0 O(^1D)}}
\def\Otwo{{\fam0 O_2}}
\def\O{{\fam0 O}}

\def\cmmthree{{\fam0\, cm^{-3}}}
\def\cmmtwo{{\fam0\,cm^{-2}}}

\def\cmtwo{{\fam0 cm^2}}

\def\scinot#1.{\hbox{$\,$ $\times$ $10^{#1}$}}
\def\smone{{\fam0\,s^{-1}}}

\def\ten#1.{\hbox{$10^{#1}$}}

\def\tover#1#2{{\strut\displaystyle #1 \over \strut\displaystyle #2}}
\def\deg{\ifmmode^\circ\else$\null^\circ$\fi}
\def\spose#1{\hbox to 0pt{#1\hss}}
\def\lta{\mathrel{\spose{\lower 3pt\hbox{$\mathchar "218$}}\raise 2.0pt\hbox{$\mathchar"13C$}}}
\def\gta{\mathrel{\spose{\lower 3pt\hbox{$\mathchar "218$}}\raise 2.0pt\hbox{$\mathchar"13E$}}}
\def\lrarrow{\mathrel{\spose{\lower 1pt\hbox{$\rightarrow$}}\raise 3.0pt\hbox{$\
leftarrow$}}}

\documentclass{emulateapj}

\slugcomment{submitted to Astrophys. J.}

\shorttitle{Disequilibrium Chemistry on HD 189733b and HD 209458b}
\shortauthors{Moses et al.}

\begin{document}


\title{Disequilibrium Carbon, Oxygen, and Nitrogen Chemistry in the Atmospheres of 
HD~189733\lowercase{b} and HD~209458\lowercase{b}}

\author{Julianne I. Moses}
\affil{Space Science Institute, 4750 Walnut Street, Suite 205, Boulder, CO, 80301, USA}
\email{jmoses@spacescience.org}

\author{C. Visscher}
\affil{Lunar and Planetary Institute, Houston, TX, 77058, USA}

\author{J. J. Fortney}
\affil{Department of Astronomy and Astrophysics, University of California, Santa Cruz, CA, 95064, USA}

\author{A. P. Showman, N. K. Lewis, and C. A. Griffith}
\affil{Department of Planetary Sciences and Lunar and Planetary Laboratory, The University of Arizona, Tucson,
AZ, 85721, USA}

\author{S. J. Klippenstein}
\affil{Chemical Sciences and Engineering Division, Argonne National Laboratory, Argonne, IL, 60439, USA}

\author{M. Shabram}
\affil{Department of Astronomy, University of Florida, Gainesville, FL, 32611, USA}

\author{A. J. Friedson}
\affil{Earth and Space Sciences Division, Jet Propulsion Laboratory, California Institute of Technology,
Pasadena, CA, 91109, USA}

\and

\author{M. S. Marley, R. S. Freedman}
\affil{NASA Ames Research Center, Moffett Field, CA, 94035, USA}

%
%

\begin{abstract}
We have developed a one-dimensional photochemical and thermochemical kinetics and diffusion model 
to study the effects of disequilibrium chemistry on the atmospheric composition of ``hot Jupiter'' 
exoplanets.  Here we investigate the coupled chemistry of neutral carbon, hydrogen, oxygen, and 
nitrogen species on HD 189733b and HD 209458b, and we compare the model results with existing 
transit and eclipse observations. We find that the vertical profiles of molecular constituents are 
significantly affected by transport-induced quenching and photochemistry, particularly on cooler 
HD 189733b; however, the warmer stratospheric temperatures on HD 209458b help maintain thermochemical 
equilibrium and reduce the effects of disequilibrium chemistry. For both planets, the methane and 
ammonia mole fractions are found to be enhanced over their equilibrium values at pressures of a few 
bar to less than a mbar due to transport-induced quenching, but CH$_4$ and NH$_3$ are photochemically 
removed at higher altitudes. Disequilibrium chemistry also enhances atomic species, unsaturated 
hydrocarbons (particularly C$_2$H$_2$), some nitriles (particularly HCN), and radicals like OH, 
CH$_3$, and NH$_2$. In contrast, CO, H$_2$O, N$_2$, and CO$_2$ more closely follow their equilibrium
profiles, except at pressures $\lta$ 1 microbar, where CO, H$_2$O, and N$_2$ are photochemically 
destroyed and CO$_2$ is produced before its eventual high-altitude destruction. The enhanced 
abundances of CH$_4$, NH$_3$, and HCN are expected to affect the spectral signatures and thermal 
profiles of HD 189733b and other relatively cool, transiting exoplanets. We examine the sensitivity 
of our results to the assumed temperature structure and eddy diffusion coefficients and discuss 
further observational consequences of these models.
\end{abstract}



\section{Introduction}

The detection and characterization of extrasolar planets is one of the most exciting and 
fastest growing fields in astronomy \citep[e.g.,][]{charb07,marley07,udry07,deming09,baraffe10}.
Of the handful of tools available for characterization of exoplanet atmospheres, transit and 
eclipse observations have proven to be invaluable for inferring properties of both the planet 
and its host star
\citep[e.g.,][]{charb00,charb02,charb05,henry00,seager00,brown01,brownand01,burrows01,hubbard01,deming05b,knut07,knut09,madhu09}.
The wavelength dependence of the signal during transit 
and eclipse observations can allow identification of atmospheric constituents on 
``hot Jupiter'' exoplanets \citep[e.g.,][]{charb02,vidal03,tinetti07,grill08,swain08b,swain09a,madhu09},
allowing a better 
understanding of the chemical and physical processes operating in these exotic atmospheres.

Hot Jupiters are often divided into two classes, those that appear to have stratospheric thermal 
inversions, and those that do not
\citep[e.g.,][]{fort08a,hubeny03,harring07,burrows07,burrows08,knut08,spiegel09,madhu09}.
Although the ultimate reasons for these differences are still under investigation
\citep[see also][]{spiegel09,zahnle09,knut10,youdin10,fressin10,machalek08,machalek10,madhu10inv}, 
exoplanets with thermal inversions appear to have some source of high-altitude absorption, 
whereas exoplanets without thermal inversions are apparently missing this stratospheric 
absorber, perhaps because it is tied up in condensed phases
\citep[e.g.,][]{hubeny03,fort08a,burrows08} or is destroyed by photolysis and/or other 
mechanisms related to chromospherically active host stars \citep{knut10}.  The 
observations discussed in the the above citations suggest that HD 209458b has a strong 
stratospheric thermal inversion, whereas HD 189733b does not.  These two transiting planets, 
which have been the most extensively studied and characterized to date because of their bright 
parent stars and large transit depths, become convenient endmembers for our study of the 
chemistry of the two main classes of hot Jupiters.  

Constraints on the composition and thermal structure of HD 189733b and HD 209458b have been 
provided through (1) the identification of Na, H, and other atomic neutral and ionic species in 
the planets' upper atmospheres 
\citep{charb02,vidal03,vidal04,vidal08,benjaff07,benjaff08,ballester07,snellen08,redfield08,sing08a,ehren08,langland09,linsky10,benjaff10,lecave10},
(2) the identification of molecular species such as $\HtwoO$, $\CHfour$, 
$\COtwo$, and CO in the $\sim$1 bar to 0.1 mbar atmospheric region 
\citep{tinetti07,fort07,burrows07,burrows08,grill08,barman08,swain08a,swain08b,swain09a,swain09b,swain10,beaulieu08,beaulieu09,desert09,madhu09},
(3) observational and theoretical inferences regarding the vertical and horizontal temperature 
structure, including the suggested presence of a stratospheric temperature inversion on HD 209458b 
but not HD 189733b, and the identification of longitudinal temperature variations and phase lags 
\citep[e.g.,][]{knut07,knut09,coop06,fort06a,fort08a,fort10,richardson07,burrows07,burrows08,cowan07,showman08,showman09,madhu09,madhu10inv},
and (4) the inferred presence of hazes, clouds, 
and Rayleigh scatterers that can affect atmospheric spectral behavior 
\citep[e.g.,][]{charb02,fort03,fort05,iro05,pont08,lecave08a,lecave08b,sing08a,sing08b,sing09,shabram11}.
Note, however, that 
the data reduction and observational analyses are often difficult due to low signal-to-noise, potential 
variable stellar activity of the host star (including star spots), and/or poorly understood systematic 
instrument effects 
\citep[e.g.,][]{gibson10};
spectral models are also poorly constrained, making 
analysis of the observations difficult.  Various groups do not always agree, even when examining the 
same or similar datasets 
(cf.~\citealt{desert09} vs.~\citealt{beaulieu08} vs.~\citealt{ehren07}; 
\citealt{swain08b} vs.~\citealt{sing09} vs.~\citealt{gibson10}; 
\citealt{stevenson10} vs.~\citealt{beaulieu11}).
An additional problem with analyses is the lack of reliable molecular line parameters for hot bands
of gases other than $\HtwoO$, CO, and CO$_2$.  
Even if the data reduction and analysis were not inherently difficult, interpretation of thermal 
infrared data can be hampered by the fact that molecular bands can manifest as either absorption 
or emission features, depending on the atmospheric thermal structure, creating a degeneracy between 
temperature profiles and compositions 
\citep[e.g.,][]{burrows08,madhu09,madhu10inv,tinetti10,fort10,odonovan10}.  
Despite these complications, transit and eclipse observations 
represent our best current means for characterizing the atmospheres of hot Jupiters.  Theoretical 
models are needed to interpret these observations.

Existing chemical models for transiting exoplanets tend to fall into two groups:
thermochemical-equilibrium models and photochemical models.  Thermochemical equilibrium is
assumed for most models 
\citep[e.g.,][]{burrows99,seag00,lodders02,sharp07,burrows07,burrows08,fort08a,viss10a},  
although potential departures from equilibrium are sometimes explored 
through evaluations of simple time-scale arguments 
\citep[e.g.,][]{lodders02,fort06a,fort08b,viss06,viss10a,madhu11} or through more complex models 
with parameterized chemical kinetics \citep{coop06}.  
Thermochemical equilibrium is a reasonable starting point for exoplanet composition 
predictions, but the strong ultraviolet flux incident on close-in transiting planets like HD 
189733b and HD 209458b can drive the observable upper regions of the atmosphere out of 
equilibrium.  Moreover, some of the existing transport and kinetics time-scale arguments 
have known flaws 
\citep[for further discussion, see][]{yung88,smith98,bezard02,viss10b,moses10,viss11}  
and are often misused, leading to questionable predictions of 
quenched disequilibrium abundances.  

The second group of chemical models exploits knowledge of Jovian photochemistry to predict 
the consequences of the large ultraviolet flux incident on exoplanet atmospheres 
\citep{liang03,liang04,yelle04,garcia07,koskinen07,koskinen09,zahnle09,zahnle11,line10}.  
Of these latter models, \citet{yelle04}, \citet{garcia07}, and \citet{koskinen07,koskinen09}
focus on the uppermost regions of 
the atmosphere, {\it i.e.}, the thermosphere, ionosphere, and exosphere and do not 
consider what goes on deeper than 1 microbar.  In contrast, \citet{liang03,liang04} and 
\citet{line10} explore the photochemistry of the deeper stratosphere and troposphere using 
thermochemical equilibrium and/or transport-quenched disequilibrium as a lower boundary 
condition for their models; however, their focus on low-temperature reactions (for the former 
two studies) and their lack of fully reversed reactions (all three studies) prevents them from 
extending the kinetic models into the thermochemical regime at temperatures $\gta$ 1000 K, 
and their assumptions can result in artificial constant-with-altitude profiles for the the major 
species in the troposphere.  \citet{zahnle09,zahnle11} have published the only models to date 
that have attempted to bridge the thermochemistry and photochemistry regimes with a single model;
however, their reliance on isothermal atmospheric models can result in some unrealistic 
vertical profiles and species abundances that will not be present with more realistic atmospheric 
structures, and it is not clear whether their reaction scheme will reproduce equilibrium at 
high temperatures.  

We therefore have developed a unique one-dimensional (1-D) thermochemical and photochemical 
kinetics and transport model to circumvent some of the problems described above.  The model 
has the ability to transition seamlessly from the thermochemical equilibrium regime in the deep 
troposphere of giant planets to the ``transport-quenched'' disequilibrium and photochemical 
regimes in the upper troposphere and stratosphere to illuminate how chemical processes in 
each regime combine to influence the observational properties of giant-planet atmospheres.  
In an initial set of studies, we applied the model to the deep troposphere of Jupiter to 
ensure that the model satisfies observational constraints on a well-studied planet within 
our own solar system \citep{viss10b,moses10}.  We now apply the model to the atmospheres of 
HD 189733b and HD 209458b, as representatives of the two classes of hot Jupiters --- those with 
and without stratospheric thermal inversions.  Our main goal is to more realistically predict the 
vertical profiles of the major observable carbon-, oxygen-, and nitrogen-bearing species on 
these close-in transiting planets to aid the analysis and interpretation of primary transit 
and secondary eclipse observations.  By including both thermochemical and photochemical 
kinetics in a traditional 1-D chemistry/{\-}diffu{\-}sion model, we can avoid some of the 
shortcomings and pitfalls of previous chemical models --- we do not need to correctly identify 
in advance the mechanism and rate-limiting steps for transport quenching of species like 
CO, $\CHfour$, HCN, and $\NHthree$, we can avoid other uncertainties in the time-scale 
arguments such as the correct length scale to use for the transport time scale (e.g., Smith 
1998), we do not need to restrict the model to any particular pressure region, and we can 
consider realistic atmospheric structures rather than isothermal atmospheres.  

As we will show in the following sections, disequilibrium processes like transport-induced 
quenching and photochemistry can perturb the composition of HD 189733b and HD 209458b away 
from equilibrium.  The effects of these disequilibrium processes are more significant for 
cooler exoplanets like HD 189733b than for warmer exoplanets like HD 209458b, but there are 
potential observational consequences for both planets.  In particular, transport-induced quenching 
and photochemistry can enhance the abundance of atmospheric constituents like $\CHfour$, HCN, 
$\NHthree$, and $\CtwoHtwo$, and the presence of these disequilibrium molecules will affect 
spectral and photometric behavior.  We compare the model results with existing transit and 
secondary eclipse observations and discuss in detail the major disequilibrium mechanisms 
affecting the abundance of carbon, nitrogen, and oxygen species on HD 189733b and HD 209458b. 

\section{Model}

We use KINETICS, the Caltech/JPL 1-D pho{\-}to{\-}chemistry/diffu{\-}sion code 
(\citealt{allen81}; see also 
\citealt{yung84,glad96,moses00a,moses00b,moses05,liang03,liang04,line10}) 
to calculate the vertical distributions of tropospheric and stratospheric 
constituents on extrasolar giant planets.  This fully implicit, finite-difference code solves 
the coupled mass-continuity equations as a function of pressure for each species:
\begin{equation}
\tover{\partial n_i}{\partial t} \, + \, \tover{\partial
\Phi_i}{\partial z} \, = \, P_i \, - \, L_i 
\end{equation}
{\noindent where $n_i$ is the number density ($\cmmthree$), $\Phi_i$ is the flux ($\cmmtwo$ $\smone$), 
$P_i$ is the chemical production rate ($\cmmthree$ $\smone$), and $L_i$ is the chemical loss rate 
($\cmmthree$ $\smone$) of species $i$, all of
which are explicit functions of time $t$ and altitude $z$.  The flux transport terms include 
both molecular and eddy diffusion, the latter parameterized by an eddy diffusion coefficient 
$K_{zz}$.  The calculations are allowed to evolve until steady state is achieved, and a 
convergence criterion of one part in a thousand is used.  We consider 198 atmospheric levels 
with a vertical resolution ranging from 30 grid points per atmospheric scale height in the deep 
atmosphere to 3 grid points per scale height in the upper atmosphere.  A finer grid is needed 
in the deep atmosphere to accurately track the quench points for some of the disequilibrium 
species.  The model extends from the deep tropospheres of HD 209458b (bottom boundary at 2754 K, 
1 kbar) and HD 189733b (2600 K, 1.67 kbar) to well past the homopause region into the thermosphere (top 
boundary at pressures less than 10$^{-10}$ bar, exospheric temperature assumed to be 12,000 K).  
The upper boundary is chosen at a high-enough altitude such that we encompass the photochemical 
production and destruction regions for the main molecular species $\Htwo$, $\HtwoO$, CO, $\CHfour$,  
$\COtwo$, N$_2$, $\NHthree$, and HCN for plausible assumptions about $K_{zz}$, and 
such that assumptions about upper boundary conditions do not affect the results.  Although the 
models extend into the thermosphere, our intent is not to specifically model thermospheric 
chemistry.  Instead, the thermospheric temperature profile and escape boundary condition are 
included in an {\it ad hoc} manner based on the results of \citet{yelle04}, \citet{garcia07}, 
and \citet{koskinen10} (see below for further details).  

We use free-convection and mixing-length theories \citep[e.g.,][]{stone76,flasar77}
to define the eddy diffusion coefficients $K_{zz}$ in the deep, adiabatic 
portion of the troposphere and estimate $K_{zz}$ profiles through the bulk of the rest of 
the atmosphere from the root-mean-square (rms) vertical velocities derived from global 
horizontal averages at a given pressure level from the 3-D GCM simulations of \citet{showman09}, 
{\it i.e.}, by assuming $K_{zz}$ = $w(z) L(z)$, where $w(z)$ is the horizontally averaged global 
rms vertical velocity from the GCM simulations and $L(z)$ is assumed to be the atmospheric pressure 
scale height $H(z)$ (but could be some fraction of $H(z)$, see \citealt{smith98}).  Note that 
this procedure provides only a crude estimate of $K_{zz}$, and it would have been preferable to 
have calculated $K_{zz}$ from the eddy vertical velocity times the eddy displacement, but this 
information was not available from the GCMs.  In particular, the procedure may overestimate 
$K_{zz}$ in the $\sim$10-200 bar radiative region, where the vertical motion tends to consist 
of small-scale wave oscillations.  In the future, it would be useful to include passive tracers into
the GCMs to directly and rigorously calculate the rate of vertical mixing.  Figure \ref{eddynom} shows 
the eddy diffusion coefficient profiles adopted for our nominal models.  However, given the above
caveats, we consider $K_{zz}$ profiles to be essentially free parameters in the models, and we 
examine the sensitivity of our results to reasonable variations in the adopted $K_{zz}$ profiles.  
The molecular diffusion coefficients are described in \citet{moses00a}.

\begin{figure}
\includegraphics[angle=-90,scale=0.37]{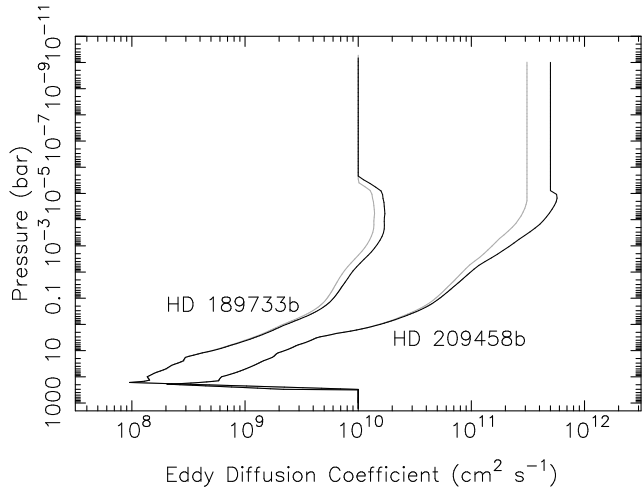}
\caption{Eddy diffusion coefficient profiles adopted for our nominal models, as estimated 
from rms vertical velocity profiles obtained from the GCMs of \citet{showman09} (see text).  
For both planets, the black line represents 
the profile for the dayside-average model, and the gray line represents the profile for 
the terminator-average model.  The high-pressure value of $K_{zz}$ = 10$^{10}$ cm$^2$ 
$\smone$ is an estimate of the eddy diffusion coefficient in the adiabatic region 
from free-convection and mixing-length theory \citep{stone76}, and $K_{zz}$'s are assumed 
to be constant at altitudes above the top level of the GCMs.\label{eddynom}}
\end{figure}

Temperature profiles are not calculated self-consistently within the KINETICS code but are
instead required as inputs to the model.  The temperatures adopted in our model are shown in 
Fig.~\ref{figtemp}.  Most of these profiles derive from the 3-D GCM simulations of \citet{showman09} 
for the region from 170 bar to a few $\mu$bar or the 1-D models of \citet{fort06a,fort10} for the 
region from 1000 to 1.0$\scinot-6.$ bar.  Because the nitrogen species in particular quench at 
very high temperatures, we needed to extend the GCM profiles to deeper levels ({\it i.e.}, to reach
temperatures of at least 2500 K).  The 1-D models of \citet{fort06a,fort10} already include the 
deep adiabatic region, but the GCM profiles do not.  For HD 189733b, the 1-D ``4$\pi$'' profiles of
\citet{fort06a,fort10} that assume the incident stellar energy is redistributed over the entire
planet compare very well with the GCM theoretical thermal profiles in the 1--100-bar region, so 
we extended the GCM profiles to deeper pressures by simply connecting smoothly to the 1-D 
profiles.  For HD 209458b, the 1-D ``4$\pi$'' profiles are somewhat
warmer than the GCM profiles in the 1--100-bar region, and we shifted the 1-D profiles by -60 K
before connecting the GCM profiles to this deeper adiabatic region.  At low pressures (high 
altitudes), we extended the profiles by assuming either an isothermal atmosphere or a 
high-temperature thermosphere, using the thermospheric profiles of \citet{yelle04} and 
\citet{garcia07} as a guide for the latter.  Once the pressure-temperature 
profiles were established, the rest of the background atmospheric grid (e.g., densities, altitudes) 
was derived through solving the hydrostatic equilibrium equation.

\begin{figure*}
\begin{tabular}{ll}
{\includegraphics[angle=-90,clip=t,scale=0.37]{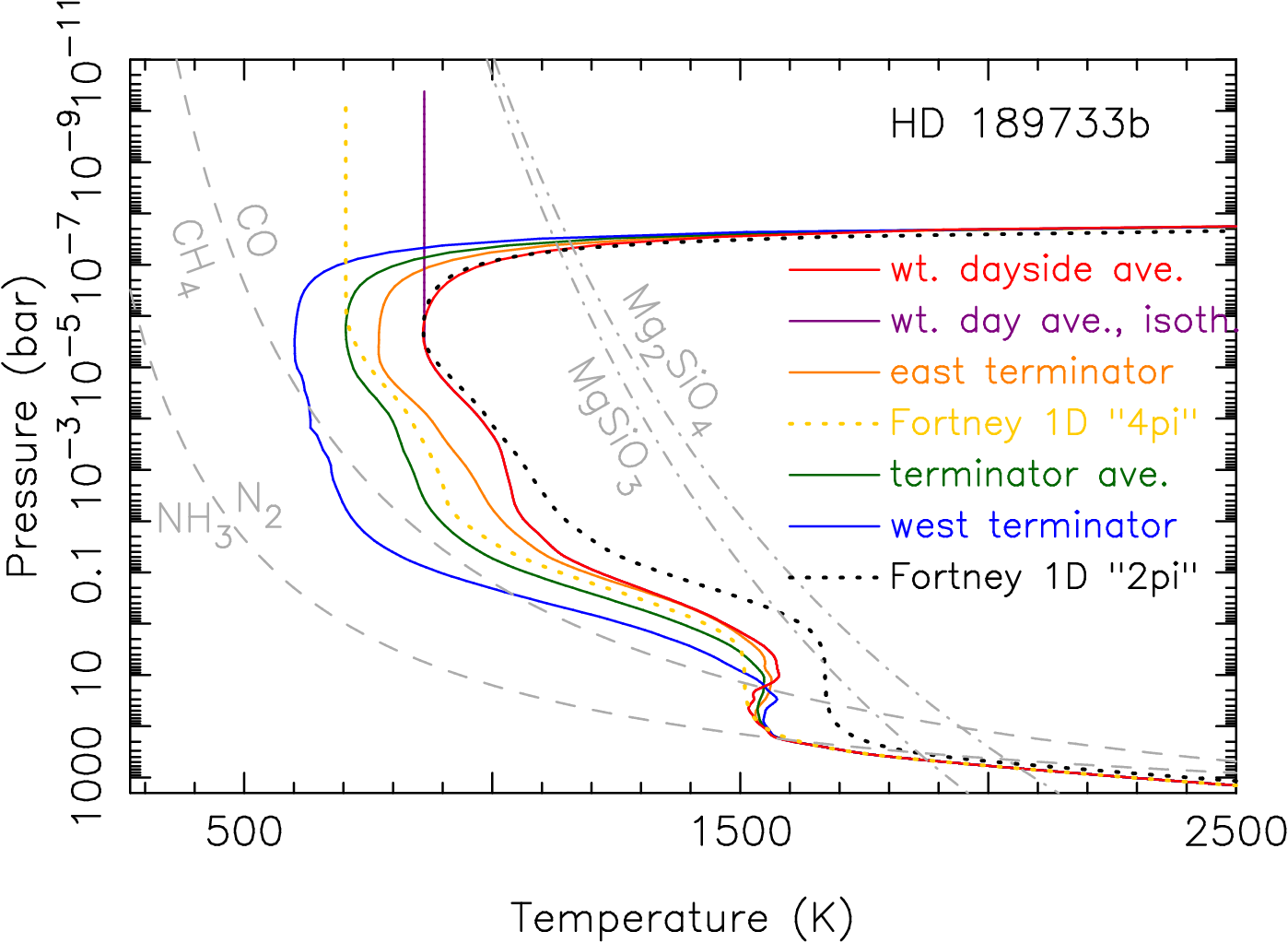}} 
& 
{\includegraphics[angle=-90,clip=t,scale=0.37]{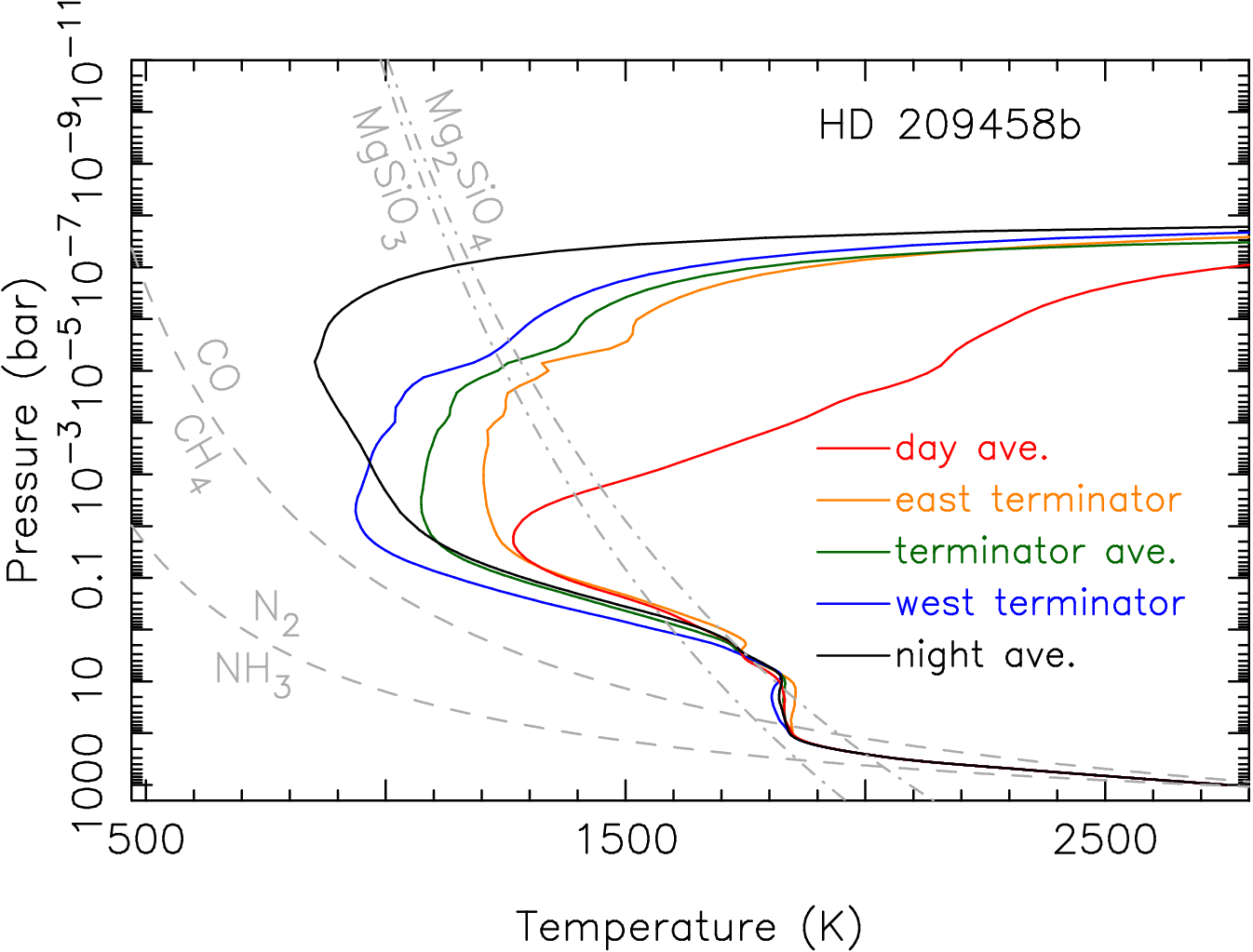}} 
\\
\end{tabular}
\caption{Temperature profiles for our models of HD 189733b (left) and HD 209458b (right), 
based on the general circulation models of \citet{showman09} (solid lines) and on the 1-D 
radiative-convective models of \citet{fort06a,fort10} (dotted lines), for an assumed 
solar-composition atmosphere.  For HD 189733b, the models include a projected-area-weighted 
profile averaged over the dayside hemisphere (to best represent the planet's apparent disk 
during the secondary eclipse), profiles averaged over the eastern and western terminators 
(to best represent the ingress and egress regions during the transits), a terminator-average 
profile, and 1-D profiles calculated assuming the incident stellar energy is distributed 
over the entire planet (curve labeled ``4pi'') or over the dayside hemisphere only (curve 
labeled ``2pi'').  For HD 209458b, the models include a straight ({\it i.e.}, not projected-area) 
average over the dayside hemisphere, along with eastern and western terminator averages, a 
terminator average, and an average over the nightside hemisphere.  Most profiles have been 
given an {\it ad hoc} high-temperature thermosphere at high altitudes based on the models of 
\citet{yelle04} and \citet{garcia07}.  
Also shown are pressure-temperature curves (dashed gray lines) at which the major nitrogen 
species, $\Ntwo$ and $\NHthree$, and the major carbon species, CO and $\CHfour$, have equal 
abundances in thermochemical equilibrium.  To the left-hand side of these $\CHfour$ = CO 
and $\NHthree$ = $\Ntwo$ curves, the reduced species ($\CHfour$ and $\NHthree$) dominate, 
while the more oxidized species (CO and $\Ntwo$) dominate to the right.  Note that both 
planets are within the CO and $\Ntwo$-dominated regimes over much of the atmosphere for all 
temperature profiles except for the coolest HD 189733b western-terminator-average profile.  
The gray dot-dashed lines represent the approximate condensation curves for the major silicates 
enstatite (MgSiO$_3$) and forsterite (Mg$_2$SiO$_4$), based on \citet{viss10a}.  
A color version of this figure is available in the online journal.\label{figtemp}}
\end{figure*}

HD 189733b orbits its host star at a distance of $\sim$0.031 AU \citep[e.g.,][]{southworth10}.
The cool but chromospherically active HD 189733 has an effective temperature in the range 
$\sim$5050--5090 K \citep{bouchy05,knut10} and has typically been classified in the literature as a 
K$1\, $V-to-K$2\, $V main-sequence star \citep[e.g.,][]{barnes10,shkolnik08}, although the 
Hipparcos catalog lists it as a G$5\, $V (\citealt{perryman97}; see also \citealt{montes01}).  
In any case, HD 189733 is less luminous than the Sun, and HD 189733b is not one of the hottest 
of the transiting exoplanets that have been discovered to date, despite its small orbital 
semimajor axis.  HD 189733 is part of a binary star system, but the M-dwarf companion to 
HD 189733 is $\sim$200 AU away and should not significantly affect the flux incident on 
HD 189733b.  Because well-calibrated ultraviolet spectra for HD 189733 are available for only 
a specific range of wavelengths \citep[e.g.,][]{lecave10}, we set up a normalized 1-AU spectrum 
using the {\it Hubble Space Telescope}\/ ({\it HST\/}) STIS spectra of epsilon Eridani (a K$2\, $V 
star) from the CoolCAT database \citep{ayers05} for the 1150--2830 \AA\ region, after correcting for
stellar distance.  At the upper end of this wavelength range, the flux from epsilon Eridani is 
much less than for typical G stars like our Sun at the same stellar distance, whereas the flux 
at the shorter EUV wavelength end of this range compares well to main-sequence G 
stars.  We therefore use the 1-AU solar flux divided by 10 for wavelengths longer than 
2830 \AA, and we use the 1-AU solar flux for low-to-average solar conditions as an analog 
for our 1-AU-normalized HD 189733 flux for wavelengths below 1150 \AA.  The photochemical 
calculations are then performed assuming that HD 189733b is located 0.031 AU from this K2$\, $V-analog 
star.  The host star for HD 209458b is a G$0\, $V stellar type (Hipparcos, Perryman et al. 
1997), and we simply use the solar flux as an analog for HD 290458, and assume HD 209458b 
orbits at a distance of 0.047 AU \citep[e.g.,][]{southworth10}. The stellar zenith angle $\theta$ 
is fixed at 48\deg\ for the dayside-average models --- where $\langle \cos \theta \rangle$ = 2/3
($\theta$ $\approx$ 48\deg) is the projected-area-weighted average of the cosine of the stellar 
zenith angle over the planetary disk at secondary eclipse conditions --- and $\theta$ is fixed at 
$\sim$84\deg--86\deg\ for the terminator models.  Multiple Rayleigh scattering by H$_2$ and He is 
considered in the model using a Feautrier radiative-transfer method \citep{michel92}.  

We consider the neutral chemistry of 90 carbon-, hydrogen-, oxygen-, and nitrogen-bearing species:  
the model contains H, H$_2$, He, C, CH, excited singlet $^1$CH$_2$, ground-state triplet $^3$CH$_2$, 
CH$_3$, CH$_4$, C$_2$, C$_2$H, C$_2$H$_2$, C$_2$H$_3$, C$_2$H$_4$, C$_2$H$_5$, C$_2$H$_6$, 
C$_3$H$_2$, C$_3$H$_3$, CH$_3$C$_2$H, CH$_2$CCH$_2$, C$_3$H$_5$, C$_3$H$_6$, C$_3$H$_7$, C$_3$H$_8$, 
C$_4$H, C$_4$H$_2$, C$_4$H$_3$, C$_4$H$_4$, C$_4$H$_5$, C$_4$H$_6$, C$_4$H$_8$, C$_4$H$_9$, 
C$_4$H$_{10}$, C$_5$H$_3$, C$_5$H$_4$, C$_6$H$_2$, C$_6$H$_3$, C$_6$H$_4$, C$_6$H$_5$, non-cyclic 
C$_6$H$_6$, C$_6$H$_6$ (benzene), O, O($^1$D), O$_2$, OH, H$_2$O, CO, CO$_2$, HCO, H$_2$CO, 
CH$_2$OH, CH$_3$O, 
CH$_3$OH, HCCO, H$_2$CCO, C$_2$H$_3$O, CH$_3$CHO, C$_2$H$_4$OH, HO$_2$, H$_2$O$_2$, N, N$_2$, NH, 
NH$_2$, NH$_3$, NNH, N$_2$H$_2$, singlet biradical H$_2$NN, N$_2$H$_3$, N$_2$H$_4$, CN, HCN, 
H$_2$CN, CH$_2$NH, CH$_3$NH, CH$_2$NH$_2$, CH$_3$NH$_2$, CH$_2$CN, CH$_3$CN, C$_3$N, HC$_3$N, 
C$_2$H$_2$CN, C$_2$H$_3$CN, NO, NO$_2$, N$_2$O, HNO, HNO$_2$, NCO, HNCO, and CH$_3$NO.  Metals,
rock-forming elements, and sulfur and phosphorus species are not considered, nor is ion chemistry.  
See \citet{zahnle09} for a discussion of the effects of sulfur photochemistry on exoplanet 
composition.  The reaction list 
derives originally from the Jupiter and Saturn models of \citet{glad96}, \citet{moses96}, 
and \citet{moses95a,moses95b,moses00a,moses00b,moses05},  
although extensive updates that account for high-temperature kinetics have been included based 
largely on combustion-chemistry literature
\citep[e.g.,][]{baulch92,baulch94,baulch05,atkin97,atkin06,smith00,tsang87,tsang91,dean00}.
See \citet{viss10b} for further discussion of 
the hydrocarbon and oxygen kinetics and \citet{moses10} for further discussion of the nitrogen 
kinetics.  The model contains $\sim$1600 reactions, with the rate coefficients of $\sim$800 of the 
reactions being taken from literature values, and the remaining $\sim$800 reactions being the 
reverse of these ``forward'' reactions, with the rate coefficients of the reverse reactions being 
calculated internally at each pressure-temperature point along the grid using the thermodynamic 
principle of microscopic reversibility (e.g., $k_{for}$/$k_{rev}$ = $K_{eq}$, where $k_{for}$ is 
the rate coefficient for the forward reaction, $k_{rev}$ is the rate coefficient for the reverse 
reaction, and $K_{eq}$ is the equilibrium constant of the reaction; see \citealt{viss11} for further 
details).  The rate-coefficient 
expressions for the forward reactions (and their literature sources) are provided in Table S1 of 
the Supplementary Material.  The rate coefficients for the reverse reactions for the 
pressure-temperature conditions of certain of our exoplanet models can be found in the full model 
outputs also presented in the Supplementary Material.  Note that \citet{miller09} demonstrate that 
detailed balance assuming microscopic reversibility is expected to be quite accurate, even for 
complicated chemical reactions that involve multiple, interconnected potential wells.

We assume a 1$\times$ protosolar composition atmosphere \citep{lodders03,lodders09,lodd09} in 
thermochemical equilibrium for our initial conditions for all the constituents, using the NASA CEA code 
\citep{gordon94} for the equilibrium calculations.  Thermodynamic parameters are taken from 
\citet{gurvich94}, the JANAF tables \citep{chase98}, \citet{burcat05}, or other literature sources.  
Note that although we do not consider rock-forming elements and their resulting effects on the 
chemistry of oxygen, we do assume that $\sim$20\% of the oxygen has been sequestered along with 
silicates and metals \citep[see][]{lodders04} in the deep atmospheres of HD 189733b and HD 209458b, 
thus depleting the available oxygen above the silicate clouds.   As such, our ``solar'' abundance of 
oxygen is slightly less than that used by many other groups, and the species profiles themselves
should not be considered accurate within and below the silicate cloud condensation regions due to our 
neglect of the effects of rock-forming elements (see Fig.~\ref{figtemp}). 

\begin{deluxetable*}{lcccccc}
\tabletypesize{\scriptsize}
\tablecaption{Column abundances above 1 bar\label{tabcolumn}}
\tablewidth{450pt}
\tablecolumns{7}
\tablehead{
\colhead{ } & \colhead{nominal K$_{zz}$} & \colhead{nominal K$_{zz}$} & \colhead{K$_{zz}$ = 10$^7$} &
\colhead{K$_{zz}$ = 10$^{11}$} & \colhead{nominal K$_{zz}$} & \colhead{nominal K$_{zz}$} \\
\colhead{ } & \colhead{terminator} & \colhead{dayside} & \colhead{dayside} & \colhead{dayside} & 
\colhead{dayside} & \colhead{terminator} \\
\colhead{Species} & \colhead{HD 189733b} & \colhead{HD 189733b} & \colhead{HD 189733b} & 
\colhead{HD 189733b} & \colhead{HD 209458b} & \colhead{HD 209458b} 
}
\startdata
H            & 2.1$\scinot20.$ & 9.3$\scinot20.$ & 1.2$\scinot21.$ & 7.9$\scinot20.$ & 2.2$\scinot22.$ & 8.9$\scinot21.$ \\
H$_2$        & 1.0$\scinot26.$ & 1.0$\scinot26.$ & 1.0$\scinot26.$ & 1.0$\scinot26.$ & 2.3$\scinot26.$ & 2.3$\scinot26.$ \\
He           & 1.9$\scinot25.$ & 1.9$\scinot25.$ & 1.9$\scinot25.$ & 2.0$\scinot25.$ & 4.5$\scinot25.$ & 4.5$\scinot25.$ \\
C            & 5.6$\scinot14.$ & 4.1$\scinot15.$ & 1.2$\scinot15.$ & 6.1$\scinot14.$ & 7.0$\scinot15.$ & 2.1$\scinot15.$\\
CH$_3$       & 2.0$\scinot17.$ & 2.3$\scinot17.$ & 7.2$\scinot16.$ & 3.7$\scinot18.$ & 1.9$\scinot17.$ & 1.7$\scinot17.$\\
CH$_4$       & 5.5$\scinot21.$ & 1.5$\scinot21.$ & 4.4$\scinot20.$ & 2.4$\scinot22.$ & 1.2$\scinot20.$ & 2.1$\scinot20.$\\
C$_2$H$_2$   & 6.6$\scinot16.$ & 3.0$\scinot16.$ & 2.8$\scinot15.$ & 4.6$\scinot18.$ & 3.4$\scinot15.$ & 2.7$\scinot15.$\\
C$_2$H$_4$   & 1.8$\scinot16.$ & 6.3$\scinot15.$ & 5.6$\scinot14.$ & 1.6$\scinot18.$ & 1.6$\scinot14.$ & 2.4$\scinot14.$\\
C$_2$H$_6$   & 2.4$\scinot11.$ & 5.4$\scinot13.$ & 4.9$\scinot12.$ & 1.3$\scinot16.$ & 3.2$\scinot11.$ & 7.7$\scinot11.$\\
C$_3$H$_2$   & 9.4$\scinot14.$ & 9.2$\scinot14.$ & 7.1$\scinot14.$ & 2.5$\scinot15.$ & 7.9$\scinot6.$  & 3.7$\scinot8.$\\
C$_4$H$_2$   & 4.3$\scinot12.$ & 1.5$\scinot10.$ & 1.5$\scinot9.$ & 6.1$\scinot13.$ & 3.3$\scinot4.$  & 1.9$\scinot5.$\\
C$_6$H$_6$   & 6.8$\scinot12.$ & 4.5$\scinot11.$ & 8.1$\scinot11.$ & 2.5$\scinot11.$ & 1.2$\scinot-4.$ & 1.0\scinot-2. \\
O            & 2.5$\scinot14.$ & 7.7$\scinot14.$ & 8.5$\scinot13.$ & 5.8$\scinot14.$ & 6.1$\scinot16.$ & 3.5$\scinot15.$\\
OH           & 1.6$\scinot15.$ & 8.3$\scinot15.$ & 7.9$\scinot15.$ & 1.3$\scinot16.$ & 5.5$\scinot17.$ & 1.5$\scinot17.$\\
H$_2$O       & 4.6$\scinot22.$ & 4.2$\scinot22.$ & 4.1$\scinot22.$ & 6.7$\scinot22.$ & 9.4$\scinot22.$ & 9.6$\scinot22.$\\
CO           & 5.1$\scinot22.$ & 5.4$\scinot22.$ & 5.5$\scinot22.$ & 2.9$\scinot22.$ & 1.3$\scinot23.$ & 1.3$\scinot23.$\\
CO$_2$       & 1.6$\scinot19.$ & 1.2$\scinot19.$ & 1.2$\scinot19.$ & 9.9$\scinot18.$ & 1.9$\scinot19.$ & 2.3$\scinot19.$\\
HCO          & 1.7$\scinot14.$ & 4.7$\scinot14.$ & 4.8$\scinot14.$ & 2.5$\scinot14.$ & 5.0$\scinot15.$ & 3.4$\scinot15.$\\
H$_2$CO      & 1.0$\scinot16.$ & 1.0$\scinot16.$ & 1.1$\scinot16.$ & 5.6$\scinot15.$ & 2.3$\scinot16.$ & 2.3$\scinot16.$\\
CH$_3$OH     & 1.8$\scinot13.$ & 7.3$\scinot10.$ & 8.7$\scinot12.$ & 6.1$\scinot13.$ & 7.3$\scinot12.$ & 9.4$\scinot12.$\\
N            & 6.4$\scinot14.$ & 2.3$\scinot15.$ & 5.6$\scinot13.$ & 9.8$\scinot14.$ & 7.1$\scinot15.$ & 2.1$\scinot15.$\\
N$_2$        & 7.3$\scinot21.$ & 7.5$\scinot21.$ & 8.2$\scinot21.$ & 4.9$\scinot21.$ & 1.9$\scinot22.$ & 1.9$\scinot22.$\\
NH$_2$       & 8.4$\scinot15.$ & 2.8$\scinot16.$ & 2.5$\scinot15.$ & 9.1$\scinot16.$ & 3.9$\scinot16.$ & 2.2$\scinot16.$\\
NH$_3$       & 1.7$\scinot21.$ & 1.2$\scinot21.$ & 1.0$\scinot20.$ & 4.1$\scinot21.$ & 1.6$\scinot20.$ & 1.8$\scinot20.$\\
HCN          & 1.8$\scinot20.$ & 1.4$\scinot20.$ & 5.0$\scinot18.$ & 2.6$\scinot21.$ & 1.4$\scinot19.$ & 1.3$\scinot19.$\\
H$_2$CN      & 1.3$\scinot13.$ & 1.8$\scinot13.$ & 8.1$\scinot11.$ & 3.6$\scinot14.$ & 5.5$\scinot12.$ & 4.1$\scinot12.$\\
CH$_2$NH     & 1.9$\scinot15.$ & 9.3$\scinot14.$ & 3.9$\scinot13.$ & 1.8$\scinot16.$ & 5.6$\scinot13.$ & 6.4$\scinot13.$\\
CH$_3$CN     & 6.2$\scinot14.$ & 1.6$\scinot14.$ & 1.9$\scinot12.$ & 4.8$\scinot16.$ & 7.7$\scinot11.$ & 1.2$\scinot12.$\\
NO           & 1.7$\scinot14.$ & 4.3$\scinot14.$ & 8.5$\scinot12.$ & 5.2$\scinot14.$ & 7.6$\scinot15.$ & 4.6$\scinot14.$\\
HNCO         & 4.5$\scinot15.$ & 5.0$\scinot15.$ & 4.4$\scinot14.$ & 9.0$\scinot15.$ & 1.1$\scinot15.$ & 1.0$\scinot15.$\\
\enddata
\tablecomments{Model column abundances have units of cm$^{-2}$; $K_{zz}$'s have units of cm$^2$ $\smone$.
``Terminator'' refers to a terminator-average and ``dayside'' refers to a dayside-average temperature 
profile for the planet in question from the GCMs of \citet{showman09}.  Our ``nominal'' $K_{zz}$ profiles 
are shown in Fig.~\ref{eddynom}.}
\end{deluxetable*}

For boundary conditions, we assume zero flux at the top and bottom boundaries so that no mass 
enters or leaves the system.  The lower boundary in our models is at a high-enough temperature 
that thermochemical equilibrium will prevail, and a zero-flux boundary condition is reasonable.  
We also checked cases with a fixed-mixing-ratio lower boundary condition for all constituents, 
set at their equilibrium values, rather than a zero-flux lower boundary condition and found no 
difference in the results.  The upper boundary in our model is at a pressure of a few times
10$^{-5}$ $\mu$bar, which typically corresponds to a planetary radius less than 3 times the 
1-bar radius $R_p$ in exoplanet hydrodynamic models (and in some cases much less than 3$R_p$, 
depending on the model, see \citealt{yelle04}, \citealt{tian05}, \citealt{garcia07}).
At these high altitudes, the hydrodynamic models and model-data comparisons show the atmosphere of 
HD 209458b to be escaping, with an overall mass loss rate that is low enough that exoplanets like 
HD 209458b would lose less than $\sim$1\% of their mass over their lifetime 
\citep[e.g.,][]{lecave04,tian05,murray09}.  As such, we also test cases with an upper escape 
boundary condition for species like atomic H, with a typical upward flux of 
$\sim(6-60)\scinot11.$ cm$^{-2}$ $\smone$ (e.g., 10$^7$--10$^8$ g s$^{-1}$ for HD 
209458b at a radius of 3$R_p$, see \citealt{yelle04}, \citealt{garcia07}, \citealt{tian05}, and 
\citealt{koskinen10}), or we investigate a range of possible upper boundary escape fluxes from 
10$^{10}$ to 10$^{16}$ cm$^{-2}$ $\smone$, but we find that this upper boundary condition has 
no effect on the stratospheric or lower-thermospheric results.  However, we have not included a 
hydrodynamic wind connecting to this escaping region, which is the main reason the escape flux 
has little effect on our model stratosphere.  Hydrodynamic models \citep{yelle04,tian05,garcia07} 
show that the atmosphere should be close to hydrostatic equilibrium below 2-3$R_p$, with the 
base of the planetary wind usually being located near the 1-nbar level (see also the discussion in 
\citealt{koskinen10}).  Thus, escape from a hydrodynamic or thermal wind is not expected to 
substantially influence the composition of the stratosphere, and hydrostatic equilibrium is a 
reasonable assumption in our models.  However, if an atmospheric bulk wind actually dominates 
down to the base of the thermosphere (or deeper) for any particular exoplanet, then hydrodynamic 
flow and atmospheric escape could conceivably affect the stratospheric composition on such 
exoplanets.  

\section{Results}

Our model results suggest that the three main chemical processes --- thermochemical equilibrium, 
transport-induced quenching, and photochemistry --- all operate effectively in the atmospheres of 
HD 189733b and HD 209458b.  Thermochemical equilibrium dominates at pressures greater than a few 
bars, transport-induced quenching can dominate for some species in the $\sim$1--10$^{-3}$ bar 
region, and photochemistry can dominate at pressures less than $\sim$10$^{-3}$ bar, except when 
the stratosphere is very hot.  All three processes combine to influence the 
vertical profiles of the important observable constituents on HD 189733b and HD 209458b.  As such, 
all three processes should be considered in investigations of the chemistry and composition of 
extrasolar giant planets.  Below we describe how the different disequilibrium processes affect the 
results, and we discuss the dominant mechanisms that control the abundances of neutral carbon, oxygen,
and nitrogen species in our thermochemical and photochemical kinetics and transport models.  
The column abundances for some of the interesting species in our models are shown in 
Table~\ref{tabcolumn}, and full model outputs are supplied in the Supplementary Material.

\subsection{Equilibrium vs.~transport-induced quenching vs.~photochemistry\label{sectequilvs}}

Fig.~\ref{equilvs} illustrates how the individual disequilibrium processes affect the mole 
fractions of some species in our dayside-average HD 189733b and HD 209458b models.  The dashed 
curves show the thermochemical-equilibrium solutions.  Note that $\CHfour$ dominates over CO, 
$\HtwoO$ dominates over CO,  and $\NHthree$ dominates over $\Ntwo$ at equilibrium in the deepest 
regions of both atmospheres, but the dominance reverses at higher altitudes such that CO and $\Ntwo$ 
are the dominant carbon- and nitrogen-bearing species in equilibrium at pressures less than 
$\sim$10 bar ({\it i.e.}, over most of the observable portion of the atmosphere; see also 
Fig.~\ref{figtemp}).  At the highest altitudes ({\it i.e.}, in the thermosphere), atomic species 
C, N, and O become the dominant neutral carbon, nitrogen, and oxygen carriers.  
Although $\HtwoO$ remains an important carrier of oxygen at all pressures on these planets, 
thermochemical equilibrium models predict that the mixing ratios of $\CHfour$ and $\NHthree$ will 
drop off dramatically with increasing altitude, such that CO $\gg$ $\CHfour$ and $\Ntwo$ $\gg$ $\NHthree$ 
in terms of the column abundance at 1 bar and above.  Thermochemical 
equilibrium predicts an even smaller column abundance for species like HCN and $\CtwoHtwo$, 
such that these species would be at most minor constituents on HD 189733b and HD 209458b if 
their atmospheres were in equilibrium.  

The situation changes significantly when vertical transport is considered.  The dotted curves in 
Fig.~\ref{equilvs} show the results of models in which thermochemical kinetics and vertical 
transport are operating but in which the planet receives no photolyzing ultraviolet radiation 
from the host star, {\it i.e.}, so that we can examine the influence of transport-induced quenching 
separately from that of photochemistry.  At deep pressure levels in these atmospheres, temperatures 
are high enough that energy barriers to the kinetic reactions can be overcome; reactions proceed in 
both the forward and reverse directions, and equilibrium is maintained.  Our kinetics models, which 
have fully reversed reactions, reproduce equilibrium predictions in these high-temperature, 
high-pressure regions.  However, as a gas parcel is lifted to cooler, higher altitudes, energy 
barriers can begin to become insurmountable over typical atmospheric residence time scales, and 
exothermic ``forward'' reactions can become favored over their endothermic reverses.  When the 
transport time scale drops below the chemical kinetic conversion time scale between different 
molecular species, the species can become quenched such that their mole fractions become ``frozen in'' 
at values representative of the quench point ({\it i.e.}, where the two time constants are equal).  
\citet{prinn77} first developed this concept analytically to explain the unexpectedly large CO 
abundance in Jupiter's upper troposphere.  Based on CO and other observations, transport-induced 
quenching appears to be ubiquitous in substellar objects, operating on our solar-system giant planets
\citep[e.g.,][]{prinn77,lewis80,prinn81,lewis84,fegley85,fegley94,lodders02,bezard02,viss05,viss10b,moses10},
on brown dwarfs 
\citep[e.g.,][]{fegley96,noll97,griffith99,saumon00,saumon03,saumon06,leggett07,hubeny07,geballe09,yamamura10,king10,viss11},
and most likely on extrasolar planets, as well (e.g., \citealt{coop06,fort06a,burrows08,line10,madhu11} 
and references therein).  Although both CO and $\CHfour$ will quench when the kinetic interconversion
between the two species becomes inefficient, the quenching is more obvious for the species that is not
expected to be abundant in equilibrium ({\it i.e.}, CO on the cooler giant planets and brown dwarfs;
$\CHfour$ on the warmer, highly irradiated, close-in transiting giant planets).

\begin{figure*}
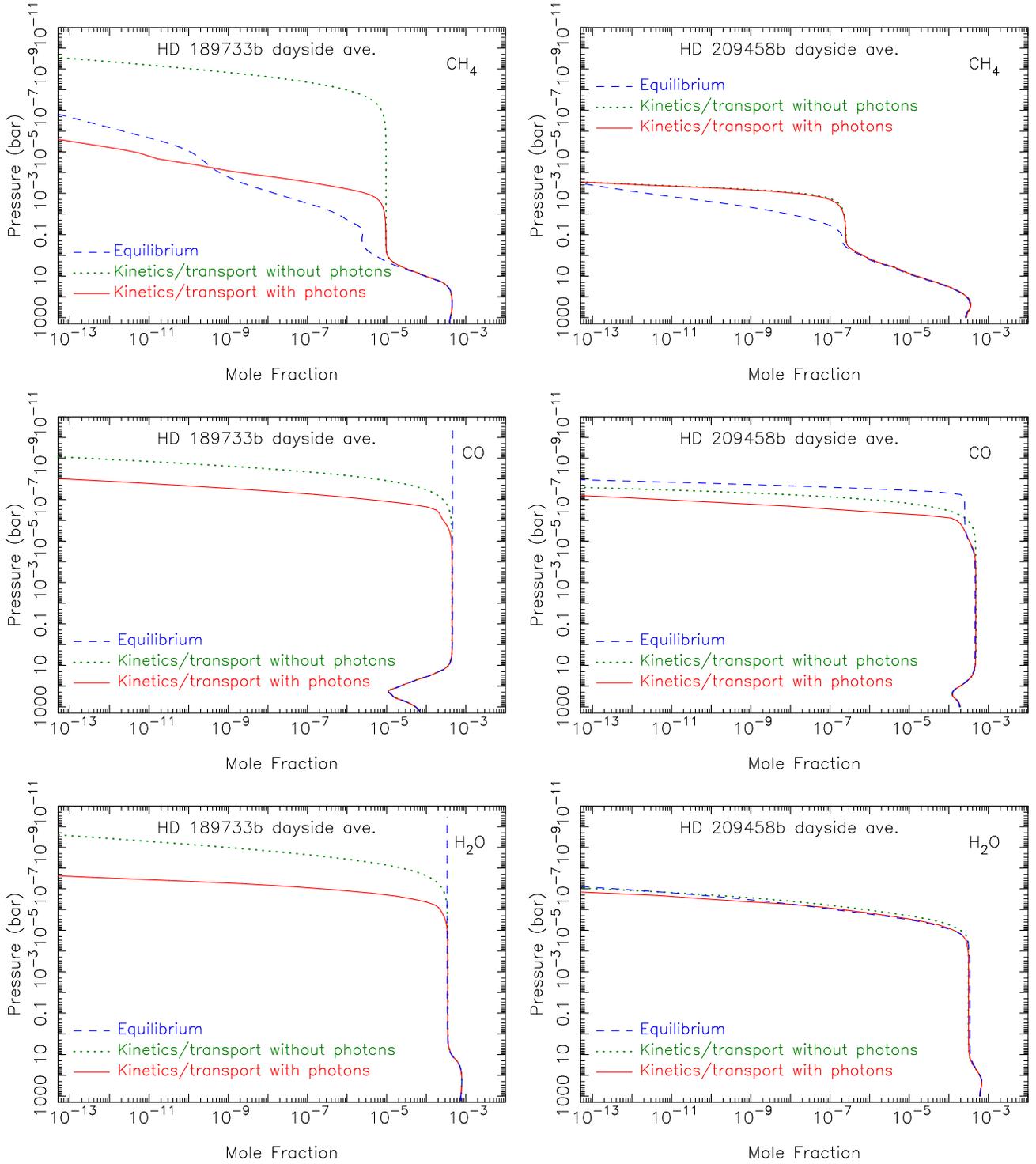

\begin{tabular}{ll}
{\includegraphics[angle=-90,clip=t,scale=0.37]{fig3a_color.ps}} 
& 
{\includegraphics[angle=-90,clip=t,scale=0.37]{fig3b_color.ps}} 
\\
{\includegraphics[angle=-90,clip=t,scale=0.37]{fig3c_color.ps}} 
& 
{\includegraphics[angle=-90,clip=t,scale=0.37]{fig3d_color.ps}} 
\\
{\includegraphics[angle=-90,clip=t,scale=0.37]{fig3e_color.ps}} 
& 
{\includegraphics[angle=-90,clip=t,scale=0.37]{fig3f_color.ps}} 
\\
\end{tabular}
\caption{Mole fractions for several species in our models for HD 189733b (left) and 
HD 209458b (right) for different assumptions about the chemistry and transport.  All 
HD 189733b models assume a projected-area-weighted dayside average thermal structure 
with an isothermal extension at high altitudes ({\it i.e.}, with a cold thermosphere), 
and all HD 209458b models assume a dayside-average thermal structure with a hot thermosphere 
(see Fig.~\ref{figtemp} for the temperature profiles 
for these models).  The dashed lines are for a purely thermochemical equilibrium model 
(no kinetics or transport), the dotted lines are for a model that has thermochemical 
kinetics and transport but no incident ultraviolet photons ({\it i.e.}, no photochemistry), 
and the solid lines represent our full model with thermochemical and photochemical 
kinetics and transport all operating.  The eddy diffusion coefficient $K_{zz}$ is fixed 
at a constant value of 10$^9$ cm$^2$ $\smone$ in the transport models.  A color version 
of this figure is available in the online journal.\label{equilvs}}
\end{figure*}

\begin{figure*}
\begin{tabular}{ll}
{\includegraphics[angle=-90,clip=t,scale=0.37]{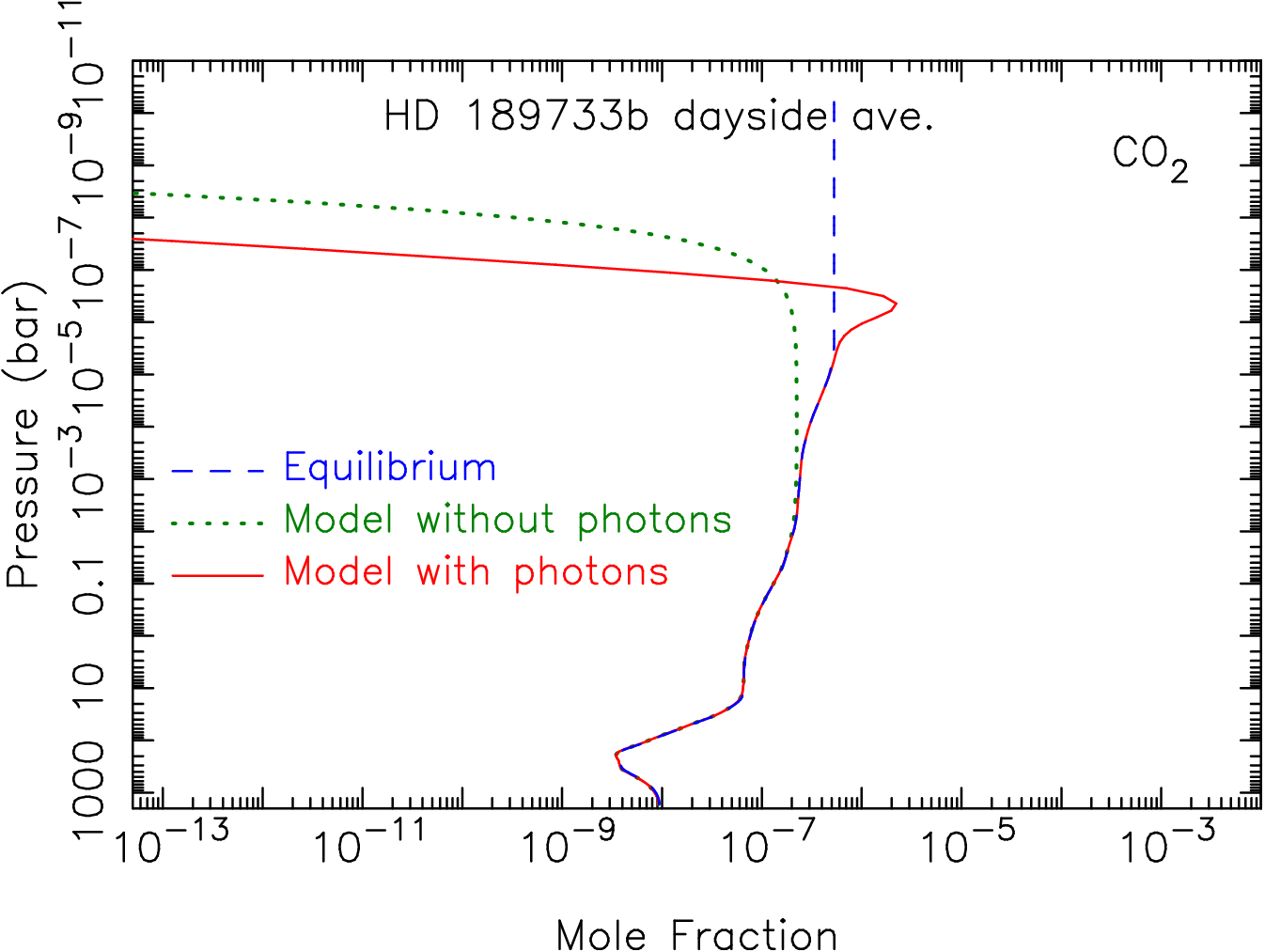}} 
& 
{\includegraphics[angle=-90,clip=t,scale=0.37]{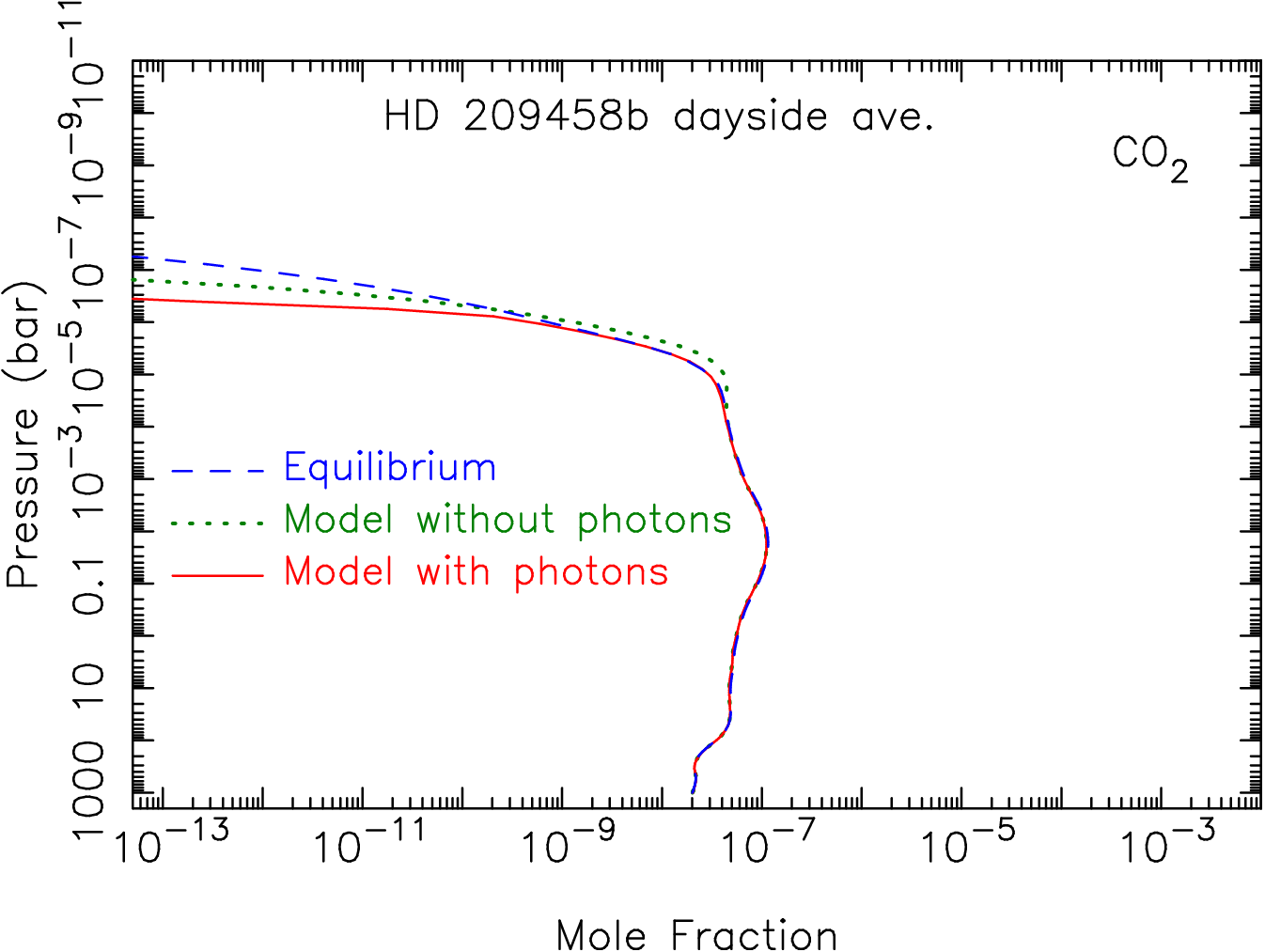}} 
\\
{\includegraphics[angle=-90,clip=t,scale=0.37]{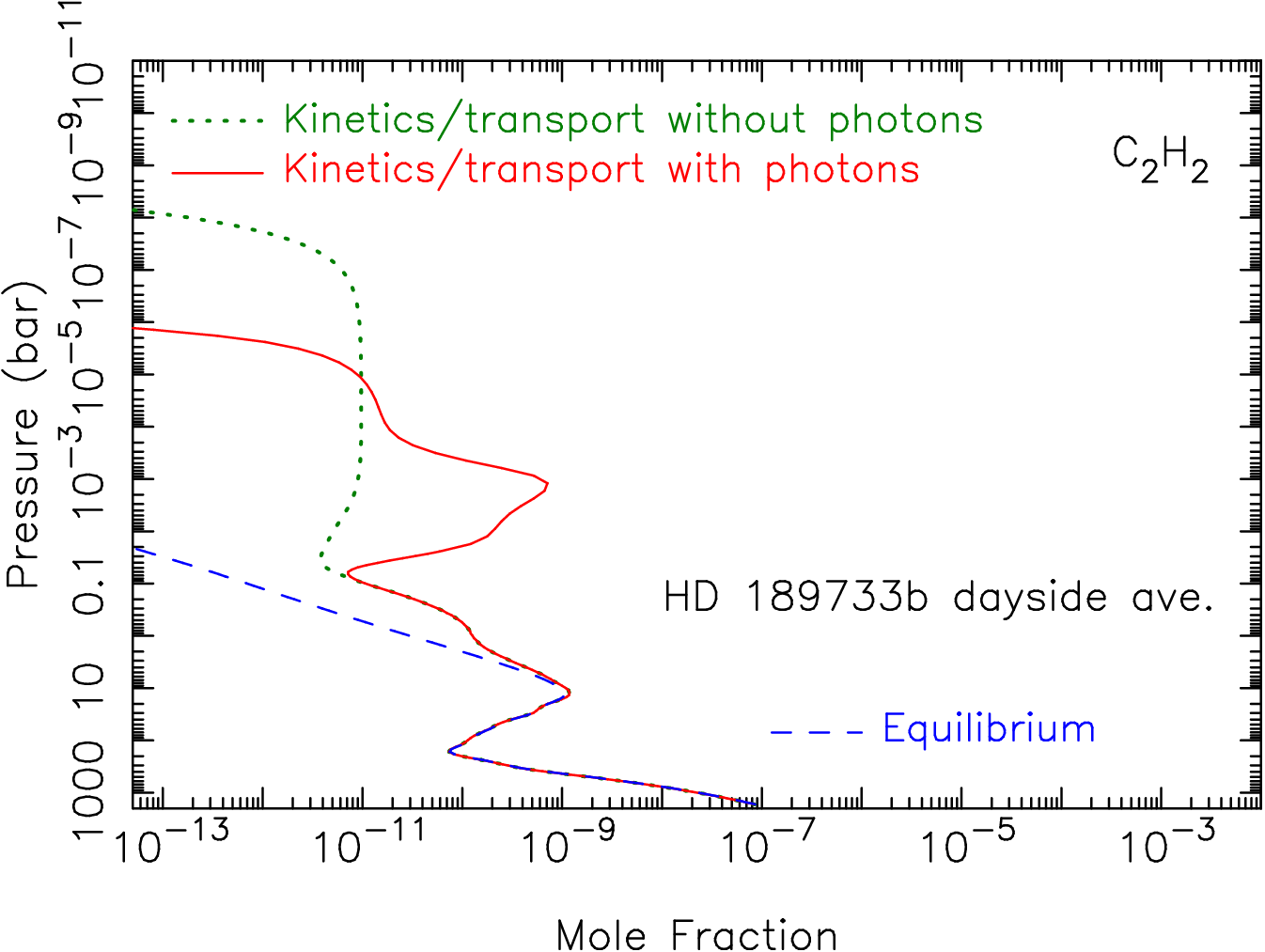}} 
& 
{\includegraphics[angle=-90,clip=t,scale=0.37]{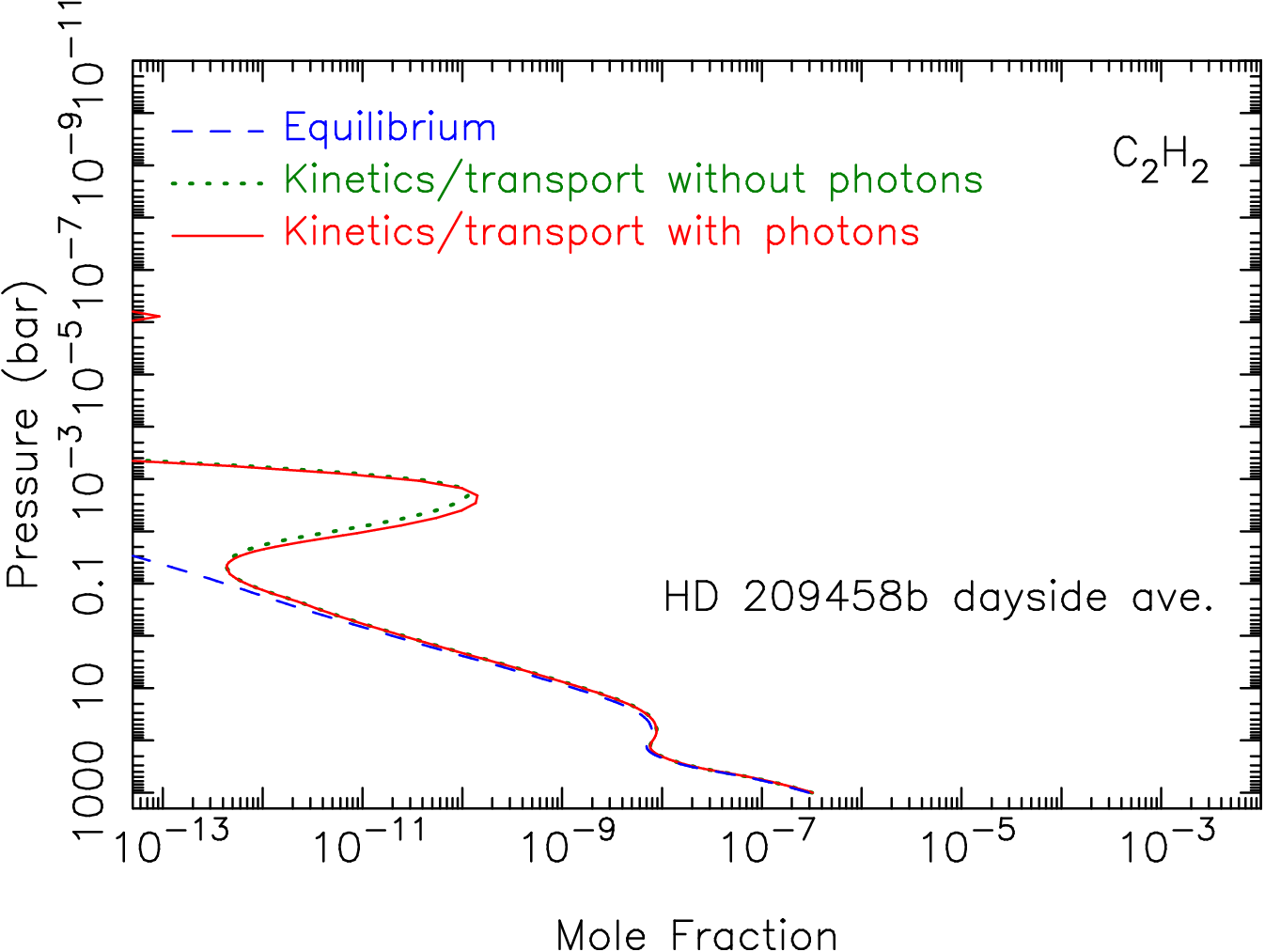}} 
\\
{\includegraphics[angle=-90,clip=t,scale=0.37]{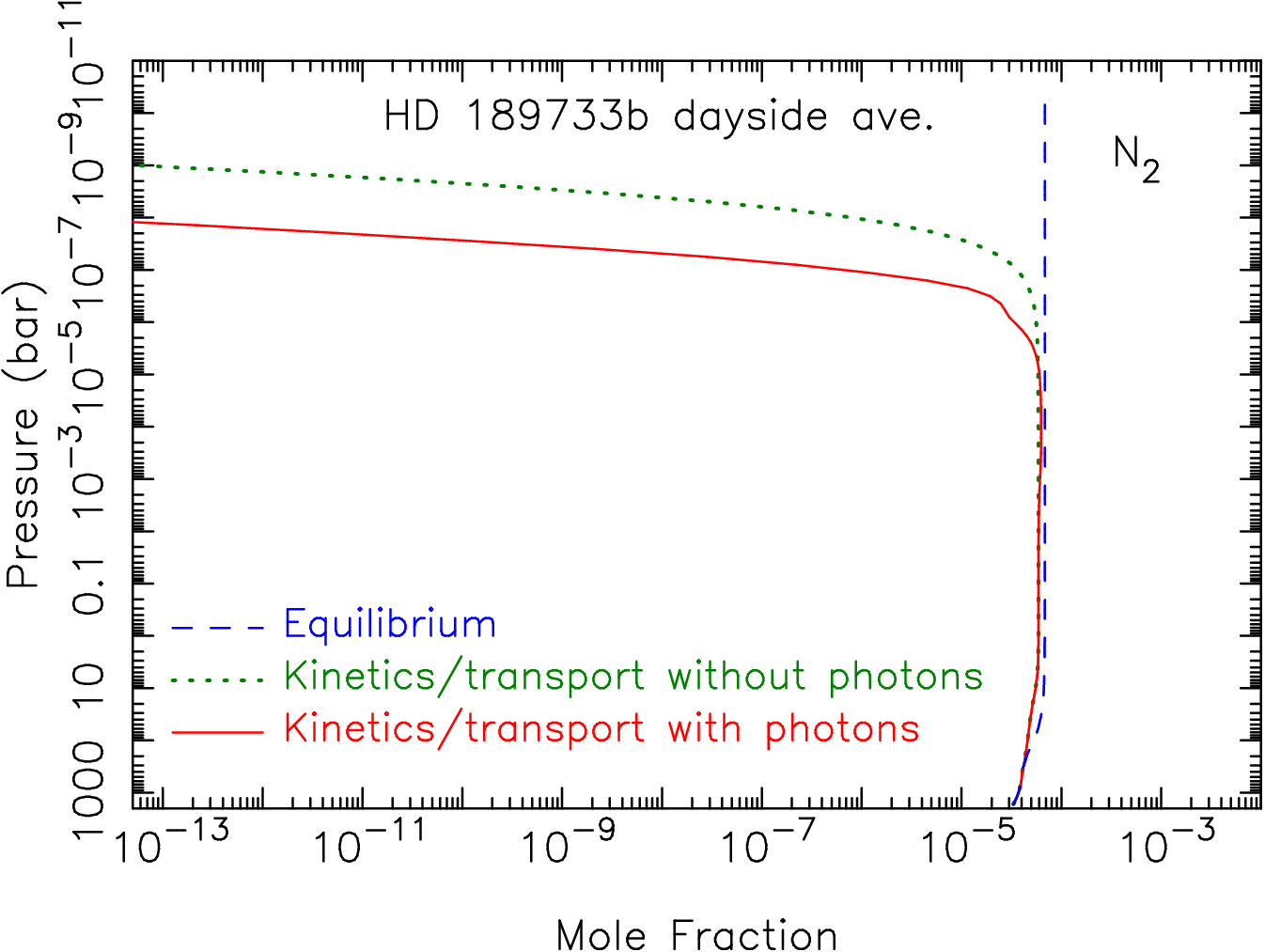}} 
& 
{\includegraphics[angle=-90,clip=t,scale=0.37]{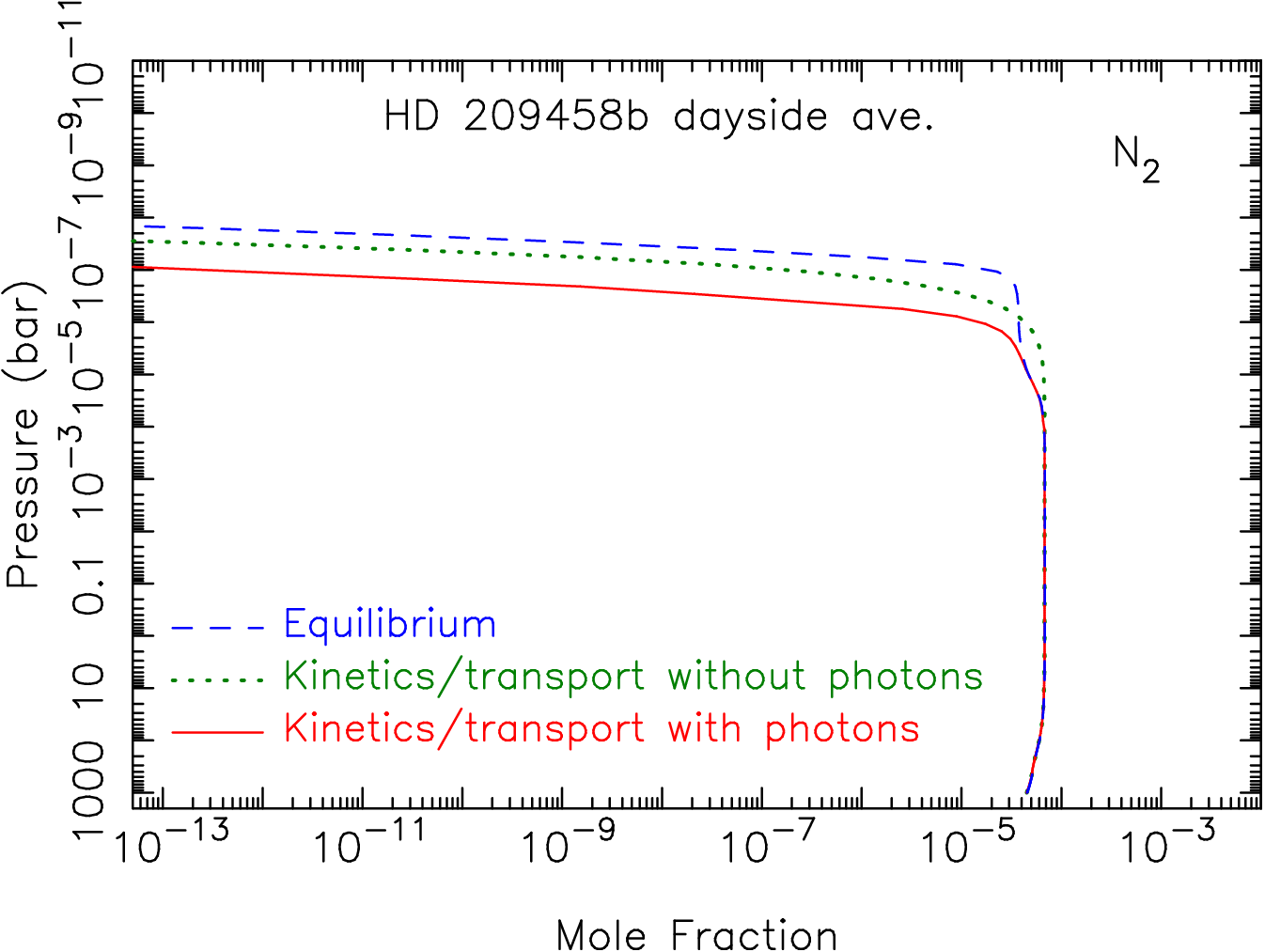}} 
\\
\end{tabular}
\vskip20pt
\noindent{Fig.~\ref{equilvs}. ({\it continued}).}
\end{figure*}

\begin{figure*}
\begin{tabular}{ll}
{\includegraphics[angle=-90,clip=t,scale=0.37]{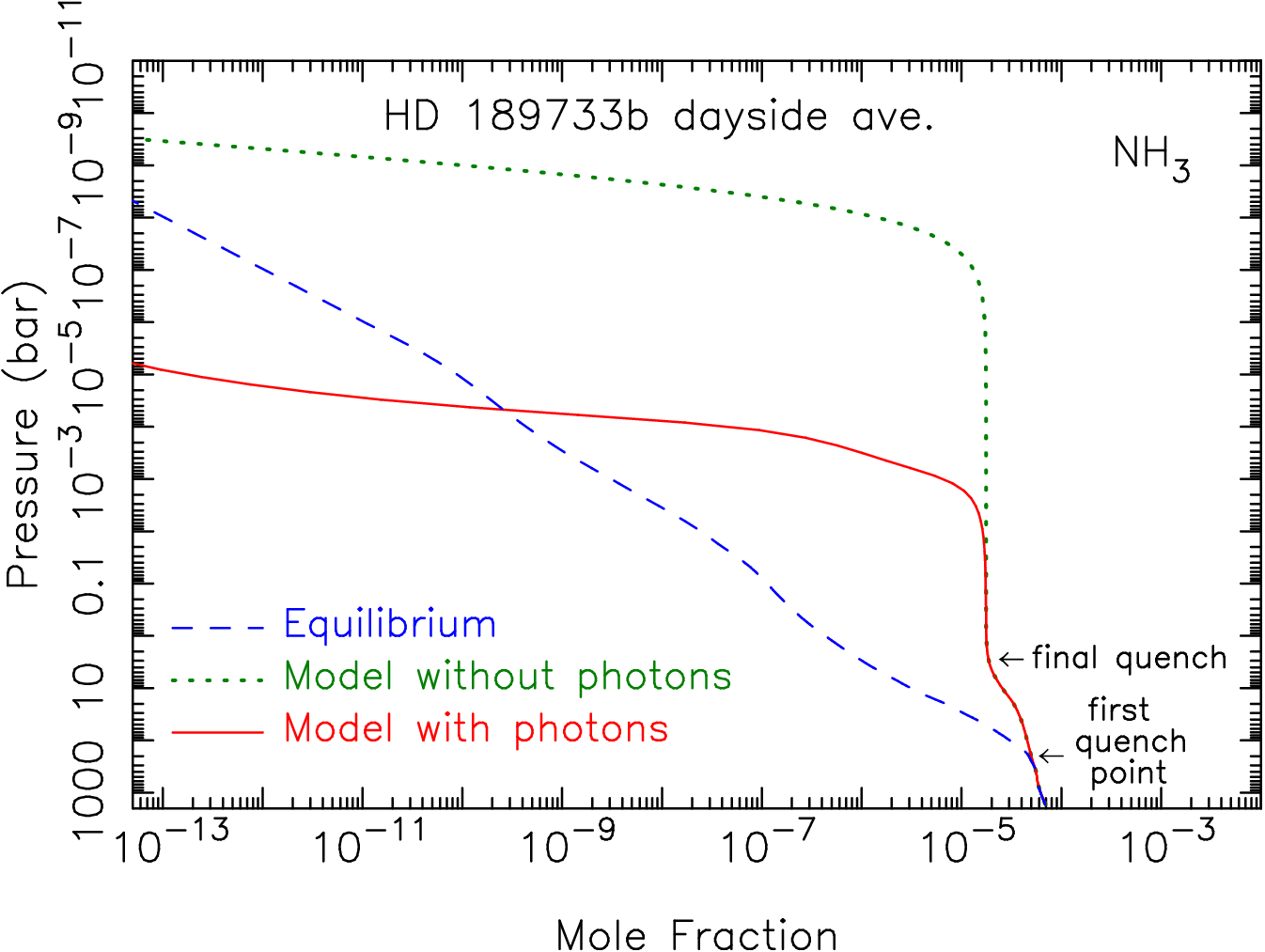}} 
& 
{\includegraphics[angle=-90,clip=t,scale=0.37]{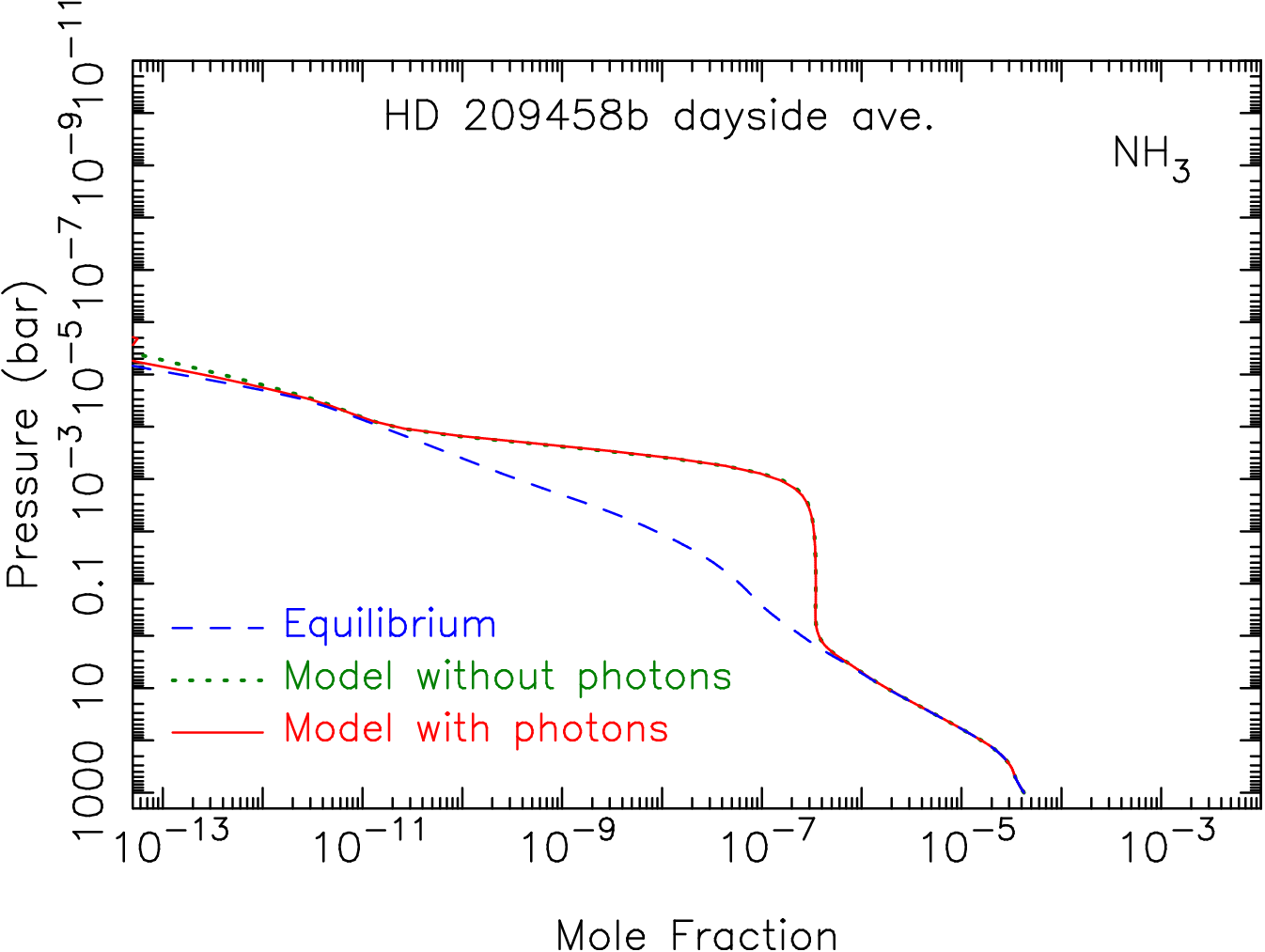}} 
\\
{\includegraphics[angle=-90,clip=t,scale=0.37]{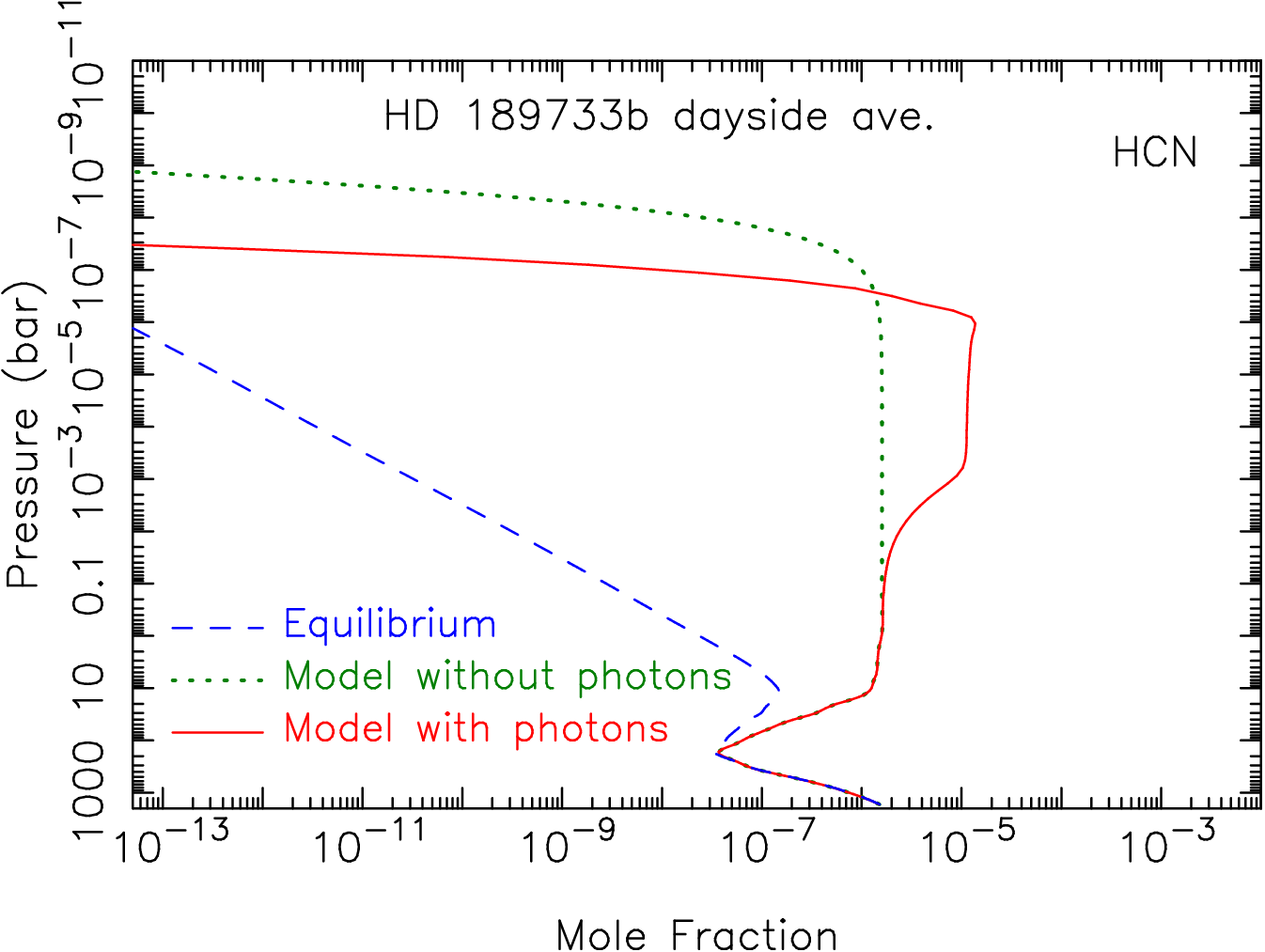}} 
& 
{\includegraphics[angle=-90,clip=t,scale=0.37]{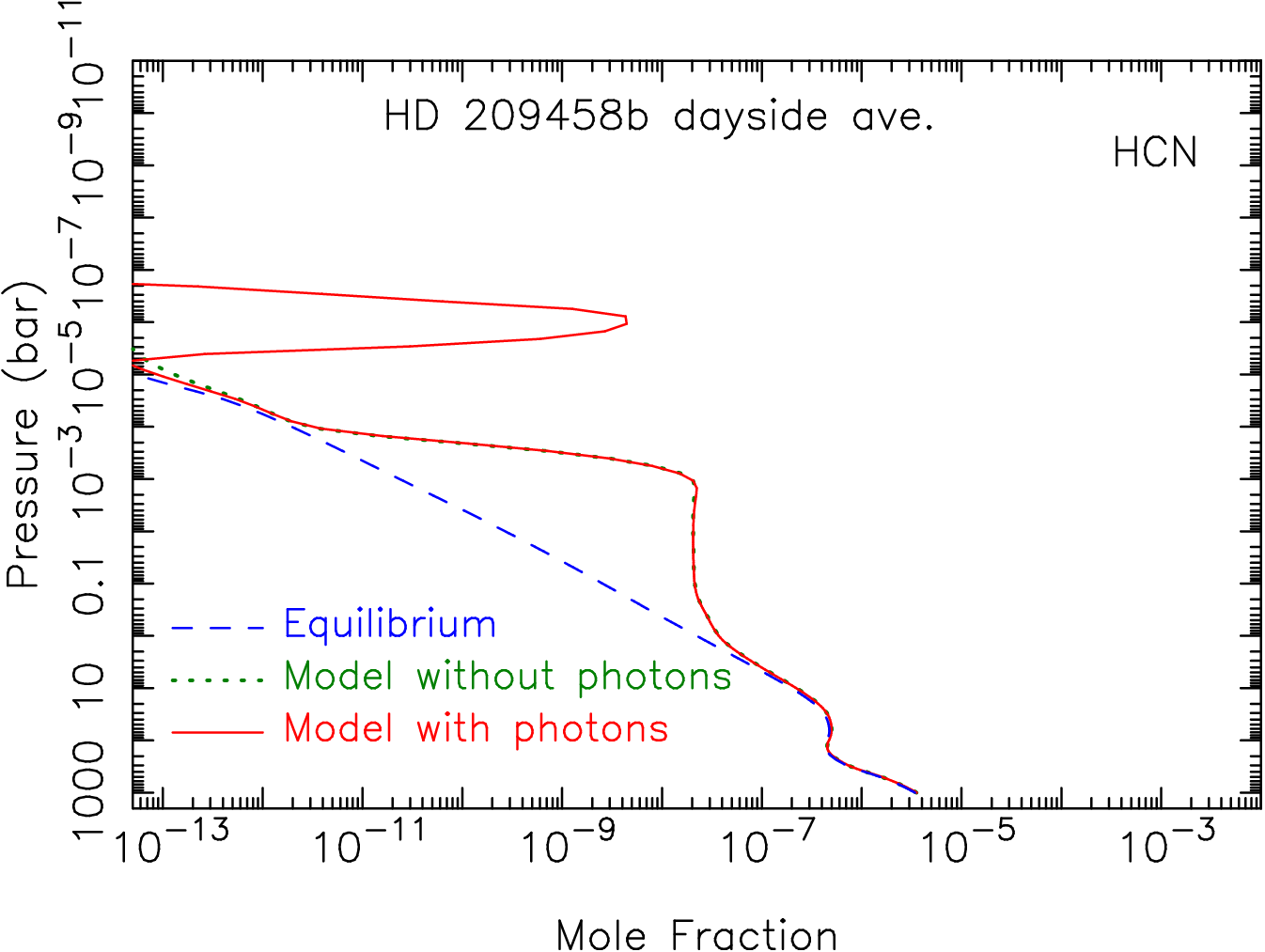}} 
\\
\end{tabular}
\vskip20pt
\noindent{Fig.~\ref{equilvs}. ({\it continued}).}
\end{figure*}

The dotted curves in Fig.~\ref{equilvs} show that transport-induced quenching can greatly 
increase the predicted abundances of several key species, including $\CHfour$, $\NHthree$, and 
HCN, over expectations based on thermochemical equilibrium (dashed curves).  The quench point 
for a major species like $\CHfour$ is caused by the breakdown of efficient kinetic interconversion 
between itself and another major species (CO in this case) and results in an obvious departure 
from the predicted equilibrium curve.  Once an abundant species like $\CHfour$ becomes quenched, 
its quenched abundance can affect other minor species like C$_2$H$_x$ that might not 
have reached their own quench points yet; these species may continue to react with the $\CHfour$ 
at higher altitudes in an equilibrium with the quenched $\CHfour$ until they too reach their own 
quench points.  Thus, minor species may depart from equilibrium when a major parent 
molecule quenches, but their mole fractions may not become constant until they themselves quench 
at higher, cooler altitudes.  As such, the mole fractions from the full kinetics and transport model 
can have complicated behavior that is not easily predicted from simple time-constant arguments 
using a single rate-limiting kinetic reaction.  Hydrogen cyanide (HCN), for example, maintains an 
equilibrium with $\Htwo$ and with the quenched abundances of $\CHfour$ and $\NHthree$ until its own 
quench point when the interconversion of $\NHthree$-$\Ntwo$-$\CHfour$ shuts down, and the resulting 
mole-fraction profile is complicated.  In addition, the quenching may not be abrupt, which can 
affect the mole-fraction profiles and make abundances more difficult to predict analytically.  
Ammonia, for example, has two quench points in the HD 189733b model.  Ammonia first diverges from 
the equilibrium curve at the $\Ntwo$-$\NHthree$ interconversion quench point, which is at relatively 
high temperatures and pressures due to the strong N$\tbond$N triple bond.  However, there is a 
pressure and temperature dependence to the different chemical pathways that allow $\NHthree$ 
$\leftrightarrow$ $\Ntwo$ interconversion, and different reactions begin to dominate at different 
pressure levels.  More importantly, the temperatures in the HD 189733b models do not drop off 
significantly between the ``initial'' $\NHthree$ quench point near $\sim$200 bar and the ``final'' 
quench point at the few-bar level, allowing the interconversion reactions to continue to operate 
at a reduced but still effective rate in the intervening region.  The final $\NHthree$ quench occurs 
when the thermal gradients become large again and the temperatures become low enough that the
$\NHthree$-$\Ntwo$ interconversion reactions cease to become important.  Ammonia quenching then operates over
an extended range on HD 189733b, as marked by the two quench-point arrows in the HD 189733b $\NHthree$ 
panel in Fig.~\ref{equilvs}.  The thermal gradients are larger in the region where $\NHthree$ initially 
quenches on HD 209458b, allowing chemical conversion reactions to shut down more abruptly, with a 
better-defined single quench level for $\NHthree$ in the HD 209458b models.

The details of the particular quenching mechanisms, and their sensitivity to temperature and 
transport assumptions, will be discussed below.  At this point, we simply want to illustrate 
that transport-induced quenching can be an important process on extrasolar giant planets that
can allow disequilibrium species to be carried to relatively high altitudes to potentially 
affect spectral behavior.  Note that transport-induced quenching is much more important on 
the cooler HD 189733b than on HD 209458b.  Although Fig.~\ref{equilvs} shows that quenching 
does occur in our dayside HD 209458b kinetics/transport models, the effectiveness of 
transport-induced quenching on HD 209458b is limited by the presence of a strong stratospheric 
thermal inversion ---  as temperatures become hot enough in the stratosphere, thermochemical 
kinetics drives the composition back toward equilibrium.  When the inversion is weak or absent 
on HD 209458b ({\it i.e.}, at the terminators or nightside), transport-induced quenching becomes 
more important.  

The stratospheric 
thermal inversion also affects the height to which the molecules can be carried.  The models for 
both planets in Fig.~\ref{equilvs} have the same eddy diffusion coefficient, yet a comparison 
of the dotted-line models for both planets shows that the species reach different pressure 
levels on each planet.  The difference is only partly due to the higher temperatures creating 
a lower homopause level on HD 209458b (where the term ``homopause'' refers to the pressure 
level at which the molecular and eddy diffusion coefficient are equal; above the homopause, 
molecular diffusion acts to limit the abundance of heavier species, whereas below the homopause 
the atmosphere can be well mixed).  This point is demonstrated by a close comparison of CO and 
$\Ntwo$ on each planet, as these thermochemically stable species reach their homopause levels, 
yet they are carried to only slightly different pressures on each planet.  The main cause of the 
limited penetration of molecular species to high altitudes on dayside HD 209458b is the high 
stratospheric temperatures --- species like $\CHfour$ and $\NHthree$ cannot survive the 
thermochemical kinetics at these temperatures and are converted in to atomic C, N, 
and H, as well as to the more stable molecular species CO and $\Ntwo$.  

Another interesting point to note from Fig.~\ref{equilvs} is that some molecules like CO, 
$\HtwoO$, $\Ntwo$, and $\COtwo$ are relatively unaffected by transport-induced quenching on 
either planet.  Although CO, $\HtwoO$, and $\Ntwo$ technically also quench in the thermochemical 
kinetics and transport models, they are already the dominant species at their quench points, 
and the quenching does not greatly affect their abundances.  As we will show below, this result 
is somewhat dependent on assumptions about the eddy diffusion coefficient and temperature 
structure.  On HD 189733b, CO$_2$ quenches eventually (see where dotted curve diverges from 
dashed curve), but only at very high altitudes, and its abundance follows equilibrium at the 
pressure levels at which the observations are the most sensitive ($\sim$1--10$^{-4}$ bar, see 
\citealt{fort08a}).  Other species like hydrocarbons, nitriles, and most nitrogen-bearing 
species are strongly affected by transport-induced quenching.

A comparison of the solid curves with the dotted curves in Fig.~\ref{equilvs} shows the influence of 
photochemistry.  When ultraviolet photons are present, molecules like $\HtwoO$, CO, $\Ntwo$, and 
$\NHthree$ are photolyzed, initiating a slew of photochemical reactions that alter the chemistry and 
composition.  As discussed by \citet{liang03}, one main result of the photochemistry in the atmospheres 
of close-in extrasolar giant planets is the production of huge quantities of atomic hydrogen from 
photolysis of $\HtwoO$, followed by catalytic destruction of $\Htwo$ from the reaction of molecular 
hydrogen with the photolytic product OH.  Other similar catalytic cycles are initiated deeper in the 
atmosphere from $\NHthree$ photolysis and from reaction of H with $\CHfour$ \citep[see
also][]{zahnle11}.  

Because many molecules are vulnerable to attack by atomic hydrogen at these temperatures, the production 
of atomic H sets the stage for the subsequent photochemistry.  Some molecules like $\HtwoO$, $\Ntwo$, and 
CO are efficiently recycled and are relatively stable against photochemistry.  As can be seen 
from Fig.~\ref{equilvs}, photochemistry only affects the profiles of these species at very high altitudes, 
where visible and infrared transit and eclipse observations have little sensitivity.  Other molecules 
like $\CHfour$ and $\NHthree$ have weaker bonds and/or are more reactive and less efficiently recycled. 
These molecules can be removed from portions of the visible atmosphere, to be replaced 
by photochemical products.  The big winners in terms of photochemistry are atomic species such as H (by 
far the dominant photochemical product), C, O, and N; some radicals like $\CHthree$, $\NHtwo$, and OH; 
unsaturated hydrocarbons like $\CtwoHtwo$; and some nitriles like HCN.  Note that photochemistry 
(and thermochemistry, if the atmosphere is warm) can keep molecular species from reaching their 
homopause levels, as species can become kinetically destroyed faster than eddy diffusion can 
transport them to high altitudes.  Below we discuss the kinetic mechanisms in more detail.

\subsection{Quench kinetics\label{sectquench}}

As shown in Fig.~\ref{equilvs} and discussed above, transport-induced quenching is an important 
process in the atmospheres of extrasolar giant planets.  For relatively warm planets like 
HD 189733b and HD 209458b, the thermal structure lies within the CO and $\Ntwo$ stability fields 
at the temperatures and pressures where quenching occurs, so that the less abundant $\CHfour$ and 
$\NHthree$ are strongly affected by quenching.  This situation differs from that of cooler planets 
like GJ 436b (which is in the $\CHfour$ and $\Ntwo$ stability fields, so that quenching of CO and 
$\NHthree$ can be important) or from cold planets like Jupiter (which is in the $\CHfour$ and 
$\NHthree$ stability fields, so that quenching of CO and $\Ntwo$ are important) or from brown 
dwarfs like Gliese 229b whose thermal structure crosses the CO = $\CHfour$ and $\Ntwo$ = $\NHthree$ 
boundaries in such a location that quenching of all four molecules --- CO, $\CHfour$, $\Ntwo$, and 
$\NHthree$ --- could be important.

The kinetic pathways leading to the quenching of CO-$\CHfour$ interconversion, $\Ntwo$-$\NHthree$ 
interconversion, and $\NHthree$-HCN-$\CHfour$ interconversion have been discussed in \citet{viss10b} 
and \citet{moses10}.  We have made several updates to the rate coefficients used in those studies, 
including adopting the $\CHthree$ + OH reaction pathways and rate coefficients of \citet{jasper07}, 
and updating the rate coefficients for the nitrogen reactions based on \citet{klipp09} and 
\citet{klipp10}.  Because of uncertainties in the kinetic rate coefficients for certain key reactions 
in the CO $\leftrightarrow$ $\CHfour$ interconversion scheme \citep[see][]{viss10b}, we have taken a 
more detailed look at the $\CHthreeOH$ + H reaction and its various pathways for this paper.  In 
particular, we have used {\it ab initio}\/ transition-state theory to predict the rate coefficients 
for the three primary channels of the reaction of H with $\CHthreeOH$:
\begin{eqnarray*}
{\rm R864} \ \ \ \ \ & \CHthreeOH \, + \, \H  & \rightarrow \, \CHtwoOH \, + \, \Htwo \\
{\rm R862} \ \ \ \ \ & \CHthreeOH \, + \, \H  & \rightarrow \, \CHthreeO \, + \, \Htwo \\
{\rm R860} \ \ \ \ \ & \CHthreeOH \, + \, \H  & \rightarrow \, \CHthree \, + \, \HtwoO  . \\
\end{eqnarray*}
We used the QCISD(T)/cc-pVTZ method \citep[e.g.][]{pople87,dunning89} to predict the rovibrational 
properties of the stationary points.  Higher-level energies were obtained from basis-set 
extrapolation of explicit QCISD(T) calculations for the cc-pVTZ and cc-pVQZ basis sets.  
These QCISD(T)/CBS//QCISD(T)/cc-pVTZ 
calculations were performed with the spin-restricted formalism using the MOLPRO software 
package.\footnote{MOLPRO is a package of {\it ab initio} programs written by H.-J.~Werner, 
P. J.~Knowles, F. R.~Manby, M.~Sch\"utz, P.~Celani, G.~Knizia, T.~Korona, R.~Lindh, 
A.~Mitrushenkov, G.~Rauhut, T. B.~Adler, R. D.~Amos, A.~Bernhardsson, A.~Berning, D. L.~Cooper, 
M. J. O.~Deegan, A. J.~Dobbyn, F.~Eckert, E.~Goll, C.~Hampel, A.~Hesselmann, G.~Hetzer, 
T.~Hrenar, G.~Jansen, C.~K\"oppl, Y.~Liu, A. W.~Lloyd, R. A.~Mata, A. J.~May, S. J.~McNicholas, 
W.~Meyer, M. E.~Mura, A.~Nickla\ss, P.~Palmieri, K.~Pfl\"uger, R.~Pitzer, M.~Reiher, T.~Shiozaki, 
H.~Stoll, A. J.~Stone, R.~Tarroni, T.~Thorsteinsson, M.~Wang, and A.~Wolf.} 
Zero-point corrected barrier heights of 11.0, 13.4, and 24.2 kcal/mol were obtained 
for channels R864, R862, and R860, respectively.  Most of the modes were treated as rigid-rotor 
harmonic oscillators, but hindered-rotor torsional and asymmetric Eckart-tunneling corrections 
were included as appropriate. 

The predicted rate coefficients for reactions R864-R860 are well reproduced over the 500 to 2500 
K region by the modified Arrhenius expressions, $k_{864}$ = 1.09$\scinot-19. \, T^{2.728} \,
\exp{(-2240/T)}$, $k_{862}$ = 6.82$\scinot-20. \, T^{2.658} \, \exp{(-4643/T)}$, and $k_{860}$ 
= 4.91$\scinot-19. \,  T^{2.485} \, \exp{(-10380/T)}$, cm$^3$ molecule$^{-1}$ s$^{-1}$.   
These predictions are expected to be accurate to within a factor of two.  For channels 
R864 and R862, the predictions are within a factor of 1.35 of the G2-based estimates from 
\citet{jodkowski99} but have lower uncertainties due to the use of higher levels of 
electronic-structure theory.  The more recent predictions of \citet{carvalho08} are 
closely related to the present ones, being only about 0.8 times lower.  The only prior 
study of channel R860 appears to be the G2-based study of \citet{lendvay97}, who did not 
provide explicit expressions for this rate coefficient.  The absence of both theoretical 
and experimental data for channel R860 was the motivating factor for the present {\it ab 
initio}\/ transition-state theory analysis, as \citet{viss10b} have suggested that 
this pathway could be important in the overall interconversion scheme between $\CHfour$ and 
CO on giant planets.

The above {\it ab initio}\/ calculations indicate, however, that the reaction R860 ($\CHthreeOH$ + H 
$\rightarrow$ $\CHthree$ + $\HtwoO$) is too slow to participate in the dominant $\CHfour$-CO 
quenching scheme in our HD 189733b and HD 209458b models.  Instead, we find that the dominant 
reaction scheme controlling $\CHfour$ $\rightarrow$ CO conversion and methane quenching depends on 
the eddy diffusion coefficient and pressure-temperature conditions at the quench point; the dominant 
scheme can even bypass methanol ($\CHthreeOH$) altogether in some cases.  For smaller eddy diffusion 
coefficients, the quench level is at lower pressures (higher altitudes), and the dominant $\CHfour$-CO 
conversion scheme becomes 
\begin{eqnarray}
\H \, + \, \CHfour \, & \rightarrow & \, \CHthree \, + \, \Htwo \nonumber \\
\H \, + \, \HtwoO \, & \rightarrow & \, \OH \, + \, \Htwo \nonumber \\
\OH \, + \, \CHthree \, & \rightarrow & \, \CHtwoOH \, + \, \H \nonumber \\
\CHtwoOH \, + \, \M \, & \rightarrow & \, \H \, + \, \HtwoCO \, + \, \M \nonumber \\
\HtwoCO \, + \, \H \, & \rightarrow & \, \HCO \, + \, \Htwo \nonumber \\
\HCO \, + \, \M \, & \rightarrow & \, \H \, + \, \CO \, + \, \M \nonumber \\
\noalign{\vglue -10pt}
\multispan3\hrulefill \nonumber \cr
\Net \ \ \CHfour \, + \, \HtwoO \, & \rightarrow & \, \CO \, + \, 3\,\Htwo  , \\
\end{eqnarray}
where M represents any third body.  For larger eddy diffusion coefficients, the quench level is at 
higher pressures (lower altitudes), and the dominant $\CHfour$-CO conversion scheme can become
\begin{eqnarray}
\H \, + \, \CHfour \, & \rightarrow & \, \CHthree \, + \, \Htwo \nonumber \\
\H \, + \, \HtwoO \, & \rightarrow & \, \OH \, + \, \Htwo \nonumber \\
\OH \, + \, \CHthree \, + \, \M & \rightarrow & \, \CHthreeOH \, + \, \M \nonumber \\
\CHthreeOH \, + \, \H \, & \rightarrow & \, \CHtwoOH \, + \, \Htwo \nonumber \\
\CHtwoOH \, + \, \M \, & \rightarrow & \, \H \, + \, \HtwoCO \, + \, \M \nonumber \\
\HtwoCO \, + \, \H \, & \rightarrow & \, \HCO \, + \, \Htwo \nonumber \\
\HCO \, + \, \M \, & \rightarrow & \, \H \, + \, \CO \, + \, \M \nonumber \\
\Htwo \, + \, \M \, & \rightarrow & \, 2\, \H \, + \, \M \nonumber \\
\noalign{\vglue -10pt}
\multispan3\hrulefill \nonumber \cr
\Net \ \ \CHfour \, + \, \HtwoO \, & \rightarrow & \, \CO \, + \, 3\,\Htwo  . \\
\end{eqnarray}

The rate-limiting step for the $\CHfour$ $\rightarrow$ CO conversion process is the slowest reaction 
in the fastest overall conversion scheme.  For the above scheme (2), the rate-limiting reaction is 
OH + $\CHthree$ $\rightarrow$ $\CHtwoOH$ + H (reaction number R670 in our list), and for the above
scheme (3), the rate-limiting reaction is OH + $\CHthree$ + M $\rightarrow$ $\CHthreeOH$ + M (reaction 
R674).  The relative importance of schemes (2) and (3) depends on which of these OH + $\CHthree$ 
pathways is faster for the pressure-temperature conditions at the quench level for the planet under 
consideration.  We adopt rate-coefficient expressions for these two potential rate-limiting 
reactions from \citet{jasper07}, who provide a thorough description of the pressure and temperature 
dependence.  Both schemes typically contribute some nontrivial fraction of the rate of interconversion 
in our models, so considering both pathways in a combined rate-limiting step can furnish a good 
description of the methane mole-fraction quenching behavior on transiting exoplanets \citep{viss11}.  

Note that although the C-H-O system has been well studied due to applications in combustion chemistry 
and terrestrial atmospheric chemistry \citep[e.g.,][]{baulch05,atkin06}, the kinetics of CO 
$\leftrightarrow$ $\CHfour$ interconversion in reducing environments remains somewhat uncertain.  
The rate coefficients for the above rate-limiting reactions are uncertain by perhaps a factor of 3,
and other alternative schemes may be competing if other rate coefficients in our full list 
are in error.  The methane mole fraction on HD 189733b and HD 209458b will depend critically on 
the kinetic rate coefficients adopted in the model, particularly for the reactions that form C-O 
bonds from species with O-H and C-H bonds.  The abundance of quenched species is always highly 
dependent on the assumed kinetic rate coefficients, as well as on the assumed strength of atmospheric 
mixing.

Despite these uncertainties, the OH + $\CHthree$ + M $\rightarrow$ $\CHthreeOH$ + M and the OH + 
$\CHthree$ $\rightarrow$ $\CHtwoOH$ + H reactions are much more likely to be the rate-limiting step 
in the conversion of $\CHfour$ to CO on exoplanets than the reaction OH + $\CHthree$ $\rightarrow$ 
$\Htwo$ + $\HtwoCO$ ({\it i.e.}, the reverse of the CO quench reaction originally proposed by 
\citealt{prinn77}), as the latter reaction is a few orders of magnitude too slow to be a likely 
player in any dominant conversion scheme \citep[see][]{dean87,yung88,griffith99,bezard02,jasper07,viss10b}.  
Note also that the dominance of $\CHtwoOH$ over $\CHthreeO$ as a product in the reaction of H with 
$\CHthreeOH$ and in the reaction of OH with $\CHthree$, combined with the rapid thermal decomposition 
of $\CHtwoOH$, suggests that the dominant pathway will involve $\CHtwoOH$ as an intermediate, not 
$\CHthreeO$, which was suggested or adopted in previous CO or $\CHfour$ quenching studies 
\citep[e.g.,][]{yung88,bezard02,coop06,viss10b,line10,madhu11}.

The transport time scale arguments of \citet{prinn77} can be used to predict the quenched abundance 
of $\CHfour$ on exoplanets (see also the updates proposed by \citealt{viss11}).  With this procedure, 
the quench point is located where the chemical kinetic conversion time scale $\tau_{chem}$ is equal 
to the vertical transport time scale $\tau_{dyn}$.  Then, the mole fraction of the quenched species 
above the quench level is simply the equilibrium mole fraction of the constituent at that quench 
point.  In our models, the chemical kinetic conversion time scale for $\CHfour$ is
\begin{eqnarray}
\tau_{chem} \ & = & \ \tover{[\CHfour]}{d[\CHfour]/dt} \nonumber \\
\ & = & \ \tover{[\CHfour]}{k_{670} [\CHthree]\, [\OH] \, + \, 
k_{674} [\CHthree]\, [\OH]\, [\M]} 
\end{eqnarray}
where [X] represents the equilibrium concentration of species X in cm$^{-3}$, k$_{670}$ is the rate 
coefficient for reaction R670 in cm$^3$ s$^{-1}$, and k$_{674}$ is the rate coefficient for R670 in 
cm$^6$ s$^{-1}$.  See \citet{jasper07}, \citet{viss11}, and Supplementary Table S1 for more details on 
k$_{670}$ and k$_{674}$.

The transport time scale is $\tau_{dyn}$ = $L^2$/$K_{zz}$, where $L$ is an effective length scale 
over which dynamical mixing operates.  As discussed by \citet{smith98}, $L$ is generally some 
fraction of the scale height $H$.  For quenching of CO on cold planets like Jupiter, $L$ $\approx$ 
0.1$H$, whereas for quenching of $\CHfour$ on close-in transiting hot Jupiters, $L$ is closer to 
0.5$H$ \citep[see][]{viss11}.  When we use the procedure suggested by \citet{smith98} to derive 
$L$ and when we calculate $\tau_{chem}$ as described in Eq.~(4) for our rate-limiting step, we 
derive a quenched mole fraction within $\sim$15\% of our kinetics/transport model results for 
HD 189733b and HD 209458b.  Therefore, as is also discussed by \citet{bezard02}, \citet{coop06}, 
\citet{viss10b}, \citet{moses10}, and \citet{viss11} the transport time-scale approach is valid 
provided that the \citet{smith98} formalism for $\tau_{dyn}$ is used and provided that the assumed kinetic 
rate-limiting step is a reasonable one.  We caution that the time-scale approach has been misused 
frequently in exoplanet and brown-dwarf literature, where errors in the calculation of the rate 
coefficients for the rate-limiting steps have arisen due to incorrect reversals of three-body 
reactions \citep{griffith99,line10}, or where numerous other misconceptions about rate-limiting reactions 
and their rate coefficients and transport time scales are ubiquitous (see \citealt{bezard02}, 
\citealt{viss10b}, \citealt{moses10}, or \citealt{viss11} for a more detailed discussion of the problem).  
Some of these problems have serious implications with respect to conclusions about the strength 
of atmospheric mixing, particularly on brown dwarfs \citep{troyer07,viss11}, where $K_{zz}$ is 
likely several orders of magnitude greater than the sluggish $K_{zz}$ $\approx$ 10$^2$--10$^4$ 
cm$^2$ $\smone$ that has often been reported 
\citep[e.g.,][]{griffith99,saumon03,saumon06,leggett07,geballe09}, but also in relation to 
the implied strength of atmospheric mixing on exoplanets \citep[e.g.,][]{madhu11}.

The time-scale approach fails for species that have more than one quench point, 
for species that have an extended quench region due to $\tau_{chem}$ remaining close to $\tau_{dyn}$ 
over an extended pressure region, or for species that otherwise exhibit complicated kinetic behavior due to
quenching of other key species.  As an example, $\NHthree$ in our HD 189733b models quenches in a region 
of the atmosphere in which the temperature does not fall off sharply with height.  The $\NHthree$ mole 
fraction departs from the equilibrium curve when $\tau_{dyn}$ drops below $\tau_{chem}$ and conversion
between $\NHthree$ and $\Ntwo$ first becomes inhibited; however, due to the nearly isothermal behavior 
of the atmosphere above the quench region, $\tau_{chem}$ remains relatively close to $\tau_{dyn}$ over an 
extended pressure range, and $\NHthree$ does not fully quench until the temperatures begin to fall off sharply
again.  Therefore, the time-constant arguments provide a good prescription for where the $\NHthree$ leaves
equilibrium but not for the final quenched mole fraction.  

Several mechanisms closely compete as the dominant mechanism involved in $\NHthree$ 
$\rightarrow$ $\Ntwo$ conversion, and again, the dominance depends on pressure-temperature conditions
near the quench point.  In our HD 209458b models, where the atmosphere is relatively warm and $\NHthree$
quenches at relatively low pressures, the dominant scheme is 
\begin{eqnarray}
2\, ( \NHthree \, + \, \H \, & \rightarrow & \, \NHtwo \, + \, \Htwo ) \nonumber \\
\NHtwo \, + \, \H \, & \rightarrow & \, \NH \, + \, \Htwo \nonumber \\
\NH \, + \, \NHtwo \, & \rightarrow & \, \NtwoHtwo \, + \, \H \nonumber \\
\NtwoHtwo \, + \, \H \, & \rightarrow & \, \NNH \, + \, \Htwo \nonumber \\
\NNH \, + \, \M \, & \rightarrow & \, \Ntwo \, + \, \H \, + \, \M \nonumber \\
\Htwo \, + \, \M \, & \rightarrow & \, 2\, \H \, + \, \M \nonumber \\
\noalign{\vglue -10pt}
\multispan3\hrulefill \nonumber \cr
\Net \ \ 2\, \NHthree \, & \rightarrow & \, \Ntwo \, + \, 3\,\Htwo  . \\
\end{eqnarray}
On the cooler HD 189733b, $\NHthree$ first quenches down deep at high pressures and slightly 
lower temperatures than at the quench point on HD 209458b, through the dominant scheme
\begin{eqnarray}
\NHthree \, + \, \H \, & \rightarrow & \, \NHtwo \, + \, \Htwo \nonumber \\
\NHtwo \, + \, \NHthree & \rightarrow & \, \NtwoHthree \, + \, \Htwo \nonumber \\
\NtwoHthree \, + \, \M \, & \rightarrow & \, \NtwoHtwo \, + \, \H \, + \, \M \nonumber \\
\NtwoHtwo \, + \, \H \, & \rightarrow & \, \NNH \, + \, \Htwo \nonumber \\
\NNH \, + \, \M \, & \rightarrow & \, \Ntwo \, + \, \H \, + \, \M \nonumber \\
\noalign{\vglue -10pt}
\multispan3\hrulefill \nonumber \cr
\Net \ \ 2\, \NHthree \, & \rightarrow & \, \Ntwo \, + \, 3\,\Htwo  . \\
\end{eqnarray}
However, this scheme (6) becomes less important with decreasing pressure, and scheme (5) is 
responsible for the final quench point on HD 189733b.  
The rate-limiting step in scheme (5) is NH + $\NHtwo$ $\rightarrow$ $\NtwoHtwo$ + H, and we adopt the 
\citet{klipp09} theoretical rate coefficient expression of 7.07$\scinot-10. \, T^{-0.272} \exp 
(39/T)$ for this reaction.  In scheme (6), the rate-limiting step 
is $\NtwoHthree$ + M $\rightarrow$ $\NtwoHtwo$ + H + M, where we take the rate coefficient from 
the theoretical calculations of \citet{dean87} and \citet{hwang03} (see Supplementary Table S1).  
Note that these two reaction schemes --- and all other potential $\NHthree$-$\Ntwo$ conversion 
schemes --- are somewhat speculative and rely on rate coefficients derived from theoretical 
calculations rather than laboratory measurements.  As a whole, the kinetics of nitrogen compounds 
is less well understood than that of the C-H-O system (see \citealt{moses10} for further discussion).

After conversion of $\NHthree$ to $\Ntwo$ is first quenched (at $\sim$1600 K in the HD 189733b 
models, and at $\gta$1700 K in the HD 209458b models), the disequilibrium abundance of $\NHthree$ 
affects the abundances of other nitrogen-bearing constituents, often at altitudes well below 
their ultimate quench points.  HCN is a prime example.  The HCN mole fraction departs 
from equilibrium at the first $\NHthree$-$\Ntwo$ quench point, but HCN remains in a pseudo-equilibrium 
with $\NHthree$, $\CHfour$, and $\Htwo$ until its final quench point at higher altitudes.  The 
dominant scheme involved in $\NHthree$ $\rightarrow$ HCN conversion in our cooler HD 189733b models 
is
\begin{eqnarray}
\NHthree \, + \, \H \, & \rightarrow & \, \NHtwo \, + \, \Htwo  \nonumber \\
\CHfour \, + \, \H \, & \rightarrow & \, \CHthree \, + \, \Htwo  \nonumber \\
\NHtwo \, + \, \CHthree \, + \, \M & \rightarrow & \, \CHthreeNHtwo \, + \, \M \nonumber \\
\CHthreeNHtwo \, + \, \H \, & \rightarrow & \, \CHtwoNHtwo \, + \, \Htwo \nonumber \\
\CHtwoNHtwo \, & \rightarrow & \, \CHtwoNH \, + \, \H \nonumber \\
\CHtwoNH \, + \, \H \, & \rightarrow & \, \HtwoCN \, + \, \Htwo \nonumber \\
\HtwoCN \, + \, \M \, & \rightarrow & \, \HCN \, + \, \H \, + \, \M \nonumber \\
\Htwo \, + \, \M \, & \rightarrow & \, 2\, \H \, + \, \M \nonumber \\
\noalign{\vglue -10pt}
\multispan3\hrulefill \nonumber \cr
\Net \ \ \NHthree \, + \, \CHfour \, & \rightarrow & \, \HCN \, + \, 3\,\Htwo  . \\
\end{eqnarray}
This scheme is the reverse of the dominant HCN $\rightarrow$ $\NHthree$ conversion scheme 
derived for Jupiter's deep troposphere \citep{moses10}, although the rate-limiting step in the above 
scheme (7) for our exoplanet modeling is $\CHthreeNHtwo$ + H $\rightarrow$ $\CHtwoNHtwo$ + $\Htwo$.  
We adopt the \citet{dean00} theoretically-derived rate coefficient of $9.3\scinot-16. \, T^{1.5} 
\exp{(-2750/T)}$ cm$^{3}$ $\smone$ for this rate-limiting reaction.  On the warmer HD 209458b, HCN 
exchanges carbon with CO, as well as with $\CHfour$, and HCN is quenched when the following scheme 
becomes ineffective:
\begin{eqnarray}
\NHthree \, + \, \H \, & \rightarrow & \, \NHtwo \, + \, \Htwo  \nonumber \\
\CO \, + \, \NHtwo \, & \rightarrow & \, \HNCO \, + \, \H  \nonumber \\
\HNCO \, + \, \H \, & \rightarrow & \, \HCN \, + \, \OH  \nonumber \\
\OH \, + \, \Htwo \, & \rightarrow & \, \HtwoO \, + \, \H \nonumber \\
\noalign{\vglue -10pt}
\multispan3\hrulefill \nonumber \cr
\Net \ \ \NHthree \, + \, \CO \, & \rightarrow & \, \HCN \, + \, \HtwoO  . \\
\end{eqnarray}

Some molecules like $\CtwoHtwo$ quench at very high altitudes or are never truly quenched, as 
kinetic processes continue to affect their abundances throughout the atmospheric column, even 
when photolyzing radiation is absent.  
The dominant mechanism controlling the interchange of $\CtwoHtwo$ and $\CHfour$ in 
our models is the following scheme or its reverse:
\begin{eqnarray}
\CtwoHtwo \, + \, \H \, + \, \M & \rightarrow & \, \CtwoHthree \, + \, \M \nonumber \\
\CtwoHthree \, + \, \Htwo \, & \rightarrow & \, \CtwoHfour \, + \, \H  \nonumber \\
\CtwoHfour \, + \, \H \, + \, \M & \rightarrow & \, \CtwoHfive \, + \, \M \nonumber \\
\CtwoHfive \, + \, \Htwo \, & \rightarrow & \, \CtwoHsix \, + \, \H  \nonumber \\
\CtwoHsix \, + \, \M \, & \rightarrow & \, 2\, \CHthree \, + \, \M \nonumber \\
2\, ( \CHthree \, + \, \Htwo \, & \rightarrow & \, \CHfour \, + \, \H \, ) \nonumber \\
2\, \H \, + \, \M \, & \rightarrow & \, \Htwo \, + \, \M \nonumber \\
\noalign{\vglue -10pt}
\multispan3\hrulefill \nonumber \cr
\Net \ \ \CtwoHtwo \, + \, 3\, \Htwo \, & \rightarrow & \, 2\, \CHfour \,  . \\
\end{eqnarray}
The rate-limiting step in this scheme is either $\CtwoHsix$ thermal decomposition or $\Htwo$ + $\CtwoHfive$
$\rightarrow$ $\CtwoHsix$ + H, depending on local pressure-temperature conditions. 

In all, transport-induced quenching can have a significant effect on the predicted abundances 
of species like $\CHfour$, $\NHthree$, HCN, and $\CtwoHtwo$ on hot Jupiters.  Given the 
overall uncertainties in the thermal structure, eddy diffusion coefficients, kinetic rate 
coefficients, and effects from horizontal transport, the quantitative predictions from our 
kinetics and transport models should not be taken too seriously.  However, the general behavior 
in the models should be heeded --- the column abundance of these species will likely be enhanced 
over equilibrium predictions (and in some cases significantly enhanced), and transport-induced 
quenching is more significant for cooler planets or atmospheric regions.

We now move on to discuss the effects of the photochemistry that is initiated when ultraviolet 
photons are absorbed by atmospheric constituents.

\subsection{Photochemistry of oxygen species\label{sectpchemoxy}}

Figure \ref{figoxy} shows vertical profiles of the major oxygen-bearing species in our nominal 
thermo/photochemical 
kinetics and transport models.  Photolysis of the major parent molecules CO and $\HtwoO$ drives 
much of the oxygen chemistry.  Water photolysis in particular is a key component of exoplanet 
photochemistry, as it sets up a catalytic cycle that destroys $\Htwo$ and produces H:
\begin{eqnarray}
\HtwoO \, + \, h\nu \, & \rightarrow & \, \H \, + \, \OH \nonumber \\
\OH \, + \, \Htwo \, & \rightarrow & \, \HtwoO \, + \, \H \nonumber \\
\noalign{\vglue -10pt}
\multispan3\hrulefill \nonumber \cr
\Net \ \ \Htwo \, & \rightarrow & \, 2\, \H  , \\
\end{eqnarray}
where $h\nu$ represents an ultraviolet photon.  Water is recycled with this efficient scheme, 
but $\Htwo$ is destroyed, resulting in a huge net production rate for atomic H, which then 
replaces $\Htwo$ as the dominant atmospheric constituent at high altitudes (see Fig.~\ref{figoxy}).  
This transition from $\Htwo$ to H likely signals a transition into the high-temperature 
thermosphere on hot Jupiters, as no potential molecular coolants survive when atomic H 
becomes the dominant constituent at high altitudes \citep[see also][]{garcia07}.  Atomic species 
therefore are expected to dominate in the thermospheres of these close-in transiting planets 
\citep[e.g.,][]{yelle04,garcia07,koskinen10}, which might help explain the lack of 
obvious $\Htwo$ emission in far-ultraviolet dayglow observations of HD 209458b \citep{france10}.
From a column-integrated sense, scheme (10) is significantly more important as a source of atomic H than 
direct $\Htwo$ photodissociation.  Atomic H flows downward from its production region and remains 
at abundances greatly in excess of equilibrium until in reaches altitudes marked by its 
mole-fraction minimum shown in Fig.~\ref{figoxy}, near 0.01-0.1 bar on both planets, at which 
point atomic H again follows its equilibrium curve to lower altitudes.

\begin{figure*}
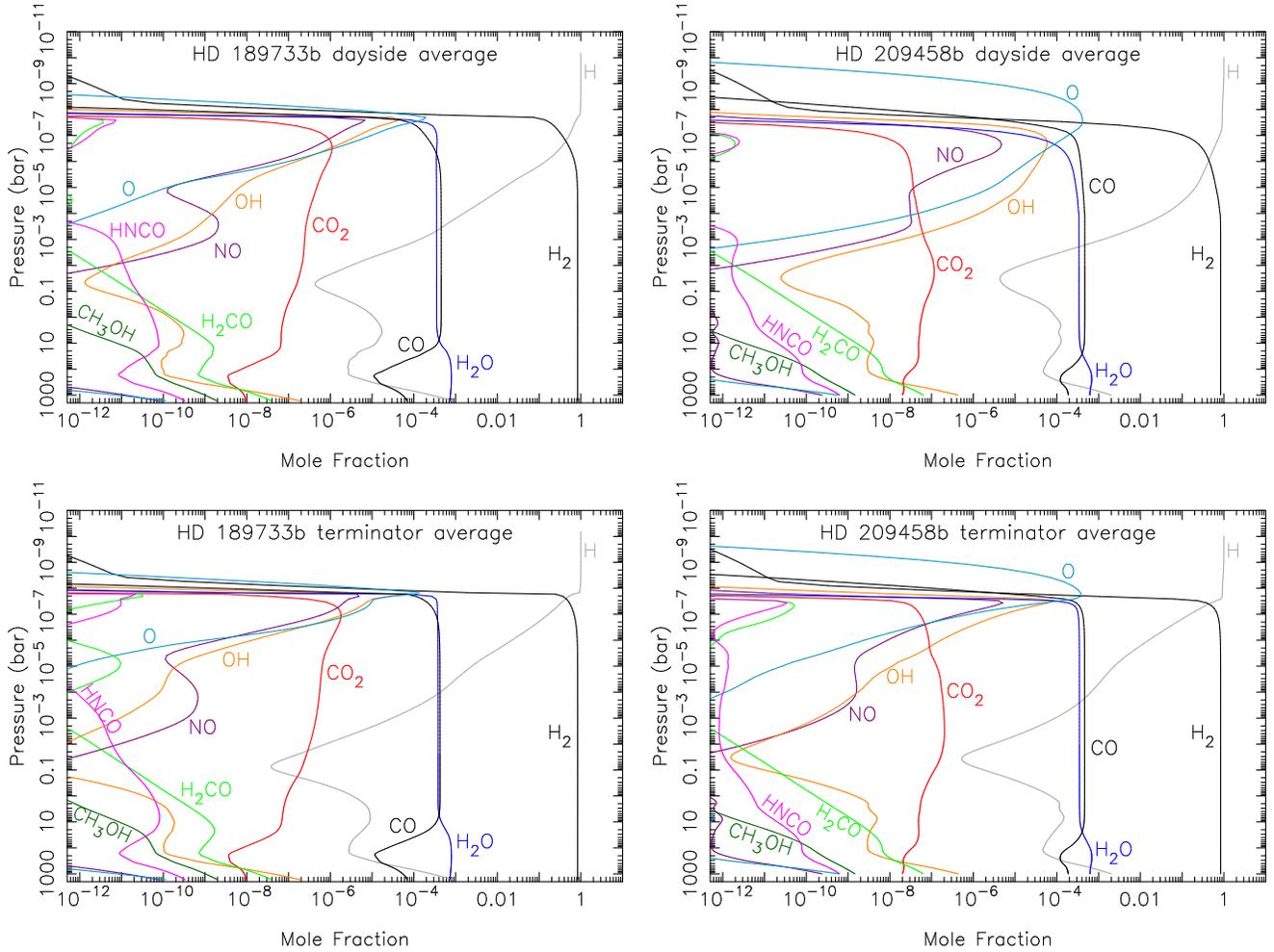

\begin{tabular}{ll}
{\includegraphics[angle=-90,clip=t,scale=0.37]{fig4a_color.ps}} 
& 
{\includegraphics[angle=-90,clip=t,scale=0.37]{fig4b_color.ps}} 
\\
{\includegraphics[angle=-90,clip=t,scale=0.37]{fig4c_color.ps}} 
& 
{\includegraphics[angle=-90,clip=t,scale=0.37]{fig4d_color.ps}} 
\\
\end{tabular}
\caption{Mole-fraction profiles for several oxygen-bearing species in our models for 
HD 189733b (left) and HD 209458b (right) for dayside-average (top) and terminator-average 
(bottom) conditions, assuming the nominal $K_{zz}$ profiles shown in Fig.~\ref{eddynom}.  
A color version of this figure is available in the online journal.\label{figoxy}}
\end{figure*}

Photolysis of CO to produce C + O 
and photolysis of $\HtwoO$ to produce $2\, \H$ + O or $\OoneD$ + $\Htwo$ also 
contribute to $\Htwo$ destruction through schemes such as 
\begin{eqnarray*}
\CO \, + \, h\nu \, & \rightarrow & \, \C \, + \, \O \\
\O \, + \, \Htwo \, & \rightarrow & \, \OH \, + \H \\
\OH \, + \, \Htwo \, & \rightarrow & \, \HtwoO \, + \H \\
\noalign{\vglue -10pt}
\multispan3\hrulefill \cr
\Net \ \ \CO \, + \, 2\, \Htwo \, & \rightarrow & \, \C \, + \, \HtwoO \, + \, 2\, \H   \\
\end{eqnarray*}
and
\begin{eqnarray*}
\HtwoO \, + \, h\nu \, & \rightarrow & \, 2\, \H \, + \, \O \\
\O \, + \, \Htwo \, & \rightarrow & \, \OH \, + \H \\
\OH \, + \, \Htwo \, & \rightarrow & \, \HtwoO \, + \H \\
\noalign{\vglue -10pt}
\multispan3\hrulefill \cr
\Net \ \  2\, \Htwo \, & \rightarrow & \, 4\, \H   \\
\end{eqnarray*}
and
\begin{eqnarray*}
\HtwoO \, + \, h\nu \, & \rightarrow & \, \Htwo \, + \, \OoneD \\
\OoneD \, + \, \Htwo \, & \rightarrow & \, \OH \, + \H \\
\OH \, + \, \Htwo \, & \rightarrow & \, \HtwoO \, + \H \\
\noalign{\vglue -10pt}
\multispan3\hrulefill \cr
\Net \ \  \Htwo \, & \rightarrow & \, 2\, \H .  \\
\end{eqnarray*}
Reactions of atomic H with $\HtwoO$ and OH can partially recycle the $\Htwo$, but key reactions 
like $2\, \H$ + M $\rightarrow$ $\Htwo$ do not operate effectively at low pressures, and the net result 
is that the loss rate for $\Htwo$ exceeds the production rate at high altitudes.
The H atoms produced in these mechanisms diffuse both upward and downward, 
where they strongly affect the abundance of other species.

Although $\HtwoO$ photolysis operates effectively down to millibar regions in these models and the 
atomic H produced from the photolysis continues to attack the $\HtwoO$, water is not permanently 
removed from this atmospheric region because of the efficient recycling by the reaction 
OH + $\Htwo$ $\rightarrow$ $\HtwoO$ + H.  Note that rapid replenishment by diffusion from other 
altitudes can also play an important role in maintaining $\HtwoO$ for our models with cooler 
stratospheric temperature profiles \citep[see also][]{line10}.  As an example, water is in 
approximate chemical equilibrium at pressures greater than $\sim$0.1 bar in the HD 189733b 
dayside-average model shown in Fig.~\ref{figoxy}; however, the transport time scales are shorter 
than the net photochemical loss time scales for $\HtwoO$ in the region from $\sim$0.1 bar to a 
few $\mu$bar, so that atmospheric mixing ``quenches'' the $\HtwoO$ mole fraction and keeps it 
roughly constant through this region.  Since the $\HtwoO$ mole fraction quenches at a value 
similar to the equilibrium value and since equilibrium would keep the $\HtwoO$ mole fraction constant 
in this region anyway, the water profile for the thermo/photochemical kinetics and transport model 
is quite similar to the thermochemical equilibrium profile for our HD 189733b models.  For our much 
warmer dayside HD 209458b model, equilibrium chemistry dominates the $\HtwoO$ profile over a much more 
expanded altitude range --- transport is only important at high altitudes, and only if the eddy 
diffusion coefficient is large.  This effect can be seen from detailed comparisons of the $\HtwoO$ 
profile from the HD 209458b dayside-average model in Fig.~\ref{equilvs} in which it is assumed that 
$K_{zz}$ = 10$^9$ cm$^2$ $\smone$, in contrast to the $\HtwoO$ profile in Fig.~\ref{figoxy} from 
an otherwise similar HD 209458b dayside-average model in which it is assumed that mixing is more 
vigorous at high altitudes (see the nominal $K_{zz}$ 
profile from Fig.~\ref{eddynom}).  The $K_{zz}$ = 10$^9$ cm$^2$ $\smone$ model has the $\HtwoO$ 
profile closely following equilibrium, whereas our higher $K_{zz}$ nominal model has water 
being carried to higher altitudes before its eventual chemical destruction.  However, this 
additional $\HtwoO$ from transport processes does not add much to the column abundance of $\HtwoO$ 
on the planet.  The net result on both the cooler and warmer classes of planets is an $\HtwoO$ profile 
that closely follows the equilibrium profile through the observable regions of the atmosphere.

Carbon monoxide also survives in the atmospheres of HD 189733b and HD 209458b despite rapid loss 
from H + CO + M $\rightarrow$ HCO + M, from OH + CO $\rightarrow$ $\COtwo$ + H, and from photolysis 
by EUV photons that continues down to mbar levels.  The first two reactions dominate the 
column-integrated CO loss in the stratosphere; the HCO produced in the first reaction efficiently 
reacts with H to reform CO, and the $\COtwo$ produced in the second reaction efficiently reacts 
with H to reform CO, so carbon monoxide is readily recycled on both HD 189733b and HD 209458b.  Even 
in the case of CO photolysis, some recycling pathways such as the following can operate:
\begin{eqnarray*}
\CO \, + \, h\nu \, & \rightarrow & \, \C \, + \, \O \\
\C \, + \, \Htwo \, & \rightarrow & \, \CH \, + \H \\
\CH \, + \, \HtwoO \, & \rightarrow & \, \HtwoCO \, + \H \\
\HtwoCO \, + \, \H \, & \rightarrow & \, \HCO \, + \Htwo \\
\HCO \, + \, \H \, & \rightarrow & \, \CO \, + \Htwo \\
\noalign{\vglue -10pt}
\multispan3\hrulefill \cr
\Net \ \ \HtwoO \, & \rightarrow & \, \O \, + \, \Htwo ,  \\
\end{eqnarray*}
allowing CO to persist.  However, these recycling schemes are not 100\% effective, and some of 
the CO photodestruction leads to the production of other species.  Atomic carbon and oxygen released 
from CO photolysis at high altitudes can remain as C and O, for example, and some of the O can 
lead to $\HtwoO$ production through reactions like O + $\Htwo$ $\rightarrow$ OH + H followed by 
OH + $\Htwo$ $\rightarrow$ $\HtwoO$ + H, while some of the C can go toward producing 
hydrocarbons and HCN (see below, and \citealt{liang04,line10,zahnle11}).  As with $\HtwoO$, the CO 
profiles are expected to closely follow the equilibrium profiles, except at high altitudes, where 
the CO will be converted to C, O, and $\HtwoO$ ({\it i.e.}, for cooler-temperature models) or where high eddy 
diffusion coefficients may allow it to be transported to higher altitudes than equilibrium models 
predict ({\it i.e.}, for our nominal dayside HD 209458b model).

The photochemistry of $\COtwo$ is not particularly interesting, as $\COtwo$ remains in equilibrium 
with CO and $\HtwoO$ throughout most of the atmosphere, via the reaction scheme:
\begin{eqnarray*}
\HtwoO \, + \, \H \, & \leftrightarrow & \, \OH \, + \Htwo \\
\OH \, + \, \CO \, & \leftrightarrow & \, \COtwo \, + \, \H \\
\noalign{\vglue -10pt}
\multispan3\hrulefill \cr
\Net \ \ \HtwoO \, + \, \CO \, & \leftrightarrow & \, \COtwo \, + \, \Htwo ,  \\
\end{eqnarray*}
Some additional $\COtwo$ in excess of equilibrium is produced at high altitudes from CO 
photochemistry on HD 189733b and/or from transport on HD 209458b, but the $\COtwo$ profiles on 
both planets remain close to equilibrium predictions.

Aside from atomic H, few photochemical products build up from the chemistry of oxygen-bearing species 
in our hot-Jupiter models.  Some O, OH, NO, and even $\Otwo$ and HNCO molecules are produced at high 
altitudes and diffuse downward; however, the photochemical lifetimes of these species are quite short, 
and they are not likely to achieve column abundances great enough to be detectable at visible or infrared 
wavelengths with current technologies unless the planet's metallicity is enhanced greatly over solar 
(see Fig.~\ref{figoxy} and Table~\ref{tabcolumn}).  Atomic 
O is produced from the reaction of H with OH to form O + $\Htwo$, with a lesser contributions from 
$\HtwoO$, OH, and CO photolysis.  Atomic O is lost predominantly from O + $\Htwo$ $\rightarrow$ OH + H.  
Because atomic O is not readily destroyed by H, it can be transported up to its homopause level in our 
HD 209458b models, unlike the situation for molecular species.  Some fraction of the O column 
will then make it into the thermosphere on planets with vigorous vertical mixing, where high thermospheric 
temperatures and hydrodynamic winds or other processes may then distribute the atomic O over a large 
radial distance \citep{lecave04,yelle04,tian05,garcia07,penz08,murray09,lammer09}, which will increase the 
likelihood of its detection at ultraviolet wavelengths \citep[e.g., see][]{vidal04,benjaff10,koskinen10}.  
Even in cases where $K_{zz}$ is small in the upper atmosphere, the dynamical wind from the
hydrodynamic escape --- which is not included in our models --- may then dominate over molecular 
diffusion in controlling the behavior of heavy atmospheric constituents, so that the concept of a 
homopause may not actually be appropriate for close-in transiting planets.  The dominant OH 
production mechanism is H + $\HtwoO$ $\rightarrow$ OH + $\Htwo$, with the atomic H coming from the 
$\Htwo$ catalytic destruction scheme (10) described above.  The reverse of this reaction is its dominant 
loss mechanism.  In general, the cooler the stratosphere, the larger the increase in the OH abundance 
over equilibrium predictions.

The main production mechanism for NO is N + OH $\rightarrow$ NO + H.  The OH derives from water, and 
the N can come from either $\Ntwo$ photolysis or from $\NHthree$ via schemes such as the following:
\begin{eqnarray*}
\HtwoO \, + \, h\nu \, & \rightarrow & \, \OH \, + \, \H \\
\NHthree \, + \, \H \, & \rightarrow & \, \NHtwo \, + \Htwo \\
\NHtwo \, + \, \H \, & \rightarrow & \, \NH \, + \Htwo \\
\NH \, + \, \H \, & \rightarrow & \, \N \, + \Htwo \\
\N \, + \, \OH \, & \rightarrow & \, \NO \, + \H \\
\noalign{\vglue -10pt}
\multispan3\hrulefill \cr
\Net \ \ \HtwoO \, + \, \NHthree \, + \, \H \, & \rightarrow & \, \NO \, + \, 3\, \Htwo  . \\
\end{eqnarray*}
The double-peak appearance in the NO mole fraction is due to the dominance of the $\Ntwo$ 
photolysis pathway in the upper stratosphere and the $\NHthree$ pathway in the middle stratosphere.  
The NO is lost through reactions with N to form $\Ntwo$ + O, with C to form N + CO and/or CN + O, 
through photolysis to form N + O, and through reaction with atomic H.  Molecular oxygen is produced 
through the reaction O + OH $\rightarrow$ $\Otwo$ + H, and is lost through the reverse of this 
reaction.

As a general rule of thumb, photochemistry in these warm, highly irradiated stratospheres 
favors small molecules and/or molecules with strong bonds.  Our results are consistent with 
other exoplanet photochemical models \citep{liang03,liang04,line10,zahnle11} in that we 
see little photochemical production of molecules like $\HtwoCO$, $\CHthreeOH$, $\HtwoCCO$, 
or $\CHthreeCHO$.

\begin{figure*}
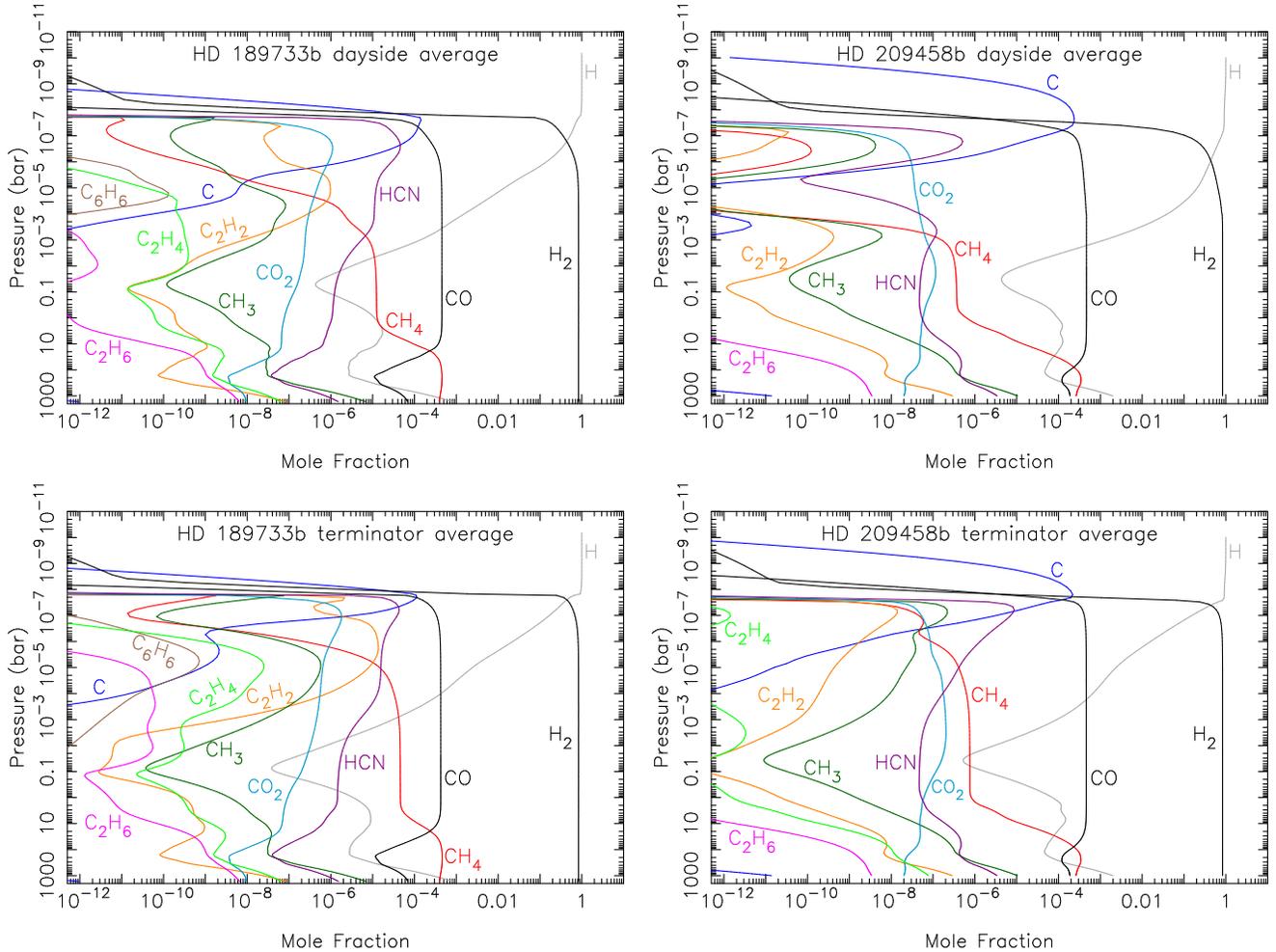

\begin{tabular}{ll}
{\includegraphics[angle=-90,clip=t,scale=0.37]{fig5a_color.ps}} 
& 
{\includegraphics[angle=-90,clip=t,scale=0.37]{fig5b_color.ps}} 
\\
{\includegraphics[angle=-90,clip=t,scale=0.37]{fig5c_color.ps}} 
& 
{\includegraphics[angle=-90,clip=t,scale=0.37]{fig5d_color.ps}} 
\\
\end{tabular}
\caption{Mole-fraction profiles for several carbon-bearing species in our models for 
HD 189733b (left) and HD 209458b (right) for dayside-average (top) and terminator-average 
(bottom) conditions, assuming the nominal $K_{zz}$ profiles shown in Fig.~\ref{eddynom}.  
A color version of this figure is available in the online journal.\label{fighc}}
\end{figure*}

\subsection{Photochemistry of carbon species\label{sectpchemhc}}

The photochemistry of carbon species is more interesting than that of oxygen species, as the 
primary parent molecule $\CHfour$ has weaker bonds.  Several carbon-bearing photochemical products 
can build up in the atmospheres of extrasolar giant planets.  Figure \ref{fighc} shows vertical 
profiles of the major carbon species in our nominal thermo/photochemical kinetics and 
transport models.  Carbon photochemistry is initiated by CO photolysis at high altitudes and 
by hydrogen abstraction from methane at lower altitudes.  At pressures less than $\sim$1 $\mu$bar, 
much of the C that is released from CO photolysis remains as atomic C, although some acetylene 
and other hydrocarbon production can occur through schemes such as
\begin{eqnarray}
2\, ( \, \CO \, + \, h\nu \, & \rightarrow & \, \C \, + \, \O \, ) \nonumber \\
2\, ( \, \C \, + \, \Htwo \, & \rightarrow & \, \CH \, + \H \, ) \nonumber \\
\CH \, + \, \Htwo \, & \rightarrow & \, \threeCHtwo \, + \H \nonumber \\
\CH \, + \, \threeCHtwo \, & \rightarrow & \, \CtwoHtwo \, + \H \nonumber \\
2\, ( \, 2\, \H \, + \, \M \, & \rightarrow & \, \Htwo \, + \, \M \, ) \nonumber \\
\noalign{\vglue -10pt}
\multispan3\hrulefill \nonumber \cr
\Net \ \ 2\, \CO \, + \, \Htwo \, & \rightarrow & \, \CtwoHtwo \, + \, 2\, \O .  \\
\end{eqnarray}
At greater pressures (and in particular where $\CHfour$ or $\CHthree$ become more abundant than 
atomic C), acetylene and other hydrocarbons are produced from methane destruction through pathways 
such as
\begin{eqnarray}
2\, ( \, \HtwoO \, + \, h\nu \, & \rightarrow & \, \H \, + \, \OH \, ) \nonumber \\
2\, ( \, \OH \, + \, \Htwo \, & \rightarrow & \, \HtwoO \, + \H \, ) \nonumber \\
2\, ( \, \CHfour \, + \, \H \, & \rightarrow & \, \CHthree \, + \Htwo \, ) \nonumber \\
\CHthree \, + \, \H \, & \rightarrow & \, \oneCHtwo \, + \Htwo \nonumber \\
\oneCHtwo \, + \, \Htwo \, & \rightarrow & \, \threeCHtwo \, + \Htwo \nonumber \\
\threeCHtwo \, + \, \CHthree \, & \rightarrow & \, \CtwoHfour \, + \H \nonumber \\
\CtwoHfour \, + \, \H \, & \rightarrow & \, \CtwoHthree \, + \Htwo \nonumber \\
\CtwoHthree \, + \, \H \, & \rightarrow & \, \CtwoHtwo \, + \Htwo \nonumber \\
\noalign{\vglue -10pt}
\multispan3\hrulefill \nonumber \cr
\Net \ \ 2\, \CHfour \, & \rightarrow & \, \CtwoHtwo \, + \, 3\, \Htwo .  \\
\end{eqnarray}
Note the importance of $\HtwoO$ photolysis in providing the atomic H used to destroy the 
$\CHfour$.  These pathways are efficient enough that $\CtwoHtwo$ becomes the dominant 
hydrocarbon at high altitudes under certain conditions in our models (see Fig.~\ref{fighc}).  

Acetylene is more effectively hydrogenated with increasing pressure, such that the 
following scheme acts to remove the $\CtwoHtwo$ and recycle the methane at middle-stratospheric 
pressures:
\begin{eqnarray}
\CtwoHtwo \, + \, \H \, + \, \M \, & \rightarrow & \, \CtwoHthree \, + \M  \nonumber \\
\CtwoHthree \, + \, \Htwo \, & \rightarrow & \, \CtwoHfour \, + \H \nonumber \\
\CtwoHfour \, + \, \H \, + \, \M \, & \rightarrow & \, \CtwoHfive \, + \M  \nonumber \\
\CtwoHfive \, + \, \H \, & \rightarrow & \, 2\, \CHthree \nonumber \\
2\, ( \, \CHthree \, + \, \Htwo \, & \rightarrow & \, \CHfour \, + \H \, ) \nonumber \\
\noalign{\vglue -10pt}
\multispan3\hrulefill \nonumber \cr
\Net \ \  \CtwoHtwo \, + \, 3\, \Htwo \, & \rightarrow & \, 2\, \CHfour .  \\
\end{eqnarray}
This scheme, rather than photolysis, is responsible for removing $\CtwoHtwo$ from the middle and 
lower stratosphere, as the $\CtwoHtwo$ photolysis products tend to simply recycle the $\CtwoHtwo$.

Other C$_2$H$_x$ hydrocarbons are less able to survive the large background H abundance.  Ethane is 
produced in the stratosphere through $2\, \CHthree$ + M $\rightarrow$ $\CtwoHsix$ + M and through 
$\CtwoHfive$ + $\Htwo$ $\rightarrow$ $\CtwoHsix$ + H.  Ethane is lost through hydrogen abstraction 
by atomic H ($\CtwoHsix$ + H $\rightarrow$ $\CtwoHfive$ + $\Htwo$), followed by H + $\CtwoHfive$ 
$\rightarrow$ $2\, \CHthree$, and eventual methane production.  The column abundance of photochemically 
produced ethane is quite small in our models.  Ethylene ($\CtwoHfour$) fares a bit better: it is 
produced predominantly from $\CtwoHthree$ + $\Htwo$ $\rightarrow$ $\CtwoHfour$ + H (see some of the 
$\CtwoHtwo$ production schemes above) and is lost from the reverse of this reaction (heading toward 
$\CtwoHtwo$ production), as well as H + $\CtwoHfour$ + M $\rightarrow$ $\CtwoHfive$ + M, to eventually 
form methane.  Although the mole fraction of $\CtwoHfour$ can exceed that of $\CtwoHtwo$ at the 
$\CtwoHtwo$ mole-fraction minimum near 0.1--10$^{-2}$ bar in some of our models, the peak $\CtwoHfour$ 
mole fraction never approaches that of the peak $\CtwoHtwo$ mole fraction, and the column 
abundance of $\CtwoHfour$ is intermediate between that of $\CtwoHtwo$ and $\CtwoHsix$ in our HD 189733b 
and HD 209458b models (see Table \ref{tabcolumn}).

Although our kinetics is far from complete for C$_3$-to-C$_6$ species, we do attempt to track the production 
and loss of some complex hydrocarbons.  The dominant C$_3$ hydrocarbon in our models is actually the 
C$_3$H$_2$ radical.  This result may be due to our insufficient knowledge of C$_3$H$_x$ kinetics, 
and we encourage further laboratory and theoretical studies of the fate of C$_3$H$_x$ species under 
the low-pressure, high-temperature, reducing conditions in the relevant regions of extrasolar-planet 
atmospheres.  Both CH and C readily insert into $\CtwoHtwo$ 
\citep[e.g.,][]{berman82,canosa97,haider93,kaiser01,casavecchia01,mebel07}, 
and these insertion reactions likely dominate the production of C$_3$ hydrocarbons on extrasolar giant 
planets, but the fate of the C$_3$H$_2$, C$_3$H, and C$_3$ radicals produced from these reactions is 
unclear.  We implicitly assume that the C$_3$ and C$_3$H radicals that are produced from the reaction  
of C with $\CtwoHtwo$ quickly end up as C$_3$H$_2$ --- an assumption that may only be correct at altitudes 
where $\Htwo$ is the dominant atmospheric constituent.  The main loss process for C$_3$H$_2$ in our models 
is H + C$_3$H$_2$ + M $\rightarrow$ C$_3$H$_3$ + M.  The resulting propargyl radicals can be photolyzed 
to recycle the C$_3$H$_2$, can react with atomic H to reform $\CtwoHtwo$, can react with $\Htwo$ or H to 
form C$_3$H$_4$, or can react with other C$_3$H$_x$ radicals to form C$_6$ hydrocarbons and eventually 
benzene.  

Although benzene is not a major constituent in our models, it can reach mole fractions of 0.1-1 ppb
in the few-microbar region in some of our HD 189733b models; however, benzene does not survive in the 
warmer HD 209458b models.  While ppb levels are by no means sufficient for C$_6$H$_6$ condensation on HD 
189733b, it does suggest that polycyclic aromatic hydrocarbon (PAH) formation could possibly occur at 
high altitudes on cooler transiting exoplanets, which could ultimately lead to the formation of a 
``soot'' aerosol layer, as has been suggested by \citet{zahnle11}.  In contrast, PAH formation will be 
strongly suppressed on planets with strong stratospheric thermal inversions \citep[see also][]{liang04}.  
Our neutral chemistry is not particularly conducive for soot formation, even under HD 189733b conditions, 
but it is possible that photoionization of $\Ntwo$ or other species could lead to the production of 
high-molecular-weight ions and photochemically produced aerosols through a Titan-like ion-chemistry 
process \citep[e.g.,][]{waite07,waite09,imanaka07,vuitton07,vuitton08}, although it is also possible 
that the high abundance of $\HtwoO$ and/or CO and the low abundance of $\CHfour$ at high altitudes on 
highly irradiated exoplanets could short-circuit this process.  The potential for ion chemistry playing a 
role in the formation of complex species in extrasolar giant planet atmospheres deserves further study.

Unlike the case for CO and $\HtwoO$, methane can be efficiently destroyed and not recycled in the 
upper atmospheres of HD 189733b and HD 209458b.  Some of the carbon ends up in $\CtwoHtwo$ and other 
hydrocarbons, as well as CO and C, but cross reactions with nitrogen species become a key mechanism for 
removing $\CHfour$ from the upper atmosphere.  Hydrogen cyanide is the dominant product of this 
coupled carbon-nitrogen chemistry.  From a column-integrated standpoint, the dominant photochemical 
scheme that permanently converts $\CHfour$ into HCN in the stratospheres of our exoplanet models is
\begin{eqnarray}
2\, ( \, \HtwoO \, + \, h\nu \, & \rightarrow & \, \H \, + \, \OH \, ) \nonumber \\
2\, ( \, \OH \, + \, \Htwo \, & \rightarrow & \, \HtwoO \, + \H \, ) \nonumber \\
\NHthree \, + \, \H \, & \rightarrow & \, \NHtwo \, + \Htwo  \nonumber \\
\CHfour \, + \, \H \, & \rightarrow & \, \CHthree \, + \Htwo \nonumber \\
\NHtwo \, + \, \H \, & \rightarrow & \, \NH \, + \Htwo \nonumber \\
\NH \, + \, \H \, & \rightarrow & \, \N \, + \Htwo \nonumber \\
\N \, + \, \CHthree \, & \rightarrow & \, \HtwoCN \, + \H \nonumber \\
\HtwoCN \, + \, \H \, & \rightarrow & \, \HCN \, + \Htwo \nonumber \\
\noalign{\vglue -10pt}
\multispan3\hrulefill \nonumber \cr
\Net \ \ \CHfour \, + \, \NHthree \, & \rightarrow & \, \HCN \, + \, 3\, \Htwo .  \\
\end{eqnarray}
Hydrogen cyanide then becomes a major disequilibrium product in exoplanet atmospheres and can 
replace methane as the second most abundant carbon-bearing species at high altitudes.  

As is the case for oxygen compounds, the photochemistry of carbon compounds is more effective 
for cooler exoplanets, both because of the larger abundance of $\CHfour$ from transport-induced 
quenching and because the lower temperatures allow disequilibrium products to survive more readily.
The dominant carbon-bearing photochemical products are HCN, $\CtwoHtwo$, C, and $\CHthree$, with 
some $\COtwo$ being produced at high altitudes.

\begin{figure*}
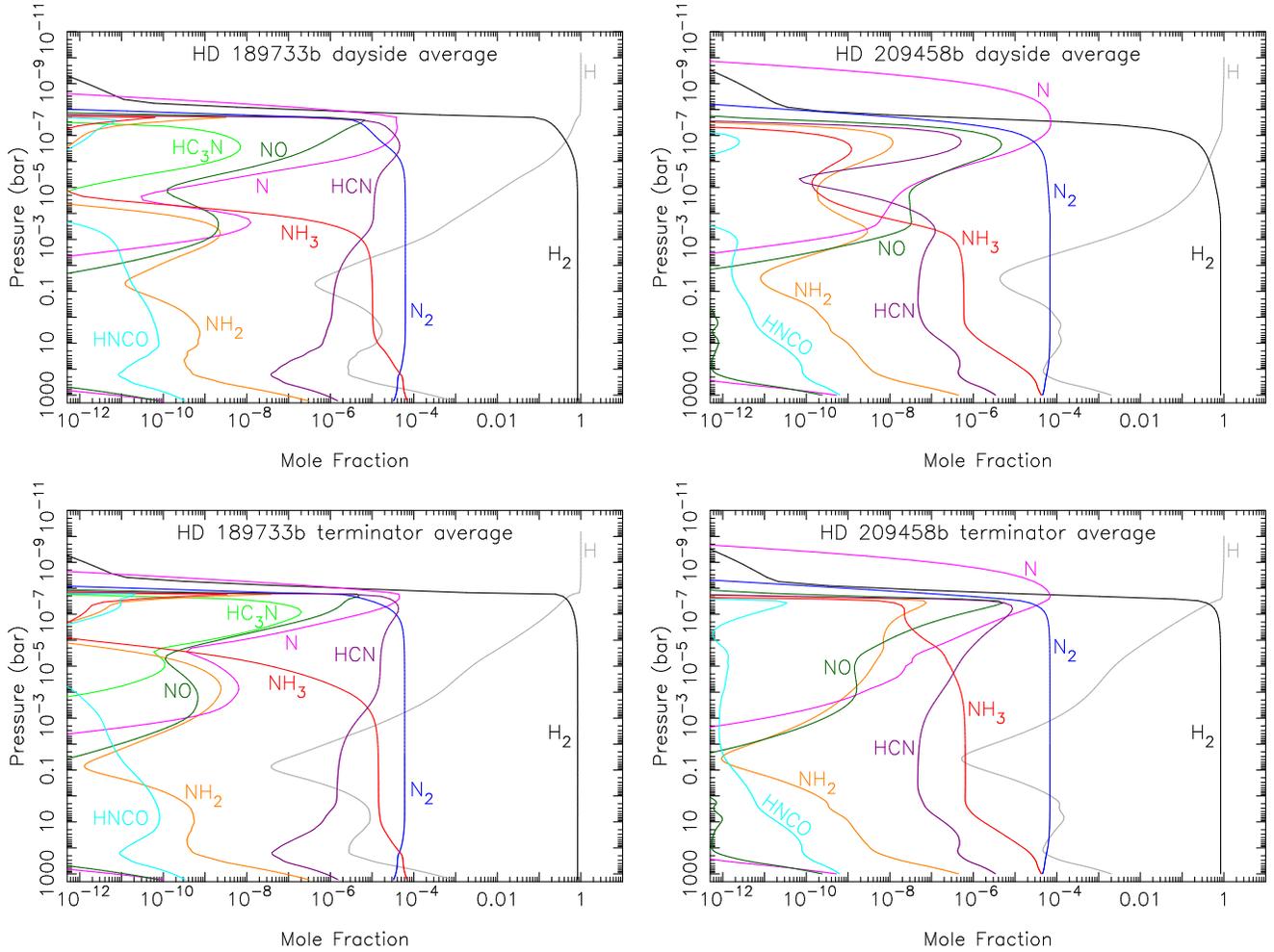

\begin{tabular}{ll}
{\includegraphics[angle=-90,clip=t,scale=0.37]{fig6a_color.ps}} 
& 
{\includegraphics[angle=-90,clip=t,scale=0.37]{fig6b_color.ps}} 
\\
{\includegraphics[angle=-90,clip=t,scale=0.37]{fig6c_color.ps}} 
& 
{\includegraphics[angle=-90,clip=t,scale=0.37]{fig6d_color.ps}} 
\\
\end{tabular}
\caption{Mole-fraction profiles for several nitrogen-bearing species in our models for 
HD 189733b (left) and HD 209458b (right) for dayside-average (top) and terminator-average 
(bottom) conditions, assuming the nominal $K_{zz}$ profiles shown in Fig.~\ref{eddynom}.  
A color version of this figure is available in the online journal.\label{fignit}}
\end{figure*}

\subsection{Photochemistry of nitrogen species\label{sectpchemnit}}

Nitrogen photochemistry parallels that of carbon chemistry: $\Ntwo$, like CO, is a high-altitude 
source, and $\NHthree$, like $\CHfour$, is a critical transport-quenched species that drives 
photochemistry throughout the bulk of the stratosphere.  Figure \ref{fignit} shows vertical 
profiles of the major nitrogen-bearing species in our nominal thermo/photochemical kinetics and 
transport models.  At high altitudes, nitrogen photochemistry is initiated by $\Ntwo$ photolysis 
through schemes such as 
\begin{eqnarray*}
\CO \, + \, h\nu \, & \rightarrow & \, \C \, + \, \O  \\
\Ntwo \, + \, h\nu \, & \rightarrow & \, 2\, \N  \\
\O \, + \, \Htwo \, & \rightarrow & \, \OH \, + \H  \\
\N \, + \, \OH \, & \rightarrow & \, \NO \, + \H  \\
\C \, + \, \NO \, & \rightarrow & \, \CN \, + \O  \\
\CN \, + \, \Htwo \, & \rightarrow & \, \HCN \, + \H \\
2\, \H \, + \, \M \, & \rightarrow & \, \Htwo \, + \M \\
\noalign{\vglue -10pt}
\multispan3\hrulefill \cr
\Net \ \ \CO \, + \, \Ntwo \, + \, \Htwo \, & \rightarrow & \, \HCN \, + \, \N \, + \, \O \, + \, \H ,  \\
\end{eqnarray*}
where the resulting HCN and N are significant high-altitude products.  On the cooler HD 189733b, 
additional speculative processes involving $\HCthreeN$ rather than NO might contribute at high altitudes:
\begin{eqnarray*}
\CO \, + \, h\nu \, & \rightarrow & \, \C \, + \, \O  \\
\Ntwo \, + \, h\nu \, & \rightarrow & \, 2\, \N  \\
\C \, + \, \CtwoHtwo \, & \rightarrow & \, \CthreeHtwo  \\
\N \, + \, \CthreeHtwo \, & \rightarrow & \, \HCthreeN \, + \H  \\
\H \, + \, \HCthreeN \, & \rightarrow & \, \CN \, + \CtwoHtwo  \\
\CN \, + \, \Htwo \, & \rightarrow & \, \HCN \, + \H \\
\noalign{\vglue -10pt}
\multispan3\hrulefill \cr
\Net \ \ \CO \, + \, \Ntwo \, + \, \Htwo \, & \rightarrow & \, \HCN \, + \, \N \, + \, \O \, + \, \H .  \\
\end{eqnarray*}

Molecular nitrogen is quite stable on HD 189733b and HD 209458b.  Although $\Ntwo$ can combine 
with atomic H to form NNH, thermal decomposition of this unstable species acts to recycle the 
$\Ntwo$ throughout the stratosphere.  At high altitudes, the strong N$\tbond$N bond can be 
destroyed by photolysis or attacked by atomic C, H, and O to form atomic N, which then can 
become the dominant nitrogen species at high altitudes.  However, $\Ntwo$ is also recycled 
through schemes such as 
\begin{eqnarray*}
\Ntwo \, + \, h\nu \, & \rightarrow & \, 2\, \N  \\
\N \, + \, \Htwo \, & \rightarrow & \, \NH \, + \H  \\
\N \, + \, \NH \, & \rightarrow & \, \Ntwo \, + \H  \\
\noalign{\vglue -10pt}
\multispan3\hrulefill \cr
\Net \ \ \Htwo \,& \rightarrow & \, 2\, \H  \\
\end{eqnarray*}
and
\begin{eqnarray*}
\Ntwo \, + \, h\nu \, & \rightarrow & \, 2\, \N  \\
\N \, + \, \OH \, & \rightarrow & \, \NO \, + \H  \\
\N \, + \, \NO \, & \rightarrow & \, \Ntwo \, + \O  \\
\O \, + \, \Htwo \, & \rightarrow & \, \OH \, + \H  \\
\noalign{\vglue -10pt}
\multispan3\hrulefill \cr
\Net \ \ \Htwo \,& \rightarrow & \, 2\, \H , \\
\end{eqnarray*}
where catalytic destruction of $\Htwo$ is the net result of both recycling schemes.
Throughout much of the stratosphere, $\Ntwo$ is shielded from photolysis by a larger 
surrounding column of $\Htwo$.

Ammonia has much weaker bonds and is less stable than $\Ntwo$, so much of the interesting nitrogen 
chemistry in our models involves $\NHthree$.  Throughout the bulk of the stratosphere, $\NHthree$ 
destruction leads to HCN production through scheme (14) discussed in section \ref{sectpchemhc}.  
This scheme is efficient enough --- and HCN is recycled readily enough --- that HCN can take over 
from $\NHthree$ and $\CHfour$ as important nitrogen and carbon carriers in the upper stratosphere. 
The main loss process for HCN in hot-Jupiter stratospheres is the reaction H + HCN $\rightarrow$ 
CN + $\Htwo$, with a much lesser contribution from HCN photolysis to form CN + $\Htwo$.  The CN 
produced by these processes reacts with $\Htwo$ to reform HCN, so HCN recycling is prevalent.

Complex nitriles are also formed from photochemistry on cooler extrasolar planets like HD 189733b.  
For example, some of the CN produced in the H + HCN reaction can react with $\CtwoHtwo$ to form 
HC$_3$N, or some of the $\CHthree$ produced from hydrogen abstraction of methane can react with HCN to 
form $\CHthreeCN$.  These nitriles tend to be unstable in a background bath of H, and their destruction 
pathways tend to reproduce the HCN, so they are nowhere near as abundant as HCN in our HD 189733b 
models.  However, as with benzene and potential PAHs, further nitrile chemistry may produce condensible 
products, particularly on colder exoplanets.  On the other hand, complex nitriles will be very 
unstable on planets like HD 209458b that have a strong stratospheric thermal inversion.  Other 
carbon-nitrogen bonded species, including imines and amines, are relatively unimportant in our 
models, although the $\HtwoCN$ that is produced 
during the course of scheme (14) can undergo three-body recombination with H to form some $\CHtwoNH$
that persists in our cooler atmospheric models.  Note that nitrogen hydrides like hydrazine
($\NtwoHfour$) are 
unimportant in our models, unlike the situation on the much colder Jupiter.

Cross reactions between nitrogen and oxygen species can also occur in exoplanet stratospheres, leading 
to the production of species like NO and HNCO.  The dominant schemes producing and destroying NO were 
discussed in section \ref{sectpchemoxy}.  HNCO is produced predominantly through the reaction 
$\NHtwo$ + CO $\rightarrow$ 
HNCO + H, where the $\NHtwo$ derives from hydrogen abstraction of ammonia by atomic H.  The dominant loss 
for HNCO is reaction with atomic H in the reverse of the main production reaction.

The dominant nitrogen-bearing photochemical products in our exoplanet models are HCN, N, NO, and 
$\NHtwo$.  As with oxygen and carbon photochemistry, cooler planets tend to have more abundant and 
more varied disequilibrium nitrogen species than warmer planets.  This tendency results from 
the larger abundance of photochemically active, transport-quenched species like $\NHthree$ on cooler 
planets and from the greater stability of the photochemical products at low temperatures.

\subsection{Sensitivity to temperature\label{secttempsens}}

The atmospheric thermal structure can strongly affect the abundance of disequilibrium constituents.  
As shown in Fig.~\ref{equilvs}, high temperatures in the stratospheres of transiting hot Jupiters can 
allow photochemically produced species to chemically react with other atmospheric constituents to 
drive the composition back toward equilibrium.  Disequilibrium compositions are thus difficult to 
maintain on planets like HD 209458b that have strong thermal inversions.  In contrast, thermochemical 
kinetics is inhibited when stratospheric temperatures are cooler, so that photochemically produced 
species are more likely to survive on HD 189733b and other exoplanets without stratospheric thermal 
inversions.  Temperatures in the deep atmosphere also affect the abundance of disequilibrium constituents 
through controlling where transport-induced quenching begins to operate.  Fig.~\ref{tempsens} illustrates 
the importance of this effect for key transport-quenched species like $\CHfour$, $\NHthree$, and HCN.

\begin{figure*}
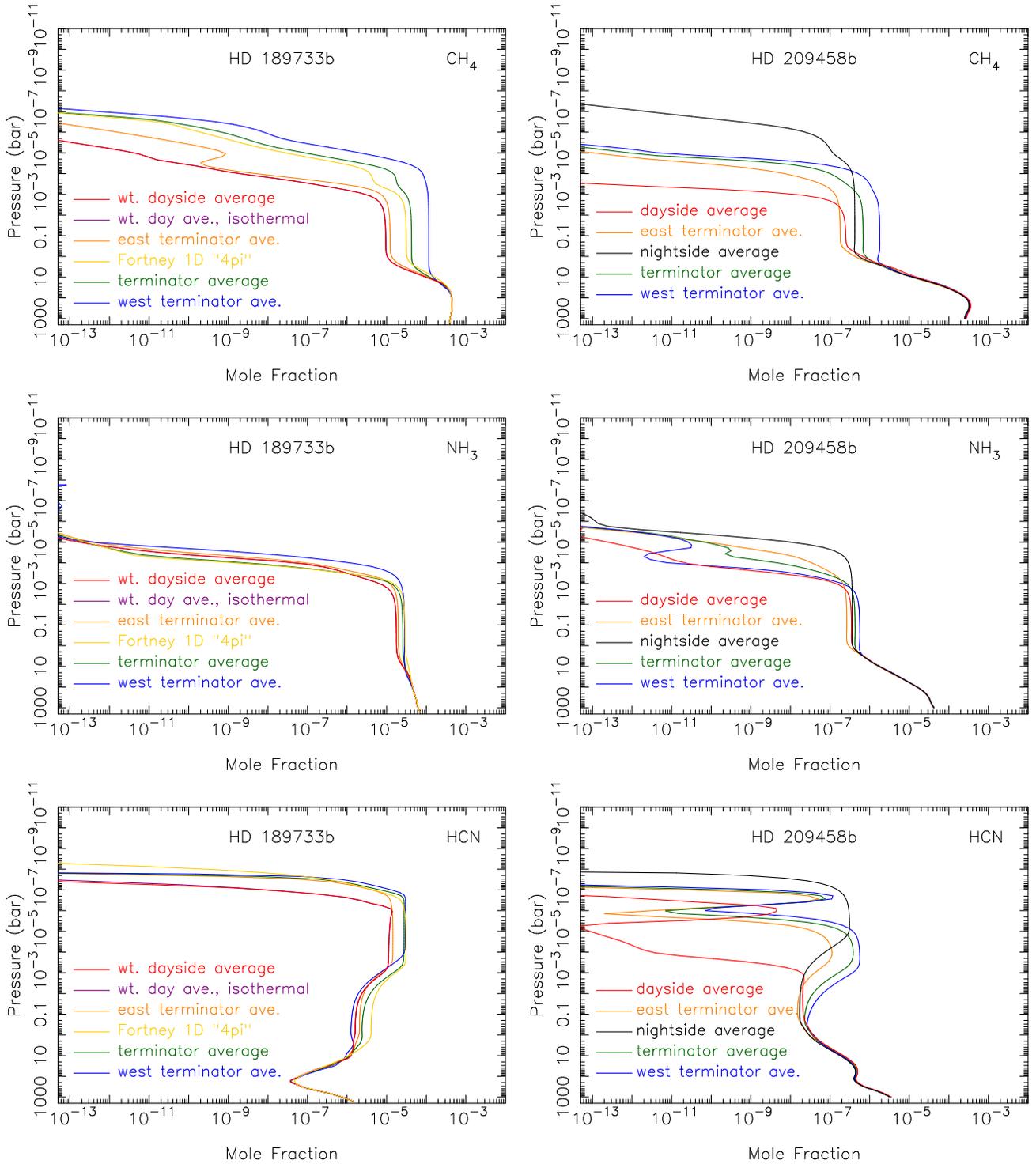

\begin{tabular}{ll}
{\includegraphics[angle=-90,clip=t,scale=0.37]{fig7a_color.ps}} 
& 
{\includegraphics[angle=-90,clip=t,scale=0.37]{fig7b_color.ps}} 
\\
{\includegraphics[angle=-90,clip=t,scale=0.37]{fig7c_color.ps}} 
& 
{\includegraphics[angle=-90,clip=t,scale=0.37]{fig7d_color.ps}} 
\\
{\includegraphics[angle=-90,clip=t,scale=0.37]{fig7e_color.ps}} 
& 
{\includegraphics[angle=-90,clip=t,scale=0.37]{fig7f_color.ps}} 
\\
\end{tabular}
\caption{Mole-fraction profiles for $\CHfour$ (top), $\NHthree$ (middle), and HCN (bottom) in our 
models for HD 189733b (left) and HD 209458b (right) for different assumptions about the 
atmospheric thermal structure (as labeled, and see Fig.~\ref{figtemp}).  All models 
assume a constant $K_{zz}$ = 10$^9$ cm$^2$ $\smone$.  A color version 
of this figure is available in the online journal.\label{tempsens}}
\end{figure*}

In Fig.~\ref{tempsens}, all models have the same constant-with-altitude eddy $K_{zz}$ = 10$^9$ 
cm$^2$ $\smone$ profile, but the thermal structure is different for each model 
(see Fig.~\ref{figtemp} for the adopted temperature profiles).  In general, colder 
temperatures at depth imply that transport will be able to compete with thermochemical kinetics at 
greater pressures; the quench point will therefore be deeper on colder planets, and the mole fraction 
of the quenched species will be larger.  This effect can be seen with the HD 189733b models shown 
in Fig.~\ref{tempsens}, where the coldest western-terminator-average model has a $\CHfour$ quench point 
near the $\sim$7 bar, $\sim$1485 K level, whereas the warmest projected-area-weighted dayside-average 
models have quench points near 1.9 bar, $\sim$1522 K.  The resulting $\CHfour$ quenched mole fraction is 
a factor of $\sim$10 greater for the cooler western-terminator model than for the warmer dayside-average 
model.  Ammonia quenches at even greater pressures and temperatures, at which point the temperature profiles 
are more uniform (see Fig.~\ref{figtemp}), so that the differences between models are less pronounced.  
The HCN profiles are more complicated because the quenched abundances of both $\NHthree$ and $\CHfour$ 
affect the HCN abundance, and HCN is photochemically produced at high altitudes, but Fig.~\ref{tempsens} 
suggests that HCN first quenches in the 10-30 bar region, where the GCM simulations of \citet{showman09} 
exhibit western-terminator-average temperatures that are actually warmer than the dayside-average 
temperatures at these pressures.  As such, the western-terminator-average model has a smaller quenched 
HCN abundance than the dayside-average models.  The 1-D ``4pi'' profile from \citet{fort06a,fort10} 
is the coolest of 
all our models at 10-30 bars, resulting in the largest quenched HCN mole fraction in this model. 
Given the lack of constraints on temperatures and eddy diffusion coefficients at depth, the 
quantitative results for the abundance of quenched species have significant uncertainty in our models; 
moreover, horizontal advection time scales may be shorter than chemical time scales at pressures
greater than a few mbar, acting to homogenize the composition longitudinally (see section
\ref{secttime}).

For HD 209458b, the presence of high stratospheric temperatures, particularly on the dayside, is the 
biggest factor in controlling the composition, as the disequilibrium molecular constituents do not 
survive in the middle and upper stratosphere.  The cooler the atmosphere is at high altitudes, 
the more likely disequilibrium species are to survive to these altitudes.  Note that the lack of UV 
photons in the nightside-average model shown in Fig.~\ref{tempsens} also helps keep disequilibrium 
molecules present at high altitudes.  Even without a strong thermal inversion at the terminators 
of HD 209458b, stratospheric temperatures are considerably higher than on HD 189733b, enabling 
thermochemical kinetic processes to destroy disequilibrium constituents.  Temperatures at depth are 
also much warmer on HD 209458b than on HD 189733b, due to the different luminosities of the host 
stars.  Transport-induced quenching is therefore less effective on HD 209458b than on HD 189733b, 
and the predicted quenched abundances of $\CHfour$, $\NHthree$, and HCN are correspondingly reduced.

The abundances of HCN and $\CHfour$ are particularly sensitive to temperature, whereas species 
like $\HtwoO$, CO, and $\COtwo$ tend to more closely follow their equilibrium profiles such that 
these species are more sensitive to metallicity than to temperature on HD 189733b and HD 209458b.
This conclusion will differ for colder planets like GJ 436b that are in the $\CHfour$ rather than 
CO stability field.

\subsection{Sensitivity to eddy diffusion coefficients\label{secteddysens}}

The model results are also sensitive to the assumed eddy diffusion coefficients, again because of 
the potential importance of transport-induced quenching on extrasolar giant planets.  Fig.~\ref{eddysens} 
demonstrates this sensitivity.  For larger eddy diffusion coefficients, transport-induced quenching 
becomes effective at deeper pressures, leading to larger mole fractions of quenched species.  
Larger $K_{zz}$ values also allow molecular constituents to be carried to higher altitudes.  Both 
these processes contribute to enhanced abundances of photochemically derived products when 
atmospheric mixing is more vigorous.  For example, when quenched species like $\CHfour$ and $\NHthree$ 
are more abundant and carried to higher altitudes, photochemical products like HCN and other nitriles, 
and $\CtwoHtwo$ and other hydrocarbons can build up over a larger column.  The $\Htwo$-to-H conversion 
level also changes with $K_{zz}$, which can have a significant effect on all molecular abundances.  
Minor heavy constituents like complex hydrocarbons and nitriles are particularly affected by $K_{zz}$, 
as these species often depend nonlinearly on the abundance of parent molecules like 
$\CHfour$ and $\NHthree$.  Note from Fig.~\ref{eddysens} that if $K_{zz}$ is sufficiently high, there 
is even a quench point for $\HtwoO$ and CO that differs from the equilibrium solution.

\begin{figure*}
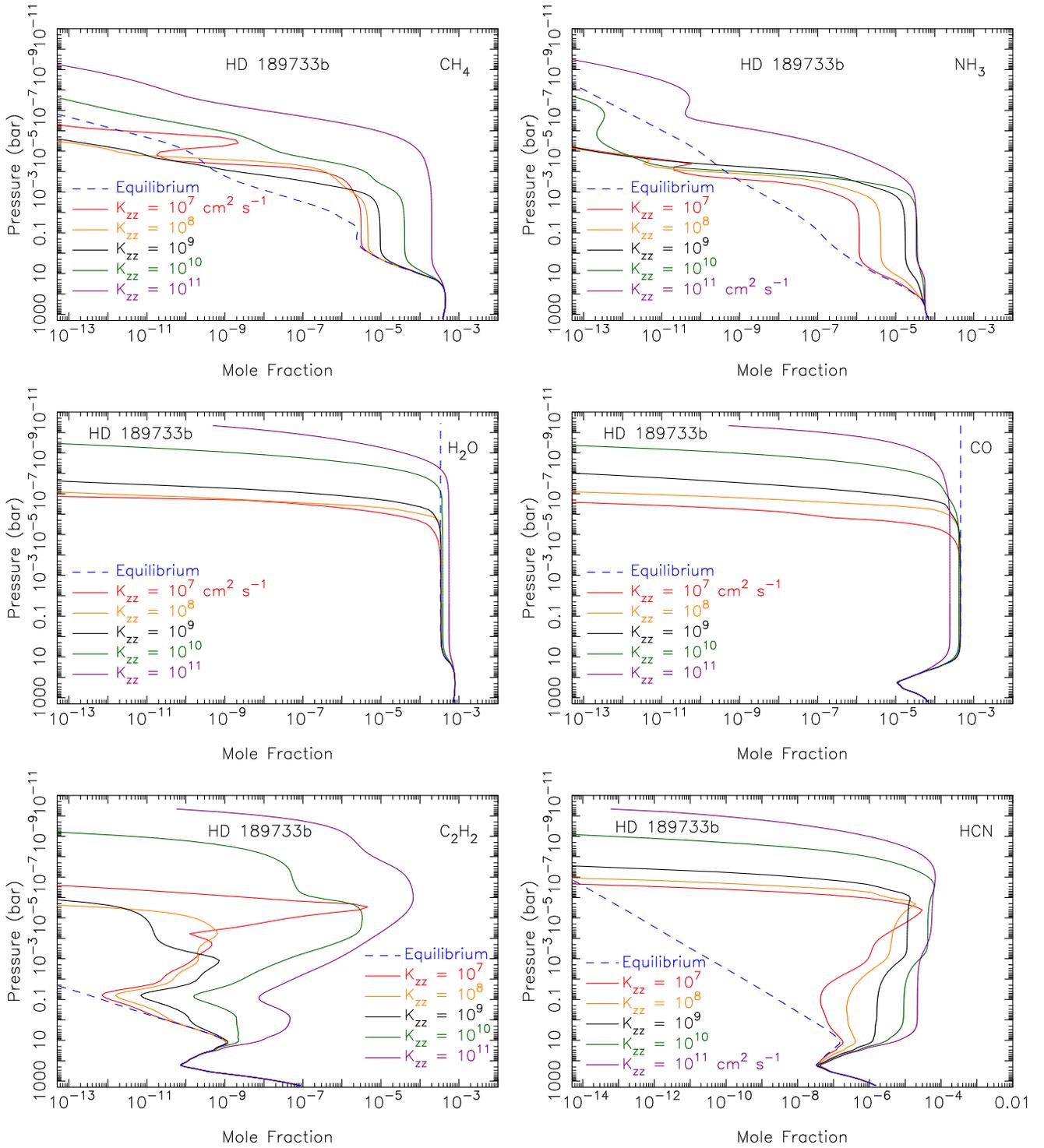

\begin{tabular}{ll}
{\includegraphics[angle=-90,clip=t,scale=0.37]{fig8a_color.ps}} 
& 
{\includegraphics[angle=-90,clip=t,scale=0.37]{fig8b_color.ps}} 
\\
{\includegraphics[angle=-90,clip=t,scale=0.37]{fig8c_color.ps}} 
& 
{\includegraphics[angle=-90,clip=t,scale=0.37]{fig8d_color.ps}} 
\\
{\includegraphics[angle=-90,clip=t,scale=0.37]{fig8e_color.ps}} 
& 
{\includegraphics[angle=-90,clip=t,scale=0.37]{fig8f_color.ps}} 
\\
\end{tabular}
\caption{Mole-fraction profiles for $\CHfour$ (top left), $\NHthree$ (top right), $\HtwoO$ (middle left), 
CO (middle right), $\CtwoHtwo$ (bottom left), and HCN (bottom right) in our dayside-average
HD 189733b model with an isothermal extension at high altitudes (see Fig.~\ref{figtemp}), for 
different assumptions about the eddy $K_{zz}$ profile (as labeled).  Larger $K_{zz}$'s imply 
greater vertical mixing, deeper quench points, and larger quenched abundances.  A color version 
of this figure is available in the online journal.\label{eddysens}}
\end{figure*}

General circulation models for highly irradiated hot Jupiters (see \citealt{showman09}, 
\citealt{showman10}, and references therein) suggest relatively strong vertical winds, which
according to our assumptions could lead to potentially vigorous mixing at pressures less 
than $\sim$100 bar ({\it i.e.}, Fig.~\ref{eddynom}).  We have assumed that convection in the 
adiabatic portion of the atmosphere at greater depth also contributes 
to vigorous vertical mixing \citep[see][]{stone76}.  Transport-induced quenching is then 
very important, and quenched disequilibrium species and their photochemical products are 
abundant.  In contrast, \citet{youdin10} suggest that weak turbulent mixing in extrasolar giant 
planet atmospheres could slow the cooling rate of the atmosphere and might explain the large, 
``inflated'' radii of some hot-Jupiter atmospheres.  \citet{youdin10} further suggest that 
$K_{zz}$ values greater than 10$^7$ cm$^2$ $\smone$ would overinflate hot-Jupiter radii beyond what 
can be supported by observations --- they favor much smaller $K_{zz}$ values of (1--5)$\scinot3.$ cm$^2$
$\smone$ at $\sim$550 bar, or in general $K_{zz} < 10^5$ cm$^2$ $\smone$ at $P$ $>$ 100 mbar for 
temperatures of $T_{deep}$ = 1500 K in the deep isothermal radiative region above the optically 
thick region of the atmosphere, similar to the few-bar-to-few-hundred-bar region for our HD 189733b 
temperature profiles.  With the \citet{youdin10} formalism, $K_{zz}$ scales as as $T_{deep}^5$, so 
that $K_{zz}$ on HD 209458b could be slightly higher by a factor of 2--3.  If eddy diffusion 
coefficients were as small as \citet{youdin10} suggest in the $\sim$2-500 bar region, then 
transport-induced quenching would be significantly suppressed in our HD 189733b and HD 209458b 
models, and the quenched abundances of $\CHfour$, $\NHthree$, and HCN would be quite low --- more 
consistent with what is predicted from thermochemical equilibrium --- and these species (including 
methane) would likely not be observable.  
Although waves propagating up from the deep interior and down from the weather layer may keep 
$K_{zz}$ from being as low as is suggested by \citet{youdin10}, the idea of a ``mechanical 
greenhouse'' is an interesting one.  Spectral observations that can unambiguously confirm the 
presence of $\CHfour$, $\NHthree$, or HCN at levels in excess of equilibrium could help constrain 
the strength of turbulent mixing in extrasolar-giant-planet atmospheres.

Given that the $\NHthree$ quench point is likely in the adiabatic region of the deep atmosphere, it is 
interesting to speculate that the $\NHthree$ abundance might eventually help provide a measure of 
the entropy of the adiabat, which is currently known only crudely from the planet's radius.  For a 
given mass, planets with larger radii have higher-entropy adiabats and lower-pressure 
radiative-convective boundaries.  Although the entropy of the adiabat will certainly affect the 
$\NHthree$ quench point, the overall uncertainties in the eddy diffusion coefficients, in the planetary 
metallicity, and in the kinetics of $\NHthree$ quenching might make such a suggestion impractical.

\subsection{Photochemistry and transport time constants and their implications\label{secttime}}

Because our models remain at a single solar zenith angle for the long duration of the simulation, 
they are not true analogs for the situation on a real planet, even if the planet is tidally 
locked and has one hemisphere always facing the star.  Real planets have meteorology, and dynamical 
models indicate that strong winds will be acting on close-in transiting exoplanets to transport 
species zonally, meridionally, and vertically (see \citealt{showman10} for a review of exoplanet 
dynamical models); zonal jets are predicted to be particularly strong, such that material is 
transported longitudinally across a hemisphere with a horizontal transport time scale $\tau_{dyn,h}$ 
$\approx$ 2$\scinot5.$ s on HD 209458b \citep{coop06}.  A parcel of gas will therefore experience 
variable external forcing from the continuously changing irradiation angles, and those effects 
can influence the chemistry.  A full 3-D model of the radiative, dynamical, and chemical coupling 
needed to describe this situation would be computationally prohibitively expensive, but 
simplifications such have been described by \citet{coop06} can be made to track the key chemical 
effects.  Although we do not attempt these kinds of simulations with our simple 1-D models, we 
can at least examine the time constants in the system and make some comments about what the 
time-constant analysis implies about the composition of more realistic planetary analogs.  

\citet{coop06} convincingly argue that the large CO abundance that is predicted for dayside 
HD 209458b will not be chemically converted to $\CHfour$ at the terminators or in the nightside 
stratosphere, despite the fact that $\CHfour$ would be favored in chemical equilibrium.  Vertical 
transport time scales are simply much shorter than the chemical conversion time scale between 
CO and $\CHfour$.  In addition, note that just as there is transport-induced quenching in the 
vertical direction, horizontal transport-induced quenching can also operate \citep{coop06}.  
Will this horizontal quenching affect photochemically produced species?  One might particularly 
want to know whether HCN will reconvert into $\CHfour$ and $\NHthree$, whether $\CtwoHtwo$ will 
reconvert into $\CHfour$, and whether H will reconvert to $\Htwo$ at high altitudes on the 
nightside of HD 189733b and HD 209458b, and whether vertical transport-induced quenching will 
lead to different $\CHfour$, $\NHthree$, and HCN mole fractions on the dayside as compared with 
the terminators (as predicted by our models) or whether horizontal transport-induced quenching 
will homogenize the quenched mole fractions in the stratosphere.  

In general, the horizontal advection time scale over a planetary radius is comparable to or 
less than the vertical advection time scale over a scale height, suggesting that if temperatures 
are low enough in the upper troposphere and lower stratosphere to shut down $\CHfour$ 
$\leftrightarrow$ CO, $\NHthree$ $\leftrightarrow$ $\Ntwo$, and $\CHfour$ $\leftrightarrow$ 
$\NHthree$ $\leftrightarrow$ HCN interconversion, then horizontal quenching will also be in 
operation in these regions.  As such, zonal winds would be expected to operate efficiently 
enough to homogenize the CO, CH$_4$, H$_2$O, $\Ntwo$, and NH$_3$ abundances in longitude, and
there should be no terminator-vs.-dayside differences in the mole fractions of these species in 
regions where those interconversion reactions dominate.  
However, in the middle and upper stratosphere, the disequilibrium that results from the absorption 
of UV photons has a different set of quench reactions operating under much different conditions.  
Photochemically produced species tend to have shorter lifetimes, and vertical transport time scales
are also expected to be shorter at high altitudes.  Zonal winds may also increase with altitude 
in low-latitude regions \citep{showman09}, but horizontal transport time scales eventually become 
longer than the photochemical time constants.  Therefore, longitudinal 
differences are more likely to be maintained at higher altitudes ({\it i.e.,} pressures less than
$\sim$0.01-0.001 mbar, depending on the species and model).  

In our nominal dayside HD 189733b 
atmosphere for instance (see Fig.~\ref{fighc}), the HCN that replaces $\CHfour$ and $\NHthree$ at 
pressures less than a few mbar is produced through scheme (14).  This scheme does not operate 
effectively at pressures greater than a few mbar, and vertical transport dominates the behavior of 
the HCN profile at these pressures such that the HCN that was produced at higher altitudes diffuses 
with a constant flux to the lower atmosphere; however, the HCN photochemical lifetime is only 
$\sim$2 hr at 10$^{-2}$ mbar and less than 3 min at 10$^{-3}$ mbar, compared to a zonal transport 
time scale that is likely greater than 20 hr at these pressures \citep{showman09}.
The photochemical lifetime of $\CtwoHtwo$ is even shorter, suggesting 
that longitudinal variations in HCN, $\CHfour$, $\NHthree$, and $\CtwoHtwo$ could be maintained at 
high altitudes on HD 189733b, such that $\CHfour$ and $\NHthree$ could recover at high altitudes 
at night.  The total column abundances of $\CHfour$ and $\NHthree$ would then be larger on the 
nightside than on the dayside, particularly at pressures less than $\sim$1 mbar.

It is interesting to speculate what might happen on HD 189733b when the high-altitude $\CHfour$ that 
comes back at night is transported by the zonal winds to the western (dawn) terminator (assuming 
atmospheric super-rotation, as is expected from the dynamical models).  As the gas flows into the
starlight again, the high-altitude CH$_4$ will strongly absorb in the ultraviolet and at certain 
visible and near-IR wavelengths where the methane bands have some optically thick lines.  Does
this expected ``pulse'' of high-altitude absorption on the dawn limb affect the high-altitude 
thermal structure or the $\CHfour$ fluorescent emission behavior enough to cause emission in these 
optically thick lines?  If the heating rates are greater than the chemical destruction rates, if 
there is sufficient methane at high altitudes, and if there is a lag in the cooling rates, then 
such a scenario might help explain the apparent 3.25 $\mu$m emission seen in the \citet{swain10} 
ground-based observations of HD 189733b.  We note that if this process were really operating, the 
emission in the $\nu_3$ methane band should be more obvious in the transit of the dawn 
terminator than the dusk terminator because more methane will be present at high altitudes on the 
dawn terminator (again, assuming atmospheric super-rotation).  Even if cooling rates were rapid, the 
sudden pulse of photons at high altitudes would provide energy needed to drive the chemical reactions 
or fluorescence and might excite any molecules present.  

\begin{figure*}
\begin{tabular}{ll}
{\includegraphics[clip=true,scale=0.40]{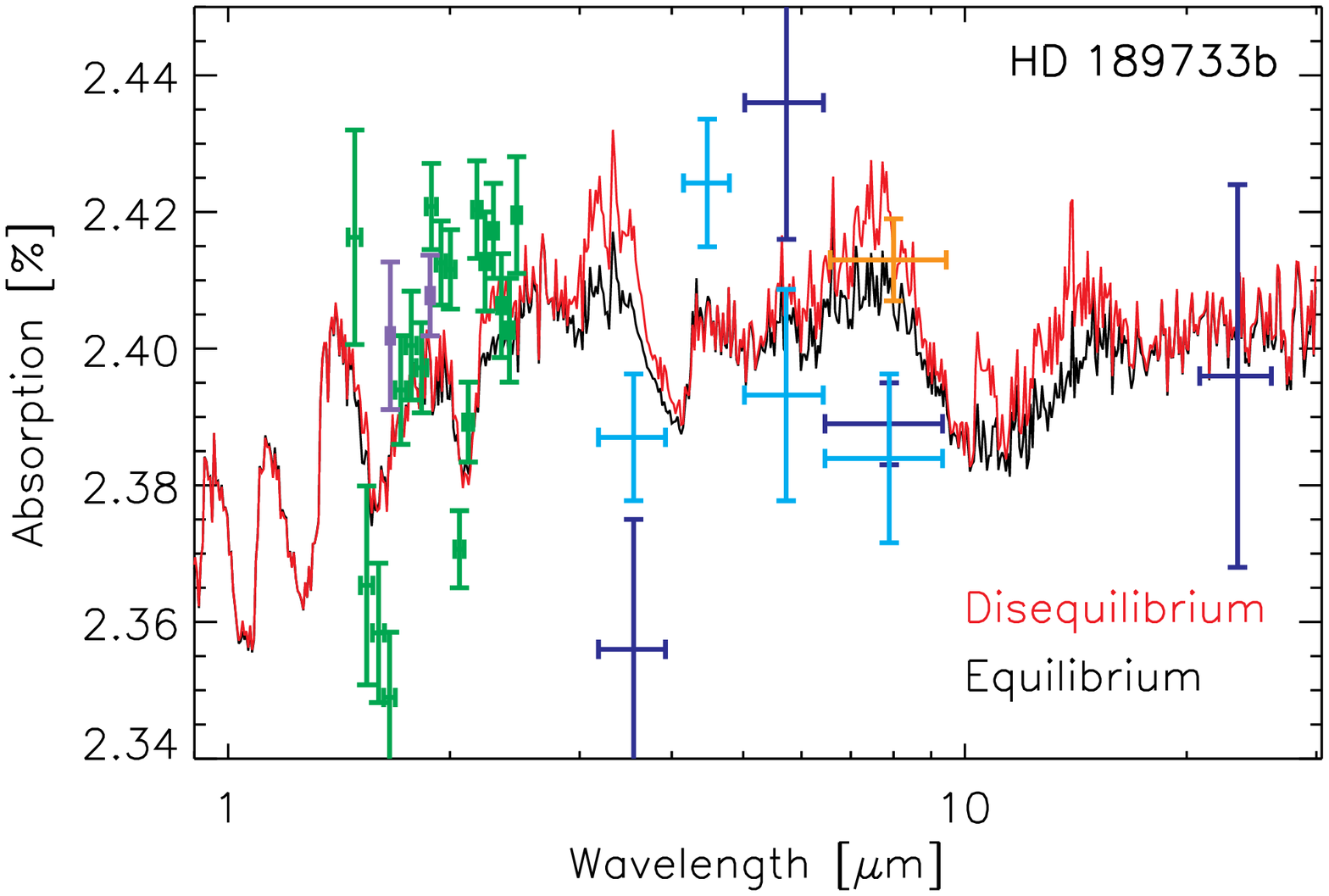}} 
& 
{\includegraphics[clip=true,scale=0.40]{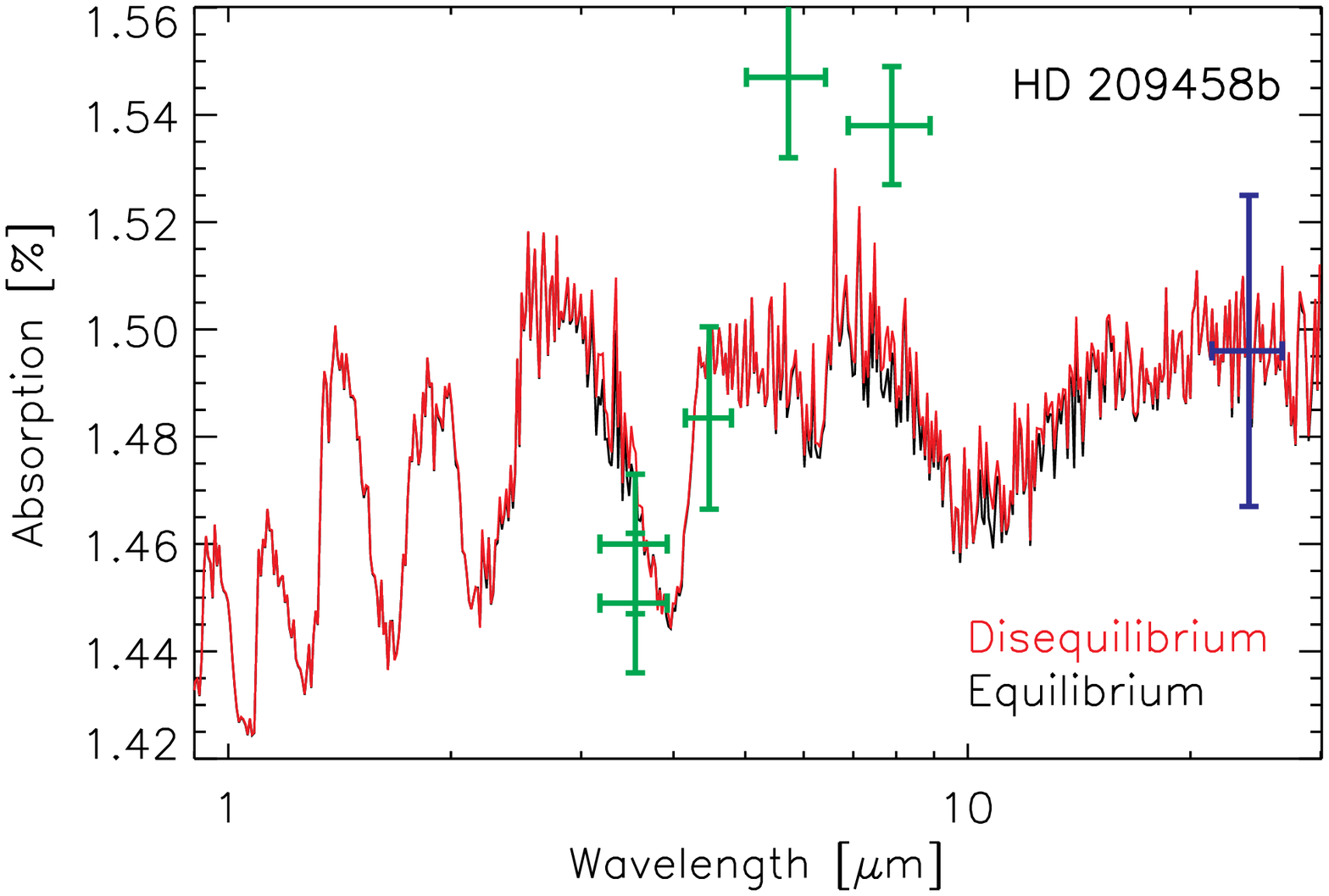}} 
\\
\end{tabular}
\caption{Synthetic transit spectra calculated for our HD 189733b (left) and HD 209458b 
(right) thermochemical and photochemical kinetics and transport models (red curves) that 
assume a terminator-average thermal structure and nominal $K_{zz}$ profile, compared 
with synthetic transit spectra from the thermochemical equilibrium models (black curves) 
for the same assumed thermal structure.  All models assume a 1x solar composition.  
Absorption depth is calculated as the square of apparent planet-to-star radius ratio.  
Observations for HD 189733b are shown as data points with associated error bars: green, 
\citet{swain08b} {\it HST\/}/NICMOS; purple, \citet{sing09} {\it HST\/}/NICMOS; light blue, 
\citet{desert09} {\it Spitzer\/}/IRAC; darker blue, \citet{beaulieu08} for {\it Spitzer\/}/IRAC 
3.6 $\mu$m and 5.8 $\mu$m, \citet{knut07} for {\it Spitzer\/}/IRAC 8 $\mu$m, and \citet{knut09} 
for {\it Spitzer\/}/IRAC 24 $\mu$m; orange, \citet{agol10} {\it Spitzer\/}/IRAC 8 $\mu$m.
For HD 209458b, the green data points represent the {\it Spitzer\/}/IRAC data of \citet{beaulieu09}
and the blue data point at 24 $\mu$m represents the average of the {\it Spitzer\/}/MIPS values from
\citet{richardson06} and H. Knutson (2009, private communication).
A color version of this figure is available in the online journal.\label{189spectrans}}
\end{figure*}

This process would work on HD 209458b as well, but it would 
be less apparent due to the overall smaller quenched abundance of $\CHfour$ and the correspondingly 
smaller column abundance of $\CHfour$ at high altitudes.  On planets like GJ 436b that are in the 
$\CHfour$ stability regime at the deep CO $\leftrightarrow$ $\CHfour$ quench point, the effect could 
be particularly important.  Because $\CHfour$ is likely relatively abundant at high altitudes on 
both terminators on GJ 436b, we suggest that high-spectral-resolution modeling of radiative absorption 
and emission on GJ 436b might be called for to determine whether portions of the $\nu_3$ band would 
be expected in emission rather than absorption on this planet (due to a high-altitude thermal inversion 
from UV, visible, and near-IR absorption from $\CHfour$), potentially explaining the apparent lack of 
$\CHfour$ 
absorption at 3.6 $\mu$m in eclipse observations of this planet \citep{stevenson10}.  
These ideas are quite speculative at this point but deserve further quantitative 
study.  The possibility of dawn-dusk differences in the transit spectra could potentially be explored 
with the {\it James Webb Space Telescope}\/ \citep[JWST; see][]{fort10}.

\section{Implications with respect to observations}\label{obsimp}

\begin{figure*}
\begin{tabular}{ll}
{\includegraphics[clip=t,scale=0.4]{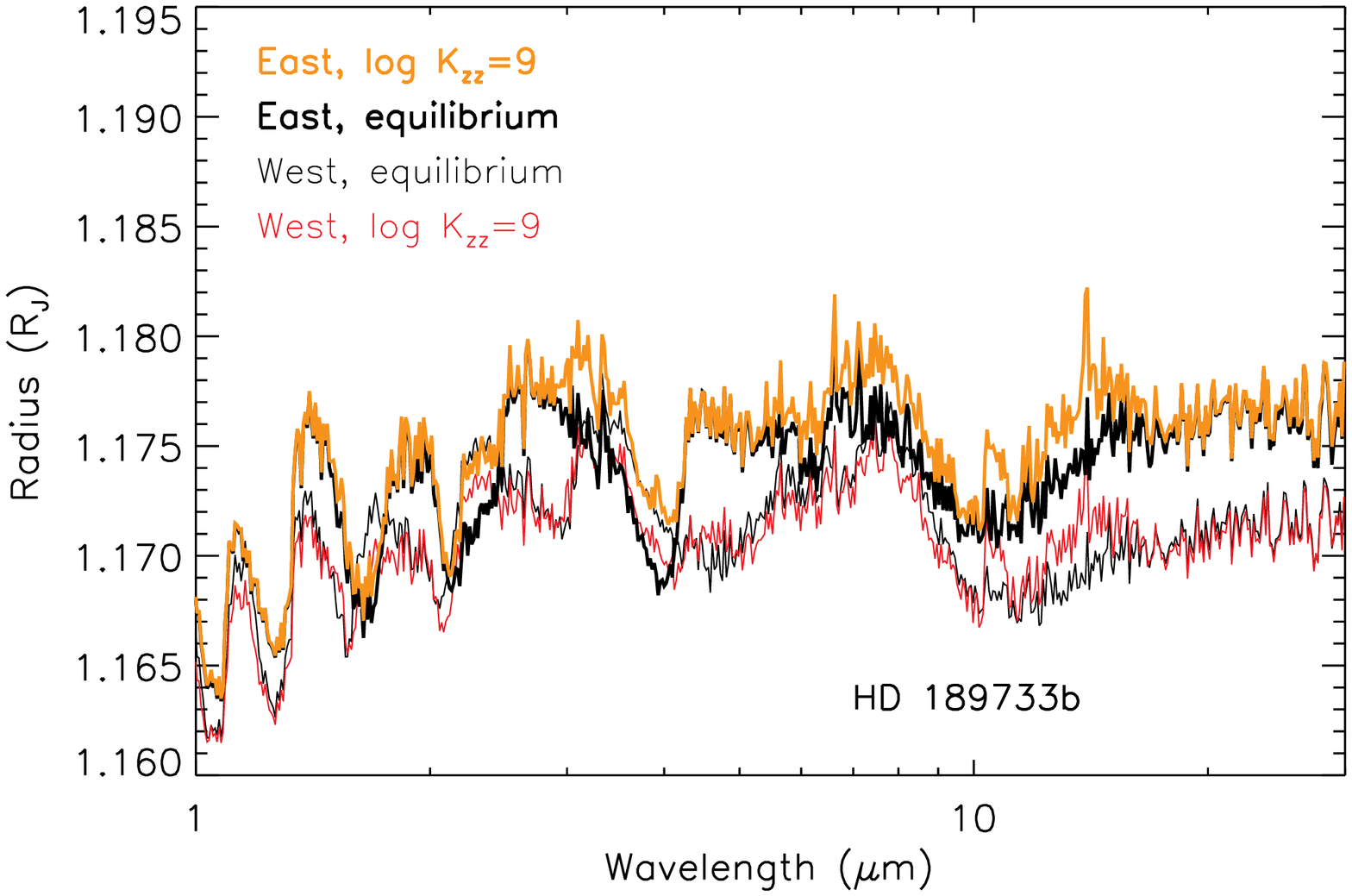}} 
& 
{\includegraphics[clip=t,scale=0.4]{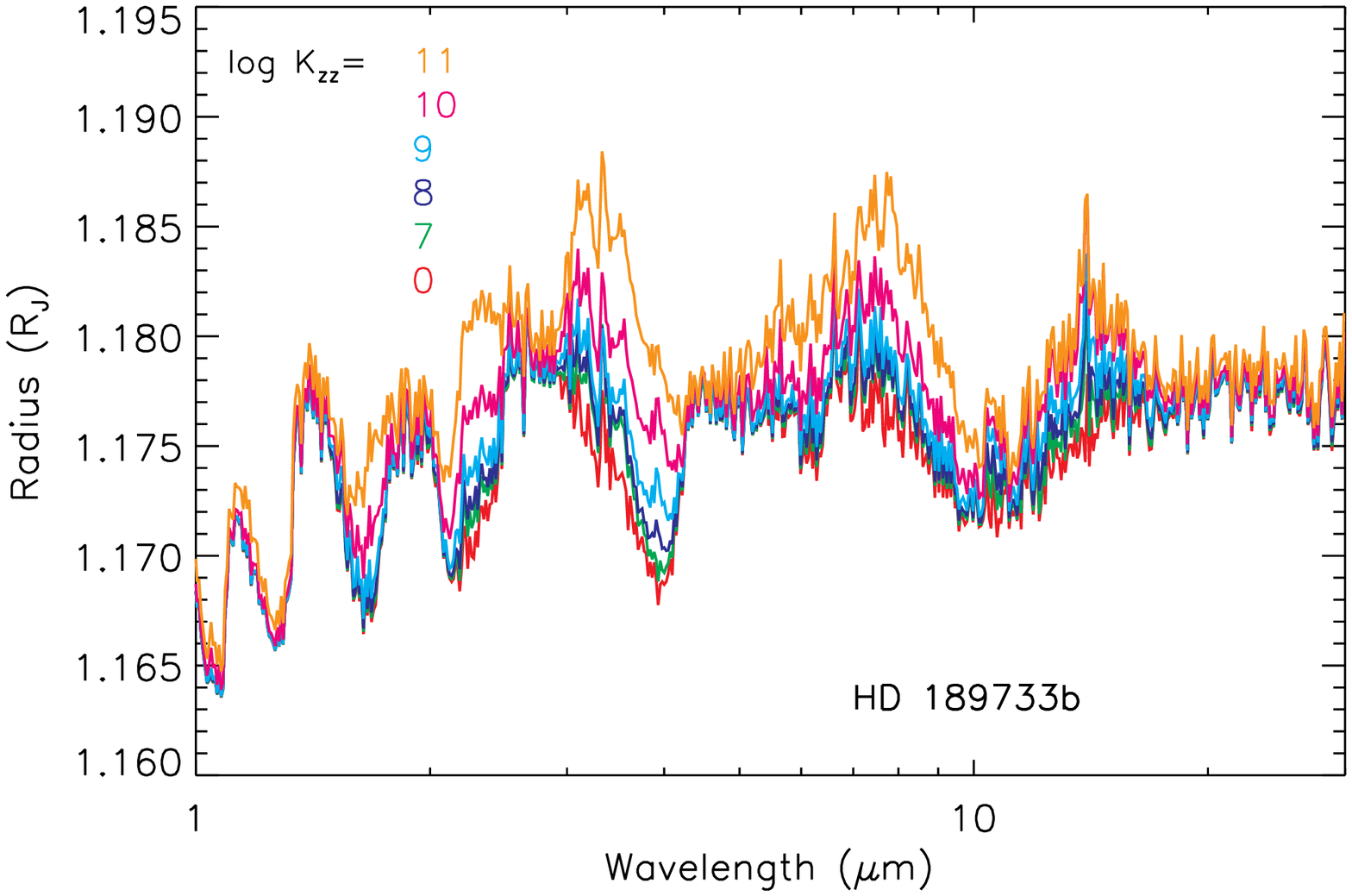}} 
\\
\end{tabular}
\caption{Synthetic transit spectra (apparent planetary radius in units of Jupiter radii) 
calculated for HD 189733b for different assumptions about the thermal structure (left) and 
the eddy diffusion coefficient profile (right).  In the left panel, the gray curve 
represents a model with the west-terminator-average temperature profile and an equilibrium 
composition, the red curve represents a model with the west-terminator-average temperature 
profile and a composition indicated by our disequilibrium model with $K_{zz}$ = 10$^9$
cm$^2$ $\smone$, the black curve represents a model with the east-terminator-average temperature
profile and an equilibrium composition, and the orange curve represents a model with the
east-terminator-average temperature profile and a composition indicated by our disequilibrium 
model with $K_{zz}$ = 10$^9$ cm$^2$ $\smone$.  In the right panel, the models assume a
dayside-average thermal structure: the red curve is the equilibrium model, and the other curves
represent disequilibrium models with assumed constant $K_{zz}$ values of 10$^7$ (green), 10$^8$
(dark blue), 10$^9$ (cyan), 10$^{10}$ (magenta), and 10$^{11}$ (orange) cm$^2$ $\smone$.
A color version of this figure is available in the online journal.\label{189transens}}
\end{figure*}

Figure \ref{189spectrans} illustrates how the predicted HD 189733b and HD 209458b spectra 
from the primary transits are altered by the presence of the disequilibrium products derived 
in our models.  Different radiative-transfer models in the literature are producing 
different predicted transmission spectra for exoplanets, for reasons that are unclear 
\citep[cf.][]{shabram11,beaulieu11}.  For Fig.~\ref{189spectrans}, the synthetic transmission 
spectra are calculated as is described in \citet{fort10} and 
\citet{shabram11} for the terminator-average temperature profiles shown in Fig.~\ref{figtemp}.  
The calculations assume either a thermochemical-equilibrium composition or use the results from 
our thermochemical and photochemical kinetics and transport models with the nominal $K_{zz}$ 
profiles.  The approximate pressure 
levels from which the spectral features originate are discussed further in \citet{fort08a}.  
For HD 189733b, Fig.~\ref{189spectrans} shows clear excess absorption from the disequilibrium 
model at wavelengths near $\sim$2.1--2.5 $\mu$m and $\sim$2.9--4 $\mu$m due to $\CHfour$ and 
to a lesser extent due to HCN near 3 $\mu$m, additional absorption in the $\sim$7--9 $\mu$m 
region due largely to $\CHfour$ (with HCN contributing near 7 $\mu$m), and additional 
absorption in the $\sim$9--15 $\mu$m region due to NH$_3$ (centered near 10.5 $\mu$m) and 
HCN (centered near 14 $\mu$m), with a much lesser contribution from $\CtwoHtwo$ (centered 
near 13.6 $\mu$m).  Note that the column abundance of $\CtwoHtwo$ in our nominal model 
is small enough that acetylene absorption features are not very prominent.  For HD 209458b, 
spectral differences between the equilibrium and disequilibrium models are much more muted, 
as the overall compositional changes are slight between the equilibrium and disequilibrium 
models at the pressure levels at which the observations are sensitive.  Our modeling 
suggests that photochemistry and transport-induced quenching will have only a minor effect 
on the spectral properties of the hottest ``hot Jupiters'' or those with strong stratospheric 
thermal inversions, as the high temperatures enable rapid kinetics that can drive the 
atmosphere back toward equilibrium.

For HD 189733b, however, the disequilibrium chemistry resulting from photochemistry and
transport-induced quenching measurably alters the composition such that the disequilibrium 
model shows additional absorption and a greater wavelength-dependent contrast overall than the 
equilibrium model.  Although the near-infrared absorption band positions predicted by the 
model are generally consistent with observations, both the equilibrium and disequilibrium 
models apparently underestimate the near-IR band depths and the overall amplitude of the 
wavelength-dependent variations --- a fact previously discussed for equilibrium models by 
\citet{fort10}.  Even when disequilibrium constituents are considered, the model fit to the 
near-IR {\it HST\/}/NICMOS data of \citet{swain08b} and \citet{sing09} is not significantly 
improved, as water absorption and $\Htwo$ collision-induced absorption dominate the spectral 
behavior at near-IR wavelengths, and these opacity sources remain roughly the same in the 
two models.  The main exceptions to this statement are wavelength regions centered near 
2.3 $\mu$m and 3.3 $\mu$m, where methane absorption can have a significant effect.  The 
additional methane in the disequilibrium models for HD 189733b improves the fit in the 
2.3 $\mu$m region \citep[see also][]{swain08b,madhu09} but degrades the fit to the IRAC 
3.6 $\mu$m data \citep[e.g.,][]{beaulieu08,desert09}.  At mid-IR wavelengths, 
the models also appear to underestimate the spectral contrast between different wavelengths, and
neither model does a good job of reproducing the {\it Spitzer\/}/IRAC photometric data, although
the data have large error bars, and different groups using different analysis procedures for the
same datasets do not always agree on the observed flux in the IRAC bands.
As was discussed by \citet{fort10}, higher metallicity models with their increased CO and $\COtwo$
abundances would improve the fit to the relative absorption strengths of the 3.6, 4.5, and 5.8
$\mu$u IRAC band data points from the \citet{desert09} analysis, but other model-data comparison
problems would still remain.  Note that we have not recalculated the thermal profile that will 
result from the additional disequilibrium opacity sources, and the expected altered thermal 
structure may also affect the observed spectrum.

As an example of the sensitivity of transmission spectra to the thermal profile, 
Fig.~\ref{189transens} shows how the synthetic spectra for HD 189733b change for different 
assumptions about the thermal structure at the terminators --- particularly the western vs.~eastern 
terminator-average profiles from Fig.~\ref{figtemp}; Fig.~\ref{189transens} also shows the 
sensitivity of the transmission spectra to different assumptions about the eddy diffusion 
coefficient profile, for assumed constant $K_{zz}$ profiles ranging from 10$^7$ to 10$^{11}$ 
cm$^2$ $\smone$.  The thermal structure affects such things as the atmospheric scale height, 
which describes how extended the atmosphere becomes, so that a colder planet will have less of an 
apparent cross section during transit for all other factors being equal.  As such, the colder 
western-terminator model exhibits a smaller apparent radius than the eastern terminator model.
Note that the spectral changes caused by the equilibrium vs. disequilibrium models have less of an 
overall impact on the apparent absorption during transit than assumptions about the thermal 
structure.  The eddy
diffusion coefficient profile affects the extent to which heavy molecular constituents can be mixed
to high altitudes on an $\Htwo$-dominated planet; in general, the larger the eddy diffusion
coefficient, the higher the altitude to which the molecular constituents can be carried, which 
allows photochemical species to be produced over a larger column of the atmosphere, leading to
larger column abundances and increased absorption by the photochemical products.  Eddy diffusion
coefficients also affect the abundances of the species controlled by transport-induced quenching, 
as is discussed in sections~\ref{sectquench} and \ref{secteddysens}.

\begin{figure*}
\begin{tabular}{ll}
{\includegraphics[clip=t,scale=0.5]{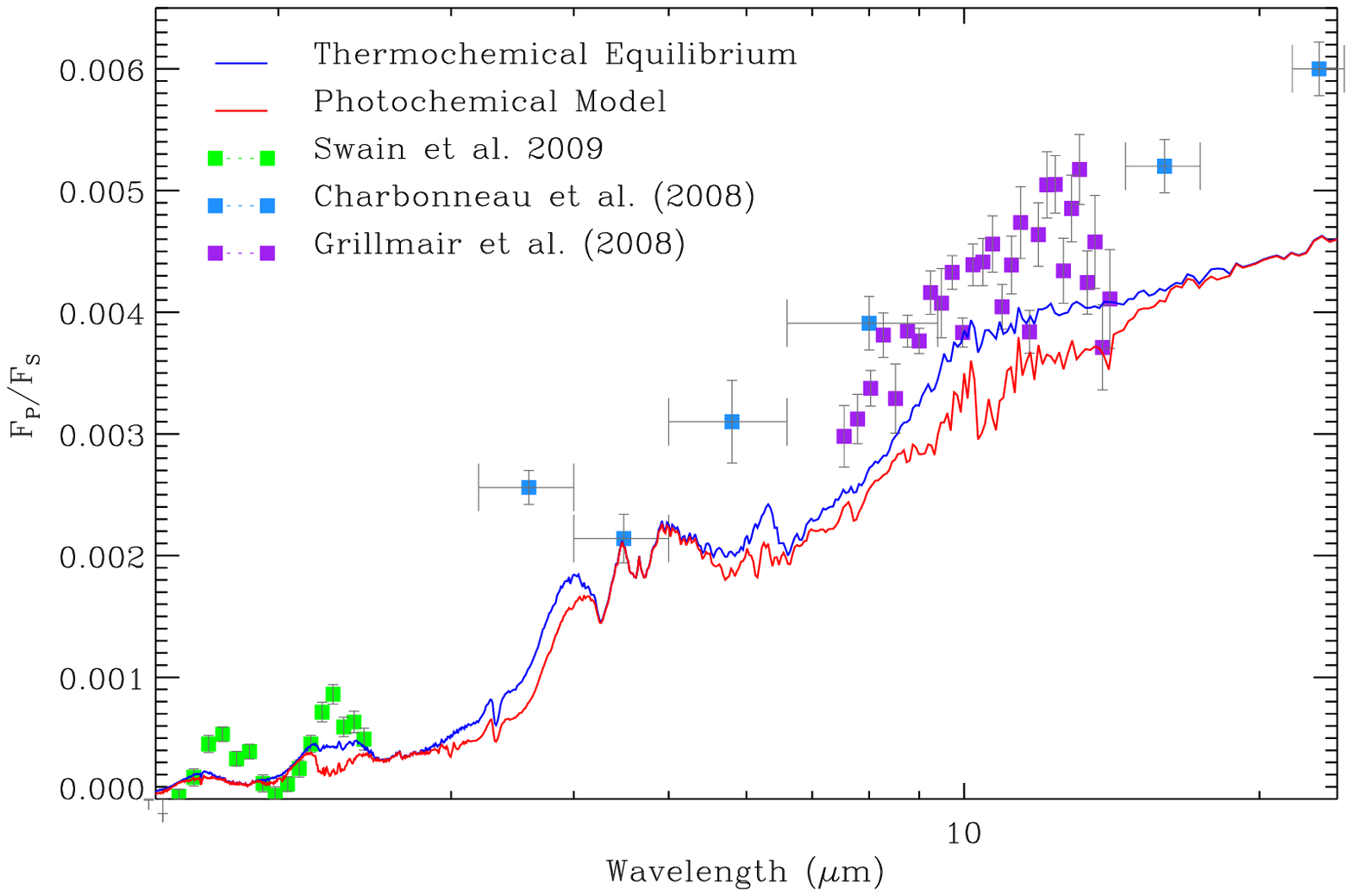}} 
& 
{\includegraphics[clip=t,scale=0.5]{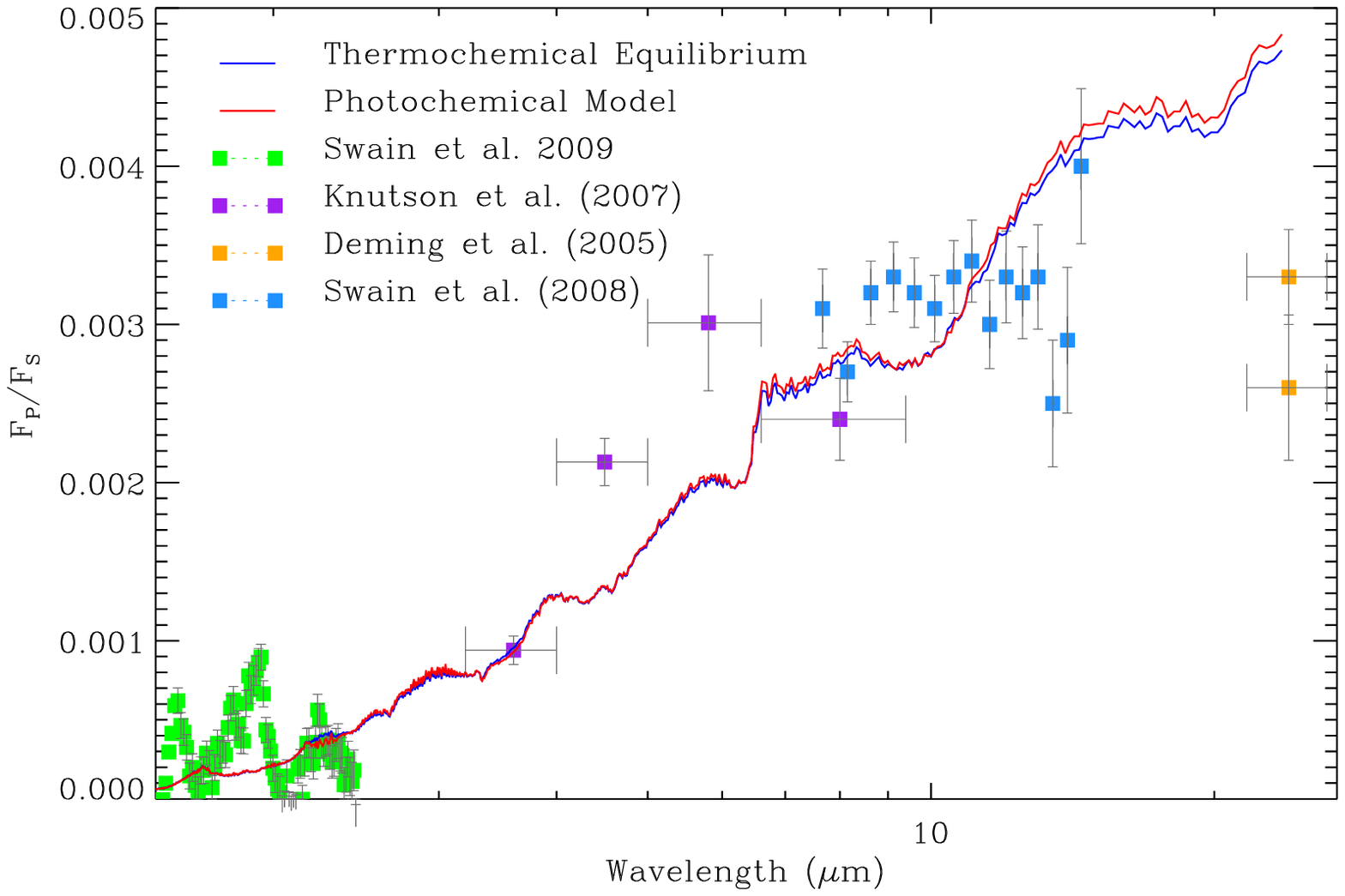}} 
\\
\end{tabular}
\caption{Synthetic emission spectra for HD 189733b (left) and HD 209458b (right) for our 
thermochemical and photochemical kinetics and transport models (red curves) that assume 
a dayside-average thermal structure and corresponding nominal $K_{zz}$ profile for the 
planet in question (see Figs.~\ref{eddynom} and \ref{figtemp}), compared with synthetic 
spectra from the thermochemical equilibrium model (blue curves) for the same assumed thermal 
structure.  All models assume a 1x solar elemental composition.  Spectra are calculated as 
the ratio of the flux of the planet to the flux of the 
star.  Also shown 
are data points from various secondary-eclipse observations, as labeled.
A color version of this figure is available in the online journal.\label{189specemit}}
\end{figure*}

\begin{figure*}
\begin{tabular}{ll}
{\includegraphics[clip=t,scale=0.5]{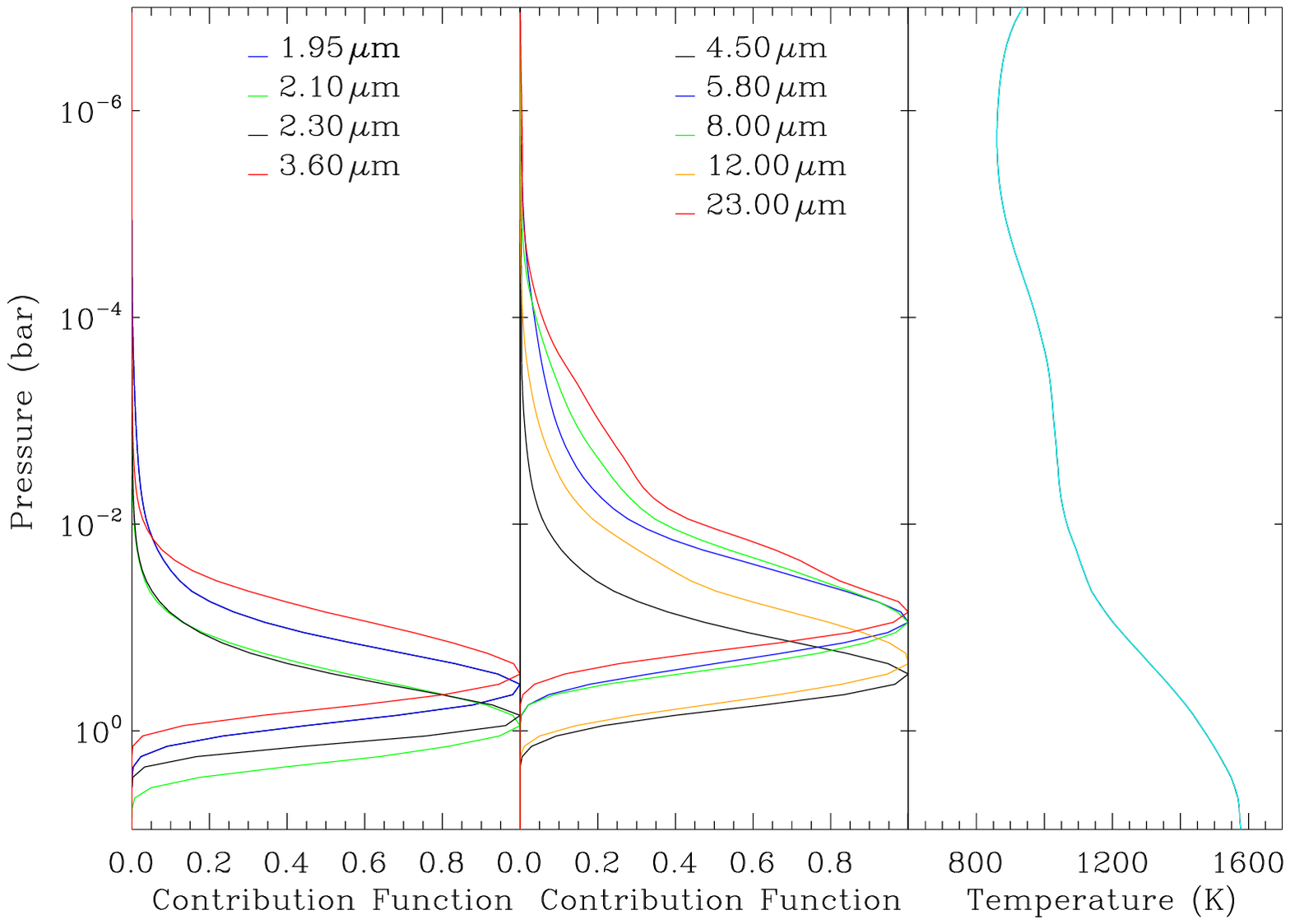}} 
& 
{\includegraphics[clip=t,scale=0.5]{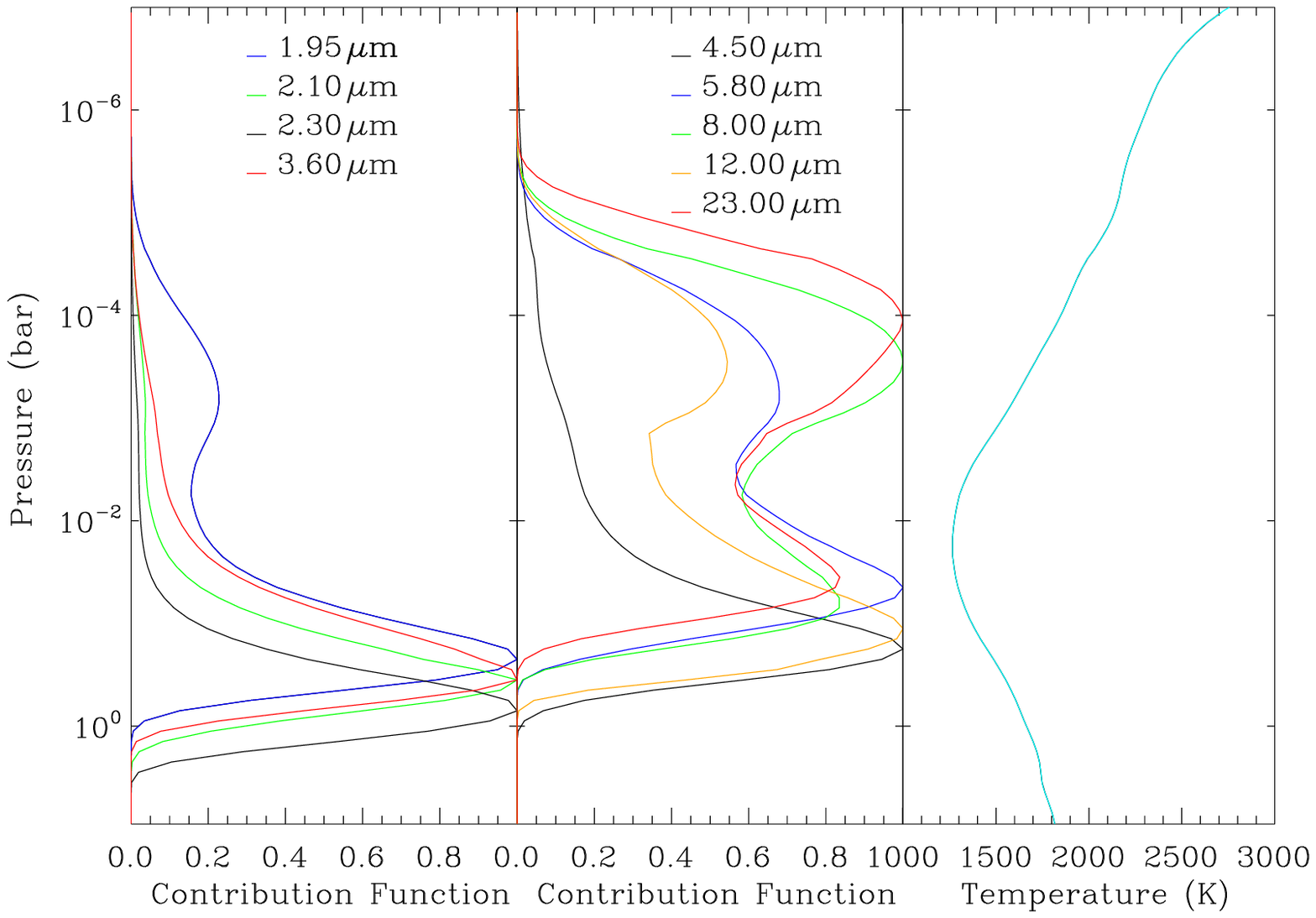}} 
\\
\end{tabular}
\caption{Contribution functions, which give the relative contribution of a given pressure level to
the observed emission at any particular wavelength, are shown alongside the thermal structure from
our dayside-average chemical models for HD 189733b (left) and HD 209458b (right). 
A color version of this figure is available in the online journal.\label{contribfct}}
\end{figure*}

Synthetic emission spectra of HD 189733b and HD 209458b for dayside-average, secondary-eclipse 
conditions are shown in Fig.~\ref{189specemit}.  For these calculations, local thermodynamic 
equilibrium (LTE) conditions are assumed, and the effects of the important spectrally active 
gases $\HtwoO$, CO, CH$_4$, CO$_2$, HCN, $\NHthree$, $\CtwoHtwo$, and $\CtwoHsix$ are 
considered.  The opacities of these gases are garnered from different sources.  The hot water 
line parameters from the latest November 2010 version of the HITRAN line list \citep{rothman09} 
are adopted, using the original data of \citet{barber06} and \citet{zobov08}.  Absorption by 
$\COtwo$ is calculated from the HITRAN hot $\COtwo$ line list \citep{tashkun03}, as is that of 
CO and $\NHthree$ \citep{rothman09}, and the methane line parameters derive from the hot 
$\CHfour$ line parameters of \citet{freedman08}.  Note that for both the transmission and 
emission modeling, the opacities of HCN and $\CtwoHtwo$ --- and really all species other than 
$\HtwoO$, $\COtwo$, and CO --- are likely underestimated because the line parameters were derived 
from cool, low-temperature 
conditions.  Absorption coefficients are calculated using a line-by-line technique every 0.004 
cm$^{-1}$ wavenumbers to provide the k-coefficients that are included in the radiative-transfer 
calculations.  For the calculations of the emission spectra, the atmosphere is divided into 80 
vertical grid points that extend from 10$^{-7}$ to 10 bar.  The effects of particulates are 
excluded.  The calculations explore the spectroscopic effects of the compositions that are 
indicated by the different chemical models, while the temperature structure is held fixed.  
The depths at which the various emission signatures originate can be seen from the contribution 
functions shown in Fig.~\ref{contribfct}.

The synthetic emission spectra shown in Fig.~\ref{189specemit} illustrate that spectral 
signatures from the disequilibrium model are noticeably altered from the equilibrium 
predictions for the cooler HD 189733b but not for the warmer HD 209458b.   The predicted 
eclipse depths are only slightly modified at long wavelengths due to the disequilibrium 
chemistry on HD 209458b.  That conclusion changes for HD 189733b, as the the key 
disequilibrium constituents CH$_4$, HCN, and NH$_3$ are more abundant and have a larger 
impact on the predicted spectrum.  On HD 189733b, the disequilibrium-model spectra diverge
from the equilibrium spectra in the 7--9 $\mu$m range due to the $\nu_4$ $\CHfour$ band at 7.7
$\mu$m, in the $\sim$3--4 $\mu$m and $\sim$2.1--2.5 $\mu$m ranges due to the 3.3 and 2.3 $\mu$m bands
of CH$_4$, in the 5.5--6.6 $\mu$m range due to the $\nu_4$ band of $\NHthree$ at 6.15 $\mu$m, in the
8--12 $\mu$m range due to the $\nu_2$ band of $\NHthree$ at 10.5 $\mu$m, and in the 13--15 $\mu$m
range due to the $\nu_2$ band of HCN at 14 $\mu$m.  The individual contributions from the different
disequilibrium species are better illustrated in Fig.~\ref{189emitwgas}.

Our emission spectra for both HD 189733b and HD 209458b do not compare well with the observed fluxes
during secondary eclipse
\citep[e.g.,][]{deming05b,richardson07,knut08,charb08,grill08,swain08a,swain09a,swain09b,madhu09}.
For HD 209458b, apparent emission features at 4.5 and 5.8 $\mu$m (see, e.g., \citealt{knut08}) are
not present in the model spectra, which could be due to an inaccurate temperature profile adopted
for our model (e.g., a thermal inversion beginning at higher altitudes in our model than on the 
real planet), or some other inappropriate assumptions for the spectral or chemical models.  For 
HD 189733b, the model eclipse depths are much smaller than has been observed at all wavelengths, 
perhaps resulting from inaccurate model assumptions such as the atmospheric thermal structure of 
the planet, the planet's metallicity and radius, or the spectrum of 
the host star (note that we have used the models of \citealt{castelli04} to simulate the spectrum 
of the host stars, with other stellar and planetary parameters as described in \citealt{torres08}).
In particular, the model-data comparisons shown in Fig.~\ref{189specemit} suggest that our assumed 
dayside atmospheric temperatures may be too cold in the 0.01-1 bar region.  We therefore also
calculate synthetic spectra for the warmer 1-D ``2$\pi$'' model of \citet{fort06a,fort10} (see
Fig.~\ref{figtemp}); the results are shown in Fig.~\ref{189emithot}, along with the corresponding
contribution functions.
Note that the magnitude of the eclipse depths are better predicted and the disequilibrium effects 
are reduced with this warmer model.  

\begin{figure}
{\includegraphics[clip=t,scale=0.52]{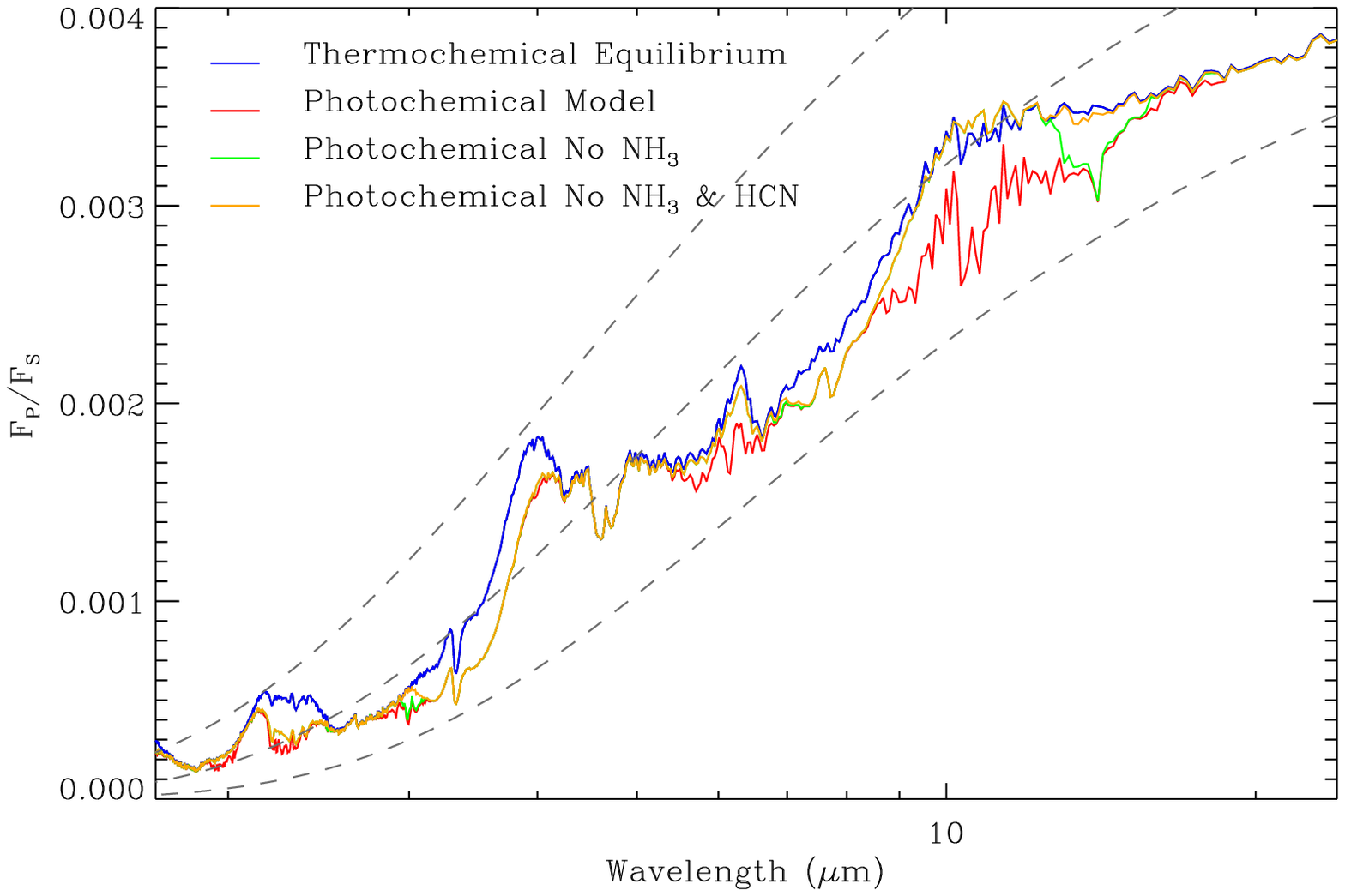}} 
\caption{Synthetic emission spectra for HD 189733b, as described in
Fig.~\ref{189specemit}, except the green curve shows how the results change when the 
contribution of $\NHthree$ is removed from the kinetics and transport model spectra, and the 
orange curve shows how the results change when the contributions of both $\NHthree$ and HCN 
are removed from the kinetics and transport model spectra.  Note that methane opacity is the 
primary source of the divergence of the orange curve from the the blue equilibrium curve.
The dashed lines are blackbody curves for particular temperatures: top curve is for 1400 K, 
middle curve is for 1200 K, bottom curve is for 1000 K.
A color version of this figure is available in the online journal.\label{189emitwgas}}
\end{figure}

\begin{figure*}
\begin{tabular}{ll}
{\includegraphics[clip=t,scale=0.52]{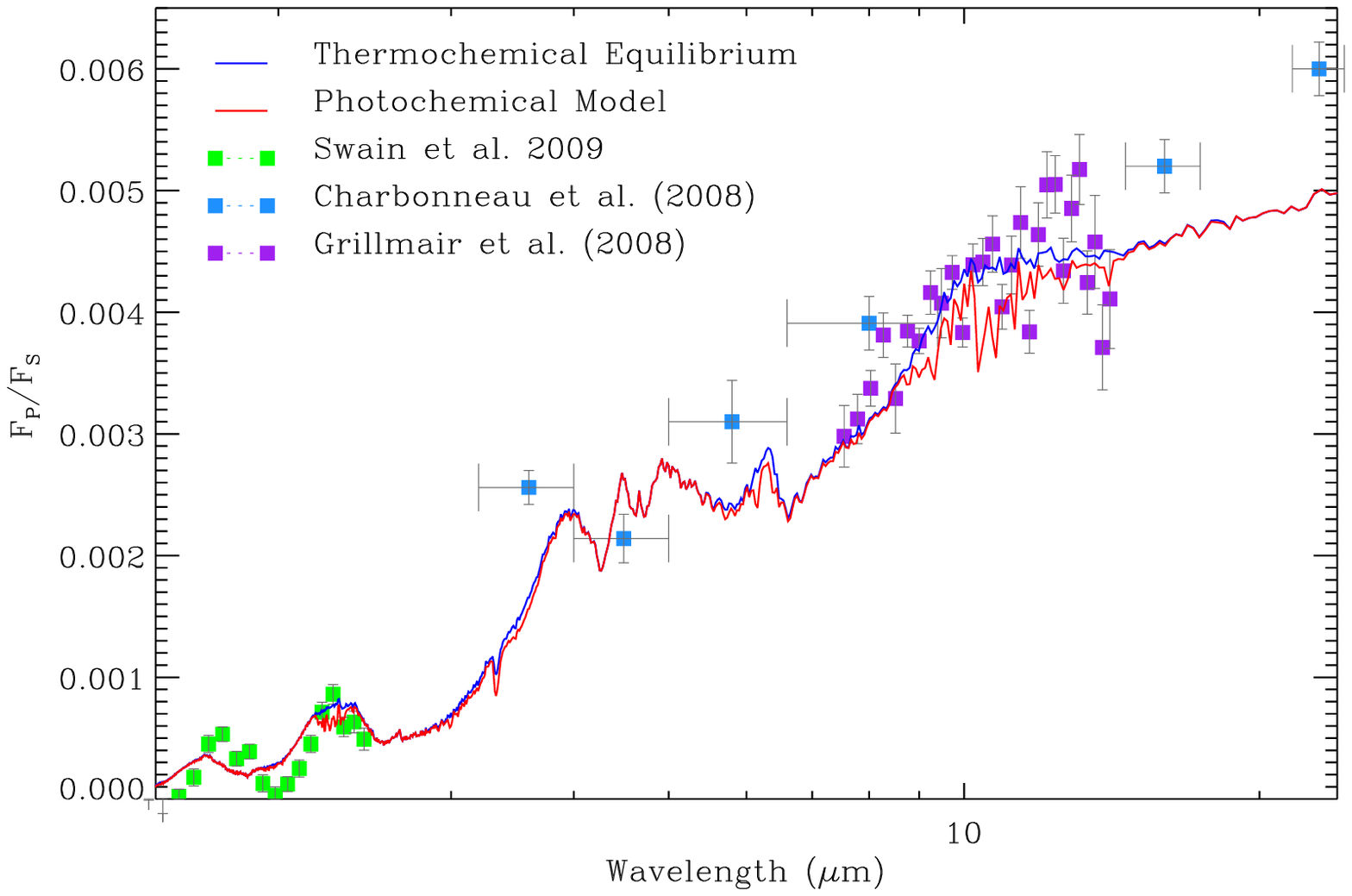}} 
& 
{\includegraphics[clip=t,scale=0.5]{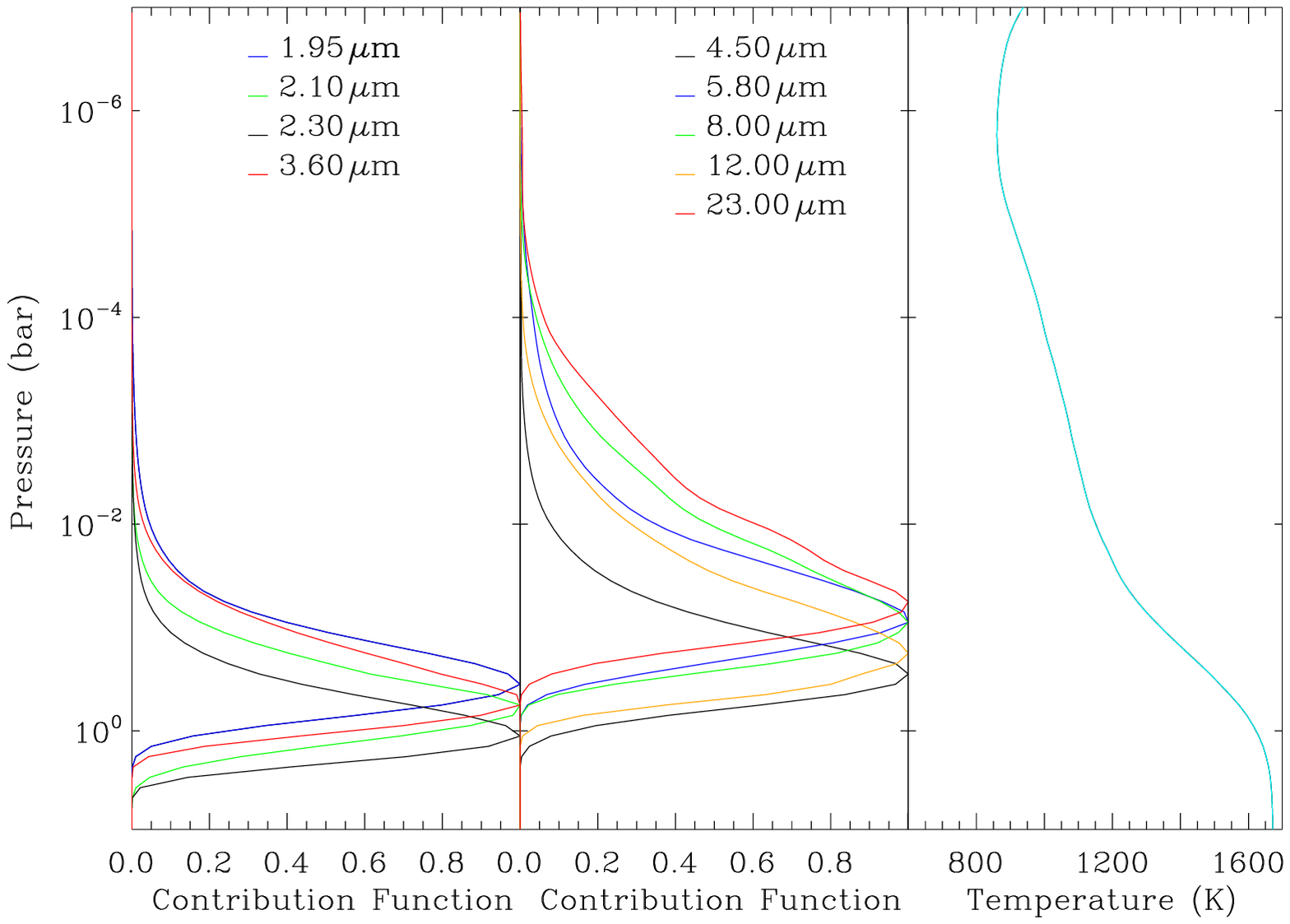}} 
\\
\end{tabular}
\caption{Synthetic emission spectra for HD 189733b (left) for the 1-D ``2$\pi$'' thermal structure 
of \citet{fort06a,fort10} ({\it i.e.}, the temperature structure is from a 1-D radiative-convection 
equilibrium model where it is assumed that the incident stellar energy is distributed over the 
dayside hemisphere only), along with the contribution functions for this model (right).  The 
model shown by the red curve uses results from our thermochemical and photochemical kinetics 
and transport model with the nominal $K_{zz}$ profile, whereas the model shown by the blue curve 
assumes a thermochemical-equilibrium composition.  Both models assume a 1x solar elemental abundance.  
Also shown are data points from various secondary-eclipse observations, as labeled.
A color version of this figure is available in the online journal.\label{189emithot}}
\end{figure*}

Our emission spectra for both HD 189733b and HD 209458b do not compare well with the observed
wavelength-dependent fluxes.  Rather than dwelling on possible reasons, we have focused on the 
expected differences in the spectral signatures of equilibrium and disequilibrium compositions 
in this section.   Detailed comparisons of our predicted mole fractions with inferences 
from published analyses of transit and eclipse observations are deferred to sections 
\ref{sectobs189} and \ref{sectobs209}.   

\section{Comparisons with Other Photochemical Models\label{sectmodcomp}}

We now compare our results with those from other photochemical models. \citet{liang03} were the first 
to develop a photochemical model for hot-Jupiter exoplanets.  Their main conclusion from their HD 209458b 
modeling was that atomic H is produced in large quantities from photolysis of $\HtwoO$ and from the 
subsequent reaction of OH with $\Htwo$ (see our scheme (10) in section \ref{sectpchemoxy}).  We agree 
with this general result, although some quantitative differences exist between our models.  For instance, 
atomic H never replaces $\Htwo$ as the dominant constituent at high altitudes in the \citet{liang03} 
models, for reasons that are unclear.
Our HD 209458b models are more comparable to the \citet{garcia07} and \citet{koskinen10} models with 
regard to this H-abundance issue.

Other quantitative differences are apparent between the \citet{liang03} models and ours.  Because 
\citet{liang03} do not consider thermochemical kinetics and do not fully reverse their reactions, 
they take a reasonable approach in beginning their model at a base pressure of $\sim$1 bar, where 
equilibrium effects will be less prevalent.  \citet{liang03} do not consider transport-induced 
quenching of $\CHfour$, and their assumed 1-bar methane abundance is quite a bit lower than ours; 
however, even ignoring this difference, the shape of their methane profiles throughout the atmosphere 
is quite different from our HD 209458b models.  In the 
\citet{liang03} models, the $\CHfour$ abundance increases from 1 bar to 0.1 mbar because of production 
of $\CHfour$ from CO photolysis, whereas our $\CHfour$ mole-fraction vertical slopes above the quench 
point are more constant or even negative before methane becomes sharply depleted when schemes (14) and 
(12) become effective, indicating a net loss rather than net production of methane in the 1--10$^{-4}$-bar 
region of our models.  However, both our models and those of \citet{liang03} suggest that $\CHfour$ does 
not survive at high altitudes on HD 209458b.  In our models, $\CHfour$ is destroyed by reaction with 
atomic H (see schemes (12) and (14), for example) and the carbon ends up in species like HCN and 
$\CtwoHtwo$.  In contrast, \citet{liang03} state that $\CHfour$ photolysis rather than reaction with 
atomic H causes the high-altitude 
depletion, but it is not clear from their discussion where the carbon ends up.

Some of these issues are clarified in the follow-up model of \citet{liang04}, where it seems that 
hydrocarbons like $\CtwoHtwo$ may replace $\CHfour$ as the second most abundant carbon bearing 
species (behind CO) at high altitudes in their HD 209458b models.  The \citet{liang04} ``Model A'' 
is closest to our HD 209458b west-terminator-average model in terms of the $\CHfour$ abundance 
at 0.1 mbar, so these two models can be compared in detail.  Our model has some similar 
general behavior, such as the fact that $\CtwoHtwo$ is the dominant C$_2$H$_x$ hydrocarbon from 
a column-integrated standpoint and that $\CtwoHtwo$ can survive to relatively high altitudes, as 
compared with $\CtwoHsix$.  We also agree that hydrogenation is the major loss mechanism for 
C$_2$H$_x$ hydrocarbons that prevents complex hydrocarbons from becoming major 
constituents on HD 209458b.  However, there are also some large differences between our models in 
terms of the overall abundances and vertical profiles of C$_2$H$_x$ hydrocarbons.  For example, our 
$\CtwoHtwo$ profiles tend to peak at high altitudes and fall off sharply with both increasing and 
decreasing altitude, 
whereas their $\CtwoHtwo$ profiles are roughly constant above $\sim$0.1 mbar.  In our models, 
three-body reactions with atomic H that act to destroy $\CtwoHtwo$ become increasingly effective 
with increasing pressure, and $\CtwoHtwo$ ends up being more efficiently converted to $\CHfour$ in 
the 1--10$^{-3}$-mbar region (see scheme (13) in section \ref{sectpchemhc}) than is apparently the 
case in the \citet{liang04} models.  Also, $\CtwoHtwo$ cannot survive the large H abundance above 
10 nbar in our model, whereas $\CtwoHtwo$ readily survives at those pressures in the \citet{liang04} 
model, suggesting that H is much less abundant or that $\CtwoHtwo$ conversion into atomic C is much 
less effective in this region of the \citet{liang04} models.  In general, C$_2$H$_x$ hydrocarbons are 
much less abundant in the 1--10$^{-6}$-bar region of our HD 209458b models, suggesting that our 
kinetics more efficiently converts these species to $\CHfour$ than in the \citet{liang04} models 
(see scheme (13) in section \ref{sectpchemhc}).

The \citet{liang03,liang04} models were updated significantly by \citet{line10}, who consider 
methane quenching for the first time and expand the range of high-temperature reactions in 
their kinetics.  \citet{line10} use the time-scale arguments developed by \citet{prinn77} 
(see section \ref{sectquench}) to estimate the quenched $\CHfour$ mole fraction on HD 189733b.  
Their resulting quenched mole fraction for $\CHfour$ --- 4$\scinot-5.$ --- is reasonably 
consistent with the results of our kinetics/transport model (see Fig.~\ref{fighc}), but their 
reasoning in deriving this value is marred by the fact that they incorrectly reverse their 
assumed rate-limiting CO $\rightarrow$ $\CHfour$ conversion step (see \citealt{bezard02} and
\citealt{viss10b,viss11} for information on correct reversal procedures for this assumed 
rate-limiting reaction) and by the fact that the reversal was not necessary in the first 
place since it is $\CHfour$ $\rightarrow$ CO quenching that matters for HD 189733b.
Given these issues and the fact that \citet{line10} 
have a different rate-limiting step for $\CHfour$ quenching than we have derived in our 
model, as well as the fact that they do not use the \citet{smith98} procedure when 
calculating the transport time scale, it is remarkable that their quenched mole fraction 
for $\CHfour$ is so close to that in our models.

In any case, the procedure used by \citet{line10} once the $\CHfour$ quench point has been established 
is perfectly valid --- it may even be preferable to full thermochemical and photochemical kinetics 
and transport models in some instances.  As an example, the quench kinetics for $\CHfour$ has some 
uncertainty associated with it due to uncertainties in the kinetic rate coefficients, the eddy 
diffusion coefficients, and the atmospheric thermal structure.  If the $\CHfour$ abundance for a 
transiting exoplanet were unambiguously 
determined from transit or eclipse observations, it might be better to take this information into 
account and begin photochemical calculations at some point above the quench level ({\it i.e.}, put 
the base of the model just above the $\CHfour$ quench level, as in the \citealt{line10} models).  
This procedure would work best for C-H-O chemistry investigations, as nitrogen-bearing and 
sulfur-bearing constituents have yet to be detected in hot-Jupiter atmospheres.

The quenched $\CHfour$ mole fraction in the \citet{line10} daytime model is most similar to our
terminator-average HD 189733b model (see Figs.~\ref{figoxy} \& \ref{fighc}), and the two can be 
directly compared.  In contrast to 
equilibrium predictions, both of our models show a greatly enhanced H abundance in the 
middle and upper stratosphere due to scheme (10), both show $\CHfour$ depletions at high altitudes 
with corresponding significant $\CtwoHtwo$ enhancements at these altitudes, both show small enhancements 
in the CO$_2$ mole fraction at high altitudes due to CO and $\HtwoO$ photochemistry, and both show that 
photochemical processes have little net effect on the $\HtwoO$ and CO profiles.  However, there are 
also some significant differences between our models that are worth mentioning.  As with the 
\citet{liang03,liang04} photochemical models, the H production rate in the \citet{line10} models is 
large, but H never exceeds the $\Htwo$ abundance at pressures $\lta$ 1 nbar, as it does in our 
HD 189733b models.  Again, the reason for this difference between our models is unclear, but it seems to
have something to do with the efficiency of scheme (10) and its reverse or with other $\Htwo$ destruction 
and recycling mechanisms.  The fact that we have considerably more O and OH at high altitudes in our 
model as compared with \citet{line10} is consistent with our greater depletion of $\Htwo$ and is consistent 
with this suggested cause.  Our models also show less net production of $\COtwo$ from photochemical 
processes than in the \citet{line10} model, as thermochemical processes in our model drive the $\COtwo$ 
abundance back toward equilibrium.

Another difference between our model and that of \citet{line10} is the more abrupt and more severe 
depletion in the $\CHfour$ mole fraction above $\sim$1 $\mu$bar in our HD 189733b model as compared with 
their model.  This difference is likely due to coupled carbon-nitrogen chemistry (see 
scheme (14), for example), as HCN takes over from $\CHfour$ at high altitudes in our model, whereas 
nitrogen chemistry is not considered in the \citet{line10} models.  A more minor difference between 
our model and theirs is the dominant scheme to form C$_2$H$_x$ hydrocarbons in the stratosphere.  In 
\citet{line10} and in the \citealt{liang03,liang04} models on which these models were based, the 
dominant reaction that produces C$_2$ species from CH$_x$ species is apparently $^3$CH$_2$ + $^3$CH$_2$ 
$\rightarrow$ $\CtwoHtwo$ + $2\, \H$, whereas we find the reaction $^3$CH$_2$ + $\CHthree$ $\rightarrow$ 
$\CtwoHfour$ + H to be more important from a column-integrated standpoint.  Qualitatively, however, the 
main behavior in the \citet{line10} models is similar to ours, other than some differences that results 
from our inclusion of nitrogen photochemistry.

We can also compare our models to the thermochemical and photochemical kinetics and transport models 
of \citet{zahnle11}.  Note that \citet{zahnle09} discuss sulfur photochemistry only, which we do not 
consider, so we do not compare our results with \citet{zahnle09}.  Direct comparisons between our 
models and those of \citet{zahnle11} are complicated by the fact that they show results only for models 
with isothermal atmospheres.  Comparisons are also hampered by the fact that \citet{zahnle11} are 
working on a revision of their model as we are preparing this paper here, so that we can only compare 
with their ``version 2'' in the arXiv archive, rather than with a published version of the models.  
Although we cannot do anything about the second problem, we can resolve the first problem --- which 
is a significant one --- by developing an isothermal model of our own to better compare with the 
\citet{zahnle11} results.  The isothermal model that we develop is for a 1$\times$ solar composition 
atmosphere, with $T$ = 1000 K, and $K_{zz}$ at a constant 10$^9$ cm$^2$ $\smone$, for a planet located 
0.1 AU from a Sun-like star, at a fixed solar zenith angle of 30\deg\ (see Fig.~\ref{figzahnle}).  
This model can be directly compared with the results shown in the top right figure of Fig.~4 of the 
arXiv version 2 of \citet{zahnle11}.  Since Zahnle et al.'s 100-bar lower boundary condition of a 
1$\scinot-12.$ mole fraction for species other than H, $\HtwoO$, CO, $\CHfour$, $\NHthree$, 
$\Ntwo$, and H$_2$S in the \citet{zahnle11} model leads to odd behavior for some of the minor 
constituents near the lower boundary, we simply assumed a constant concentration gradient for 
minor species (including H 
and CO, which are minor species under these conditions at the 100-bar lower boundary), which means 
that these species will flow through the lower boundary at a maximum possible rate.  We also 
assumed equilibrium mole fractions for $\Htwo$ and He at the lower boundary.

\begin{figure}
\includegraphics[angle=-90,clip=t,scale=0.37]{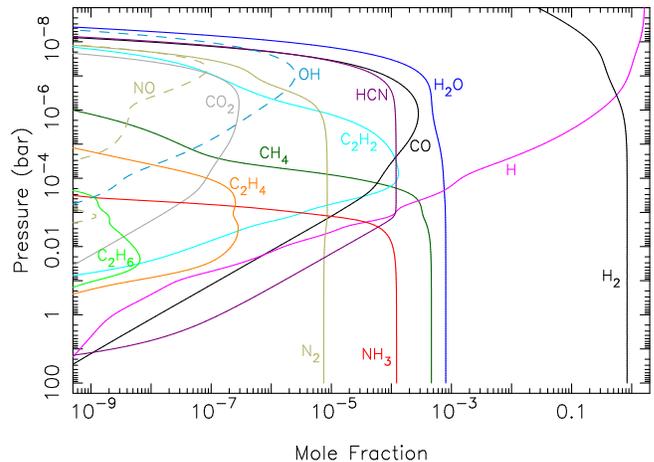}
\caption{Mole-fraction profiles for major species in a 1$\times$ solar composition model that assumes 
temperatures are constant at 1000 K, $K_{zz}$'s are constant at 10$^{9}$ $\cmtwo$ $\smone$, and 
the planet is located at 0.1 AU from a Sun-like star, for a stellar zenith angle fixed at 30\deg.  
These model results can be compared with isothermal models described in \citet{zahnle11}, as 
well as with our HD 189733b and HD 209458b models results shown in Figs.~\ref{figoxy}--\ref{fignit}.  
A color version of this figure is available in the online journal.\label{figzahnle}}
\end{figure}

A detailed comparison of Fig.~\ref{figzahnle} with Fig.~4b of \citet{zahnle11} reveals some 
similarities and differences between our models.  The $\HtwoO$ mole fraction remains just below 
10$^{-3}$ in both models and is fairly unperturbed by photochemistry.  Atomic H is very abundant 
at high altitudes and greatly affects the chemistry of molecular species in both models.  Although 
our results for atomic H are qualitatively similar here, the quantitative results differ, likely 
because of differences in chemical reaction rate coefficients for hydrocarbon reactions.  Methane
has a sharp mole-fraction cutoff in both models at pressures of 10$^{-3}$ or 10$^{-4}$ bar, although the 
location of the cutoff and shape of the $\CHfour$ profile is different in both models.  This difference 
appears to be related to $\CtwoHfour$ chemistry, as the ethylene profiles are the most obvious difference 
between the models.  The \citet{zahnle11} results have $\CtwoHfour$ replacing $\CHfour$ at a relatively 
deep $\sim$ 0.1 bar, causing a large ``bite out'' in the $\CHfour$ profile from 0.1--0.001 bar before 
$\CtwoHtwo$ takes over as the dominant hydrocarbon at high altitudes.  The $\CtwoHfour$ in our model 
peaks at a somewhat 
higher altitude, and its abundance never approaches that of $\CHfour$, so methane is carried up to 
pressures of 10$^{-4}$ mbar before its eventual replacement by $\CtwoHtwo$, HCN, and CO --- note that 
both models agree that CO, HCN, and $\CtwoHtwo$ will be the dominant carbon carriers at high altitudes.  
The differences in the hydrocarbon profiles and some portion of the H profiles seem to stem 
from a different low-pressure limiting rate coefficient for H + $\CtwoHfour$ + M $\rightarrow$ 
$\CtwoHfive$ + M between our models.  Due to an inadvertent data-entry error, that rate coefficient 
in the \citet{zahnle11} model is three orders of magnitude smaller than the expression we have adopted, 
effectively cutting off an important $\CtwoHfour$ destruction pathway.  When this expression is changed 
to better reflect literature values \citep[e.g.,][]{baulch94}, the \citet{zahnle11} results are more 
in line with ours (K.~Zahnle, personal communication, 2010).  

Another apparent difference between the \citet{zahnle11} models and ours is the CO and $\COtwo$ mole 
fractions at lower altitudes.  Although the CO mole fraction remains at 10$^{-3}$ to 10$^{-4}$ at high 
altitudes in both models, the drop off with decreasing altitude has a much different gradient in our 
model compared with theirs.  Carbon monoxide is produced photochemically at high altitudes in both 
models, and CO then flows down to the lower atmosphere where in our model the large flux boundary condition 
allows it to be transported out of the system.  \citet{zahnle11} do not discuss their dominant methane 
oxidation pathways in detail, so it is not clear how methane is converted to CO in their model, although 
they do mention that C + OH is important and they mention that they include three-body OH addition 
reactions that can form C-O bonds.  In our model, methane is oxidized to CO through processes such as 
\begin{eqnarray}
\CHfour \, + \, \H \, & \rightarrow & \, \CHthree \, + \, \Htwo  \nonumber \\
\H \, + \, \CHthree \, & \rightarrow & \, \threeCHtwo \, + \Htwo  \nonumber \\
\H \, + \, \threeCHtwo \, & \rightarrow & \, \CH \, + \Htwo  \nonumber \\
\CH \, + \, \HtwoO \, & \rightarrow & \, \HtwoCO \, + \H \nonumber \\
\HtwoCO \, + \, \H \, & \rightarrow & \, \HCO \, + \Htwo \nonumber \\
\HCO \, + \, \H \, & \rightarrow & \, \CO \, + \Htwo \nonumber \\
\noalign{\vglue -10pt}
\multispan3\hrulefill \nonumber \cr
\Net \ \ \CHfour \, + \, \HtwoO \, + \, 4\, \H \, & \rightarrow & \, \CO \, + \, 5\, \Htwo ; \\
\end{eqnarray}
however, this mechanism becomes ineffective once the H abundance drops significantly, and mechanisms like 
schemes (2) and (3) in section \ref{sectquench} take over but do not operate effectively at 1000 K.  
Therefore, in our model, diffusion controls the CO profile at low altitudes.  If the \citet{zahnle11} 
model is fully converged, then it appears that either the above CO production scheme (15) or alternative 
ones that produce the C-O bond remain effective at low altitudes in the \citet{zahnle11} model --- 
perhaps due to their much larger low-altitude H abundance --- allowing CO to be kinetically produced in 
the lower atmosphere.  Whatever the cause, their increased CO abundance in the lower atmosphere 
allows an increased abundance of $\COtwo$ compared to our model, as CO and $\COtwo$ are kinetically linked.

The qualitative results for the nitrogen-bearing species appear to be similar between our model and 
that of \citet{zahnle11}.  Ammonia has a sharp cutoff at $\sim 10^{-3}$ bar in both models, with HCN 
replacing $\NHthree$ as the dominant nitrogen carrier at higher altitudes.  Molecular $\Ntwo$ is 
relatively unreactive in both models and has a constant mole fraction throughout most of the 
atmosphere.  The \citet{zahnle11} model apparently has an effective mechanism that converts HCN to 
$\Ntwo$ at high altitudes, whereas HCN is more efficiently recycled at high altitudes in our model.
Similarly, NO is produced at high altitudes more efficiently in the \citet{zahnle11} model, or we 
have more efficient loss processes in our model.

In general, our main conclusions from this isothermal modeling exercise are similar to those of 
\citet{zahnle11} in terms of the major carbon-, oxygen-, and nitrogen-bearing species and their 
relative importance in different regions of the atmosphere, with the exception of the $\CtwoHfour$ 
abundance, as discussed above.   However, we disagree with the assertion of \citet{zahnle11} that 
such isothermal atmospheres are good analogs for extrasolar planets, and we feel the likelihood of 
soot formation on cooler exoplanets has been overemphasized as a result of Zahnle et al.'s reliance 
on isothermal models.   Isothermal model results do have some resemblance to those from models with more realistic 
temperature profiles (compare Figs.~\ref{figoxy}--\ref{fignit} with Fig.~\ref{figzahnle}).  However, 
one critical difference is the presence of transport-induced quenching in the models with more 
realistic temperature profiles.  Like the real HD 189733b atmosphere, the isothermal 1000-K profile 
starts in the $\CHfour$ stability regime in the deep atmosphere but crosses over to the CO stability 
field in the upper atmosphere.  However, unlike the real HD 189733b atmosphere, temperatures are 
low enough throughout the 1000-K isothermal atmosphere that interconversion between CO and $\CHfour$ 
(and between $\Ntwo$ and $\NHthree$, etc.) is inhibited everywhere.  Therefore, whatever molecule is dominant at 
the base of the isothermal model will continue to be the dominant molecule throughout most of the 
atmosphere in the isothermal photochemical models.  Transport-induced quenching will not occur, and 
thus the abundance of species like CO will be underpredicted (as its only source will be photochemistry, 
not thermochemistry), and the $\CHfour$ abundance will be overpredicted as compared to kinetic-transport 
models that have higher temperatures at depth.

This problem is particularly acute if the real planet has a temperature profile such that the composition 
would remain in thermochemical equilibrium as the atmosphere transitions into the CO stability field, as 
seems to be the case for both HD 189733b and HD 209458b.  In that case, the $\CHfour$ $\rightarrow$ CO 
interconversion quench point will occur while CO is dominant, and CO will be the main carrier of 
carbon in the stratospheres of these planets, in stark contrast to the isothermal model results.  Even 
if the stratospheres become quite cold on these planets with more realistic temperature profiles, methane 
is not going to be as abundant as the isothermal models would indicate.  Because $\CHfour$ is the 
source of the complex hydrocarbons in the \citet{zahnle11} models (and in our own models), a greater 
CO/$\CHfour$ ratio in 
the upper atmosphere implies fewer hydrocarbons and less chance of soot formation.  The inability to form 
complex hydrocarbons is also exacerbated by the efficient hydrogenation of unsaturated hydrocarbons as the 
pressure increases, so that species like $\CtwoHtwo$ and $\CtwoHfour$ are efficiently converted back to 
methane at pressures greater than 1 $\mu$bar in our model (and probably would be in the
\citet{zahnle11} 
model for more realistic considerations of the low-pressure limiting rate coefficient for H + $\CtwoHfour$ 
+ M $\rightarrow$ $\CtwoHfive$ + M).  Complex hydrocarbons are thus several orders of magnitude less 
important in our HD 189733b models than the isothermal \citet{zahnle11} models would suggest, and soot 
formation is much less of a factor.  That is not to say that soot formation cannot still occur on these 
cooler exoplanets, as is discussed in section \ref{sectpchemhc}, particularly if ion chemistry is conducive 
to the formation of heavy neutrals, but the soot precursors would be confined to high altitudes, and the 
overall column abundance of heavy hydrocarbons will be many orders of magnitude smaller than is 
suggested by the Zahnle et al. isothermal models.

\section{Comparisons with observed mole fractions on HD 189733\lowercase{b}}\label{sectobs189}

Our model results can also be compared with the mole fractions inferred from the transit and eclipse
observations of HD 189733b (see Table \ref{tabhd189}).  The suggestion that C-, O-, or 
N-bearing molecules are affecting the observational signature of HD 189733b during the primary
transit and secondary eclipse came originally from mid-infrared photometry from the Infrared Array 
Camera (IRAC) and from mid-infrared spectra from the Infrared Spectrograph (IRS) on the {\it Spitzer\ 
Space\ Telescope}\/ (\citealt{tinetti07,grill08,beaulieu08,knut07,knut08,charb08,desert09}; see also 
the theoretical modeling and model-data comparisons of \citealt{fort07,barman08,burrows08,madhu09}). 
The mid-IR data have provided relatively loose constraints on molecular abundances for HD 189733b 
(see Table \ref{tabhd189}), whereas the near-infrared wavelength region is potentially more useful 
for constraining abundances because of the presence of potentially diagnostic absorption features 
in this region.  In fact, near-IR transit and eclipse data from the NICMOS instrument on the {\it
HST}\/ do seem to provide tighter constraints on molecular abundances
\citep{swain08b,swain09a,sing09,madhu09}.  However, the state of affairs in the analysis of transit 
and eclipse observations is clearly not ideal, as contradictory data sets and/or analyses exist, 
both within the mid-IR and near-IR regions and between these two regions 
\citep[e.g.,][]{madhu09,desert09,beaulieu08,sing09,swain08b,gibson10,fort10}.  If systematic 
instrument effects are not completely inhibiting the ability to acquire meaningful spectral and 
photometric information \citep[e.g.,][]{gibson10}, we can at least attempt to compare our model 
results with the available observations (see Table \ref{tabhd189}) and discuss some general issues.

\begin{deluxetable*}{lcccc}
\tabletypesize{\scriptsize}
\tablecaption{Volume mixing ratios from HD 189733\lowercase{b} observations and models\label{tabhd189}}
\tablewidth{400pt}
\tablecolumns{5}
\tablehead{
\colhead{Source} & \colhead{$\HtwoO$} & \colhead{CO} & \colhead{$\CHfour$} & \colhead{$\COtwo$}
}
\startdata
\sidehead{Secondary eclipse: {\it Spitzer} IRS}
\ \ \ \citet{grill08}  & 1$\times$ solar & \nodata & \nodata & \nodata \\
\ \ \ \citet{madhu09}  & 10$^{-6}$ to 0.1   & \nodata & $<$ 10$^{-2}$ & \nodata \\
\sidehead{Secondary eclipse: {\it Spitzer} IRAC}
\ \ \ \citet{charb08}  & 1$\times$ solar & 1$\times$ solar & \nodata & \nodata \\
\ \ \ \citet{madhu09}  & (0.1-10)$\, \times 10^{-4}$ & \nodata & $\leq$ $2 \times 10^{-6}$ & (7-700)$\, \times 10^{-7}$ \\
\sidehead{Secondary eclipse: {\it HST}\/ NICMOS}
\ \ \ \citet{swain09a} & (0.1-1)$\, \times 10^{-4}$  & (1-3)$\, \times 10^{-4}$ & $\leq$ $1 \times 10^{-7}$ & (1-10)$\, \times 10^{-7}$ \\
\ \ \ \citet{madhu09}  & $\sim 1 \times 10^{-4}$ & (0.2-20)$\, \times 10^{-3}$ & $\leq$ 6$\, \times 10^{-6}$ & $\sim 7 \times 10^{-4}$ \\
\sidehead{Our dayside-average thermal-structure models at 0.1 bar ({\it i.e.}, for secondary-eclipse conditions)}
\ \ \ nominal $K_{zz}$ & $3.5 \times 10^{-4}$ & $4.5 \times 10^{-4}$ & $1.2 \times 10^{-5}$ & $1.3 \times 10^{-7}$ \\
\ \ \ $K_{zz}$ = 10$^7$ cm$^2$ $\smone$ & $3.4 \times 10^{-4}$ & $4.6 \times 10^{-4}$ & $2.9 \times 10^{-6}$ & $1.3 \times 10^{-7}$ \\
\ \ \ $K_{zz}$ = 10$^{10}$ cm$^2$ $\smone$ & $3.9 \times 10^{-4}$ & $4.1 \times 10^{-4}$ & $3.9 \times 10^{-5}$ & $1.3 \times 10^{-7}$ \\
\ \ \ nominal $K_{zz}$, 10$\times$ solar & $3.4 \times 10^{-3}$ & $4.6 \times 10^{-3}$ & $6.1 \times 10^{-6}$ & $1.3 \times 10^{-5}$ \\
\sidehead{Our warmer ``2$\pi$'' \citet{fort06a,fort10} model at 0.1 bar ({\it i.e.}, for secondary-eclipse conditions)}
\ \ \ nominal $K_{zz}$ & $3.4 \times 10^{-4}$ & $4.7 \times 10^{-4}$ & $6.7 \times 10^{-7}$ & $9.7 \times 10^{-8}$ \\
\sidehead{Primary transit: {\it Spitzer} IRAC}
\ \ \ \citet{tinetti07}  & $\sim 5 \times 10^{-4}$ & 1$\times$ solar & \nodata & \nodata \\
\ \ \ \citet{desert09}  & upper limit & CO/$\HtwoO$ $\sim$ 5-60 & \nodata & \nodata \\
\sidehead{Primary transit: {\it HST}\/ NICMOS}
\ \ \ \citet{swain08b}  & $\sim 5 \times 10^{-4}$ & \nodata & $\sim 5 \times 10^{-5}$ & \nodata \\
\ \ \ \citet{madhu09}  & $5 \times 10^{-4}$ to 0.1   & \nodata & 10$^{-5}$ to 0.3 & \nodata \\
\ \ \ \citet{sing09}  & upper limit   & \nodata & \nodata & \nodata \\
\sidehead{Our nominal $K_{zz}$ terminator-average model ({\it i.e.}, for primary-transit conditions)}
\ \ \ at 10$^{-4}$ bar & $3.8 \times 10^{-4}$ & $4.2 \times 10^{-4}$ & $3.2 \times 10^{-5}$ & $5.5 \times 10^{-7}$ \\
\ \ \ at 10$^{-3}$ bar & $3.8 \times 10^{-4}$ & $4.2 \times 10^{-4}$ & $4.3 \times 10^{-5}$ & $4.5 \times 10^{-7}$ \\
\ \ \ at 10$^{-2}$ bar & $3.8 \times 10^{-4}$ & $4.2 \times 10^{-4}$ & $4.5 \times 10^{-5}$ & $3.2 \times 10^{-7}$ \\
\ \ \ at 10$^{-1}$ bar & $3.8 \times 10^{-4}$ & $4.2 \times 10^{-4}$ & $4.6 \times 10^{-5}$ & $2.0 \times 10^{-7}$ \\
\enddata
\tablecomments{For the observations in which ``1$\times$ solar'' or ``\nodata'' (no data) is listed, detailed
constraints on the abundances were not presented.  Our models assume 1$\times$ solar abundance, unless 
otherwise indicated.  The ``nominal'' $K_{zz}$ profiles are shown in Fig.~\ref{eddynom}.}
\end{deluxetable*}

Our 1$\times$-solar-composition kinetics-transport models for HD 189733b appear consistent with the loose 
constraints supplied by the {\it Spitzer\/} IRS secondary eclipse observations of $\HtwoO$ and $\CHfour$ 
\citep{grill08,madhu09}, with the tighter constraints from {\it HST\/} NICMOS secondary eclipse 
observations of CO and $\COtwo$ as reported by \citet{swain09a}, and with the $\CHfour$ and $\HtwoO$ 
abundances from the {\it HST\/} NICMOS transit observations analyzed by \citet{madhu09} and 
\citet{swain08b}.  The models cannot, however, account for the much smaller $\CHfour$ mixing ratios 
inferred for the secondary eclipse as opposed to the primary transit of HD 189733b 
\citep[cf.][]{swain08b,swain09a,madhu09}, unless large temperature differences between the dayside
and terminators can be maintained in the 1-100 bar region and unless zonal transport does not 
homogenize the $\CHfour$ abundance in the 0.1-1 bar region.  As was first demonstrated by \citet{coop06}, 
vertical transport-induced quenching is quite effective in the upper atmospheres of hot Jupiters, and 
$\CHfour$ is not readily converted to CO at dayside temperatures in our models.  Some high-altitude 
$\CHfour$ is lost to HCN and $\CtwoHtwo$ in our models, but there is still sufficient $\CHfour$ at 
transit-sensitive pressures that we would expect dayside observations to show $\CHfour$ mole 
fractions only a factor of a few smaller than at the terminators for the dayside-average and 
terminator-average temperature profiles derived from the \citet{showman09} GCMs.  If temperatures 
in the deep (1-10 bar) dayside atmosphere are much warmer than we have assumed (e.g., compare the 1-D 
2$\pi$ \citet{fort06a,fort10} profile with the area-weighted dayside-average thermal profile shown in
Fig.~\ref{figtemp}) or if $K_{zz}$'s at the deep quench points are much smaller on the dayside than 
at the terminators, then this result could change.  

The fact that the predicted infrared eclipse depths from our dayside-average model seem to be
uniformly too small compared to observations, combined with the low dayside $\CHfour$ 
abundance inferred by \citet{swain09a,madhu09}, suggest a scenario in which the 
dayside-vs.-terminator temperature differences in the 1-100 bar region of HD 189733b are larger 
than is predicted from the \citet{showman09} GCMs.  Quenched $\CHfour$ abundances are much 
smaller in the model that uses  the warmer ``2$\pi$'' 1-D \citet{fort06a,fort10} thermal profile, for 
example (see Table~\ref{tabhd189}).  
The main problem with this scenario is that the radiative time 
constants at the $\CHfour$ quench level near 1-10 bar are expected to be greater than the zonal 
transport time scales, so that one would expect the temperatures to be homogenized longitudinally 
in this region.  The strong zonal winds at would also be expected to help 
homogenize the $\CHfour$ abundance across longitudes, so that there would not be a big terminator 
vs.~dayside difference, in conflict with the observations.  

As it stands, neutral C-H-O-N chemistry does not seem to be able to account for a depletion of 
$\CHfour$ at 1--10$^{-3}$-bar pressures on the dayside of HD 189733b --- where we expect 
transport-induced quenching to dominate --- when we adopt the \citet{showman09} GCM-based 
temperature profiles.  
It is possible that sulfur photochemistry could contribute to a loss of $\CHfour$ at these 
pressures on HD 189733b \citep{zahnle09,zahnle11}, despite the low S/C ratio expected for 
solar-like compositions, if catalytic destruction cycles are in operation.  This process could 
only work effectively at low altitudes if the $\CHfour$ mole fraction were comparable to the 
H$_2$S mole fraction ({\it i.e.}, in situations where the quenched $\CHfour$ abundance is small) 
because the solar ratio of S/C $\ll$ 1 will make it unlikely that products like CS and CS$_2$ can 
make a dent in the methane abundance if $\CHfour$ is the dominant carbon carrier, and other 
potential products like C$_2$H$_x$ hydrocarbons are unstable and will not replace methane at 
these pressures.  In the \citet{zahnle11} isothermal models, for instance, CS and CS$_2$ do not 
replace $\CHfour$ at low altitudes because of the much greater $\CHfour$ abundance as compared 
with H$_2$S.  However, since $\CHfour$ is likely {\it not} the dominant carbon carrier on HD 
189733b and HD 209458b, sulfur chemistry might potentially play a larger role in methane 
destruction than the \citet{zahnle11} models indicate.  Similarly, PH$_3$ has weak bonds and 
could be an important source of atomic H at low altitudes that could affect $\CHfour$ chemistry, 
but to truly destroy $\CHfour$ in the lower stratosphere , some product like HCP must be able to 
compete with methane, which we think is unlikely given the thermodynamic stability of HCP --- 
sulfur photochemistry seems to have greater promise as a mechanism for destroying $\CHfour$ in 
the dayside lower stratosphere of HD 189733b.  

\begin{figure*}
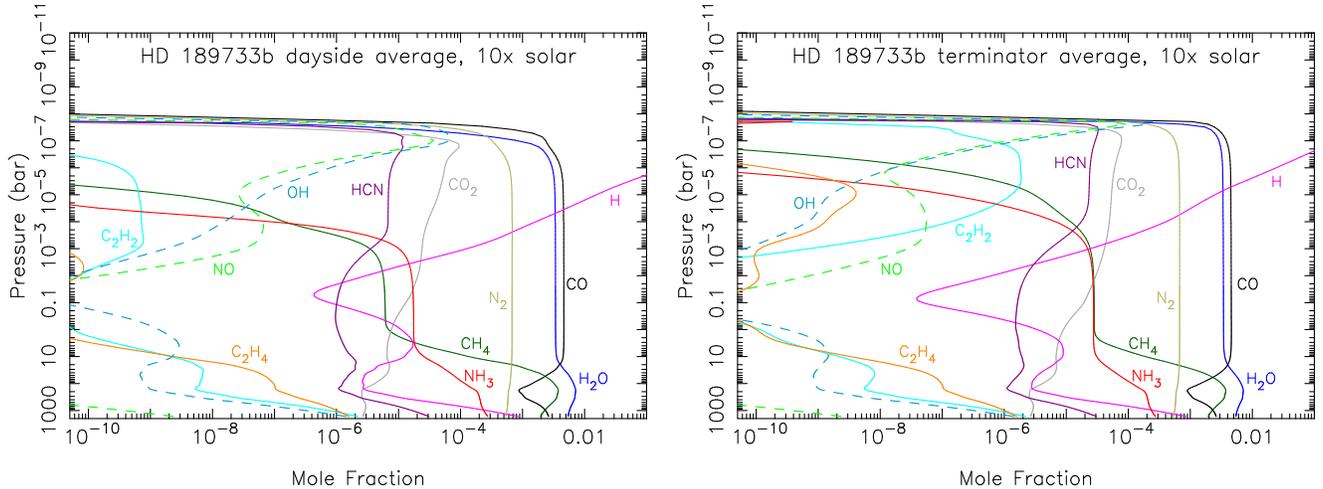

\begin{tabular}{ll}
{\includegraphics[angle=-90,clip=t,scale=0.37]{fig16a_color.ps}} 
& 
{\includegraphics[angle=-90,clip=t,scale=0.37]{fig16b_color.ps}} 
\\
\end{tabular}
\caption{Mole fraction profiles for several important species in dayside-average (left) 
and terminator-average (right) HD 189733b models that assume a 10$\times$ solar composition 
but otherwise have the same $K_{zz}$ profile, thermal structure, and background atmospheric 
grid as our nominal 1$\times$ solar HD 189733b models shown in Figs.~\ref{figoxy}--\ref{fignit}.  
From comparisons between these figures, note the linear increase in the $\HtwoO$ and CO mole 
fractions, the quadratic increase in the $\COtwo$ mole fraction, the slight increase in the quenched 
$\NHthree$ mole fraction, and the slight decrease in the quenched $\CHfour$ mole fraction as 
the metallicity is increased from 1$\times$ to 10$\times$ solar.  
A color version of this figure is available in the online journal.\label{10xsolar}}
\end{figure*}

Although our predicted $\COtwo$ abundances for HD 189733b are consistent with those from the 
{\it HST\/} NICMOS secondary eclipse observations reported by \citet{swain09a}, a much greater 
$\COtwo$ abundance is suggested from the \cite{madhu09} reanalysis of both these data and the 
{\it Spitzer\/} IRAC secondary eclipse data.  Carbon dioxide closely follows thermochemical 
equilibrium in our models.  Some additional $\COtwo$ is produced at high altitudes from 
photochemistry in our models, but this amount is not sufficient to affect the column abundance 
at pressures greater than $\sim$10$^{-2}$ bar.  Therefore, if these larger observationally 
derived $\COtwo$ values are correct, then HD 189733b likely has an enhanced metallicity 
\citep[see also][]{fort10,line10,zahnle11}.  Enhanced metallicities might also help explain the 
strong absorption in the 4.5 $\mu$m band from {\it Spitzer\/} IRAC transit observations 
\citep{desert09}.  A large CO/$\HtwoO$ ratio was invoked by \citet{desert09} to explain this 
4.5-$\mu$m feature, which would not be consistent with solar elemental ratios; however, as is 
described by \citet{fort10}, this feature could also be explained by the high $\COtwo$ 
abundances that are predicted from enhanced-metallicity models \citep{line10,zahnle11}.  A 
high metallicity for HD 189733b might then explain both the \citet{madhu09} derived high 
$\COtwo$ abundances from the {\it HST}\/ NICMOS data of \citet{swain08b,swain09a}, as well 
as the 4.5 $\mu$m absorption seen by \citet{desert09}.

Figure \ref{10xsolar} illustrates how our HD 189733b model results change if we assume a 10$\times$ 
solar metallicity, but otherwise keep the same $K_{zz}$ profile and atmospheric grid (e.g., the $T$-$P$ 
profiles and scale height stay the same despite the increased mean molecular weight and atmospheric 
opacity) as in our nominal 1$\times$ solar model.  The CO and $\HtwoO$ mole fractions increase 
by a factor of 10, but the $\COtwo$ mole fraction increases by a factor of $\sim$100, and the $\CHfour$ 
mole fraction actually decreases when the metallicity is increased from 1$\times$ to 10$\times$ solar.  
The sensitivity of the $\COtwo$ abundance to metallicity has been documented previously
\citep{lodders02,line10,zahnle11} and is the natural result of $\COtwo$'s dependence on both CO 
and $\HtwoO$.  The 
{\it decrease}\/ in the $\CHfour$ abundance with increased metallicity seems to result from the 
fact that the $\CHfour$ $\rightarrow$ CO conversion schemes (2) and (3) in section \ref{sectquench} are more 
effective when there is more $\HtwoO$ and CO present, such that $\CHfour$ follows its equilibrium curve to 
higher altitudes before the quench point occurs, resulting in a decrease in the quenched $\CHfour$ 
abundance at higher metallicities.  This higher-metallicity model seems at face value to be more 
consistent with the secondary-eclipse $\COtwo$ and $\CHfour$ abundances inferred by \citet{madhu09} 
and with the strong 4.5-$\mu$m absorption observed in the transit observations of \citet{desert09}, 
but the resulting water abundance may be higher than can be supported by the {\it HST}\/ NICMOS secondary 
eclipse data, unless some additional absorption or scattering is muting the near-IR $\HtwoO$ absorption 
bands.

In fact, the relatively low inferred $\HtwoO$ mixing ratio for HD 189733b from the 
{\it HST}\/ NICMOS secondary eclipse data \citep{swain09a,madhu09} is somewhat disconcerting.  
Water is prevalent in both equilibrium and disequilibrium models for a wide range of metallicities.  Only 
for greatly subsolar metallicities or greatly supersolar C/O ratios can the $\HtwoO$ mole fraction be much 
less than 10$^{-4}$ in the atmosphere of HD 189733b \citep[e.g.,][]{lodders02,line10}.  As is discussed 
by numerous authors, water absorption is therefore expected to dominate the infrared spectrum of this 
and most other transiting extrasolar giant planets.  Given the photochemical stability of $\HtwoO$ and the 
likelihood of $\HtwoO$ being carried to as high altitudes as any other stable molecular constituent, it 
is unlikely that clouds or hazes could obscure $\HtwoO$ absorption features at near-infrared wavelengths 
without all other molecular features being obscured as well.  However, where $\HtwoO$ does not strongly 
absorb, other molecules --- including disequilibrium molecules --- might contribute opacity that could 
mask the expected signatures of $\HtwoO$.  As such, the observed apparent radius ratios between 1.66 
$\mu$m and 1.87 $\mu$m in the {\it HST\/} NICMOS transit observations of \citet{sing09} could indicate 
the presence of an additional absorber rather than a lack of $\HtwoO$.  Alternatively, the contrast in the 
absorption bands could be washed out due to multiple Rayleigh or aerosol scattering, but such effects 
should mute the contrast for all near-IR absorption features such that the relatively strong absorptions 
reported by \citet{swain08b,swain09a} could not be explained.  In any case, spectral models need to be 
updated to include the effects of disequilibrium constituents. 

These disequilibrium products will affect not only the observed spectra of HD 189733b, but the 
energy balance of the planet as well.  A full discussion of the effect of disequilibrium 
constituents on the thermal structure of the planet are deferred to a future work.

\section{Comparisons with observed mole fractions on HD 209458\lowercase{b}}\label{sectobs209}

\begin{deluxetable*}{lcccc}
\tabletypesize{\scriptsize}
\tablecaption{Volume mixing ratios from HD 209458\lowercase{b} observations and models\label{tabhd209}}
\tablewidth{400pt}
\tablecolumns{5}
\tablehead{
\colhead{Source} & \colhead{$\HtwoO$} & \colhead{CO} & \colhead{$\CHfour$} & \colhead{$\COtwo$}
}
\startdata
\sidehead{Secondary eclipse: {\it Spitzer} IRAC (3.5, 4.6, 5.8, 8 $\mu$m), IRS (16 $\mu$m), and MIPS (24 $\mu$m)}
\ \ \ \citet{madhu09}  & $<$ 10$^{-4}$ & $>$ 4$\scinot-5.$ & 10$^{-8}$ to 0.04 & (0.02-70)$\scinot-7.$ \\
\sidehead{Secondary eclipse: {\it Spitzer} IRAC, IRS and {\it HST} NICMOS}
\ \ \ \citet{swain09b}  & 8$\scinot-7.$ to 1$\scinot-4.$ & \nodata & (0.2-2)$\scinot-4.$ & (0.1-1)$\scinot-5.$ \\
\sidehead{Our nominal $K_{zz}$ dayside-average model at 0.1 bar ({\it i.e.}, for secondary-eclipse conditions)}
\ \ \ nominal $K_{zz}$ & $3.4 \times 10^{-4}$ & $4.7 \times 10^{-4}$ & $3.7 \times 10^{-7}$ & $5.7 \times 10^{-8}$ \\
\sidehead{Primary transit: {\it Spitzer} IRAC}
\ \ \ \citet{beaulieu09}  & 1$\times$ solar & \nodata & \nodata & \nodata \\
\sidehead{Our nominal western-terminator-average model ({\it i.e.}, for primary-transit conditions)}
\ \ \ at 10$^{-4}$ bar & $3.5 \times 10^{-4}$ & $4.6 \times 10^{-4}$ & $6.3 \times 10^{-7}$ & $1.8 \times 10^{-7}$ \\
\ \ \ at 10$^{-3}$ bar & $3.5 \times 10^{-4}$ & $4.6 \times 10^{-4}$ & $7.3 \times 10^{-7}$ & $2.0 \times 10^{-7}$ \\
\ \ \ at 10$^{-2}$ bar & $3.5 \times 10^{-4}$ & $4.6 \times 10^{-4}$ & $7.6 \times 10^{-7}$ & $1.9 \times 10^{-7}$ \\
\ \ \ at 10$^{-1}$ bar & $3.5 \times 10^{-4}$ & $4.6 \times 10^{-4}$ & $7.6 \times 10^{-7}$ & $1.2 \times 10^{-7}$ \\
\enddata
\tablecomments{For the observations in which ``1$\times$ solar'' or ``\nodata'' (no data) is listed, detailed
constraints on the abundances were not presented.  Our models assume 1$\times$ solar abundance,
unless
otherwise indicated.  The ``nominal'' $K_{zz}$ profiles are shown in Fig.~\ref{eddynom}.}
\end{deluxetable*}

Although some infrared photometric and spectral data that can potentially help constrain the abundance 
of C-, O-, and N-bearing molecules on HD 209458b have been acquired and analyzed 
\citep{brown02,richardson03,richardson06,richardson07,deming05a,deming05b,deming07,knut08,beaulieu09,madhu09},
actual reported constraints on the molecular abundances are rare, due to signal-to-noise issues and to 
the fundamental degeneracy between temperatures and abundances that becomes a bigger problem for 
planets with stratospheric thermal inversions.  Available constraints are listed in Table \ref{tabhd209}, 
along with some results from our models.  
Our nominal HD 209458b dayside-average model, which is supposed to be representative of 
secondary-eclipse conditions, has a greater water mixing ratio, a smaller $\CHfour$ mixing ratio, 
and a smaller $\COtwo$ mixing ratio than the values derived by \citet{swain09b}.  Since the mid-IR 
{\it Spitzer}\/ data themselves may not provide strict constraints on the $\CHfour$ mixing ratio 
\citep{madhu09}, this model-data discrepancy for $\CHfour$ may not represent a true problem, 
unless the near-IR {\it HST}\/ NICMOS data rather than the {\it Spitzer}\/ IRAC and IRS data 
are the main source of the $\CHfour$ constraints from the \citet{swain09b} analysis.  In any case, 
it is hard to imagine how HD 209458b could be very methane-rich, especially on the dayside, as the 
high stratospheric temperatures should drive the composition toward equilibrium, where $\CHfour$ 
would then be a minor constituent.  Larger eddy diffusion coefficients might help quench $\CHfour$ 
at greater abundances than we are seeing in our nominal model, but thermochemical reactions on the 
dayside would still act to limit the total column abundance that could be built up.  Perhaps some 
other constituent is masquerading as methane in the \citet{swain09b} data.  Hydrogen cyanide is 
one possibility, although it, too, is predicted to be less abundant on the warm dayside.  Carbon 
dioxide is another species that seems to be only loosely constrained by the mid-IR photometric 
data for HD 209459b \citep{madhu09}, so our models might not be out of line with the 
\citet{swain09b} data unless their $\COtwo$ constraints come largely from the {\it HST}\/ NICMOS 
data rather than the {\it Spitzer}\/ photometric data.  However, both \citet{swain09b} and 
\citet{madhu09} constrain the $\HtwoO$ mixing ratio to be less than 10$^{-4}$, which would seem 
to favor low metallicities for HD 209458b, and the constraints on other species in relation to 
$\HtwoO$ seems to favor large (non-solar) C/O ratios \citep{madhu09}.  We have yet to investigate 
these types of models.

For the atmospheric composition at the terminators, even fewer constraints exist for the molecular 
composition.  \citet{beaulieu09} do not provide any firm conclusions regarding molecular abundances 
and simply state that models that assume $\HtwoO$ mixing ratios of 10$^{-4}$ to 10$^{-3}$ 
provide reasonable fits to the data, and that a CO mixing ratio of 10$^{-4}$, a $\COtwo$ mixing 
ratio of 10$^{-7}$, and a $\CHfour$ mixing ratio of 10$^{-6}$ are allowable.  Model-data comparisons 
by \citet{fort10} with the \citet{beaulieu09} {\it Spitzer}\/ IRAC data and the 24 $\mu$m 
{\it Spitzer}\/ MIPS point from \citet{richardson06} and H.~Knutson (private communication, 2009) 
do not add any further constraints, although \citet{fort10} note that the contrast in the 
\citet{beaulieu09} mid-IR data is hard to reproduce with their equilibrium models.  In particular, 
absorption at 5.8 and 8 $\mu$m is under-represented, or in general the contrast between the 
3.6 $\mu$m and the 5.8 and 8 $\mu$m apparent radii is under-represented by their models.  As shown
in section \ref{obsimp} and Fig.~\ref{189spectrans}, the predicted spectra from the
disequilibrium models are similar to spectra from equilibrium models, so disequilibrium chemistry of
C, H, O, and N species cannot resolve these discrepancies.

\section{Conclusions}

Our kinetics-transport models for HD 189733b and HD 209458b demonstrate that disequilibrium 
processes such as photochemistry, chemical kinetics, and transport-induced quenching can 
dramatically affect the composition of extrasolar-giant-planet atmospheres.  Although 
thermochemical equilibrium can be maintained in the deep, hot tropospheres of these planets, 
rapid transport and the photochemistry that is initiated from the absorption of ultraviolet 
photons from the host star can drive the composition away from equilibrium.  The effects of 
these disequilibrium processes are more prominent for cooler exoplanets than for warmer 
exoplanets for two main reasons.  First, the quench points for major species like $\CHfour$, 
CO, $\NHthree$, and $\Ntwo$ are dependent on temperature: the colder the deep atmosphere, 
the less effective that thermochemical kinetics will be at maintaining equilibrium, and the 
deeper the pressure level at which the quenched species will depart from their equilibrium 
composition.  Since the major transport-quenched species $\CHfour$ and $\NHthree$ on 
HD 189733b and HD 209458b have equilibrium abundances that increase with decreasing altitude, 
a deeper quench level corresponds to a greater mole fraction of the quenched constituent.  
Second, thermochemical kinetics will be less effective on planets with cooler stratospheres, 
such that disequilibrium compositions can be maintained.  When stratospheric temperatures are 
high, such as is expected for HD 209458b and other planets with stratospheric thermal 
inversions --- or that are very strongly irradiated in general --- kinetic reactions are 
more effective at driving the composition back toward equilibrium, even in the presence of 
ultraviolet photons and transport-induced quenching.  Disequilibrium processes are therefore 
expected to be more effective on the cooler HD 189733b than on the warmer HD 209458b.

Although transport-induced quenching can modify the abundances of all species, the effects are
typically more noticeable for the species that are not expected to be abundant in equilibrium at
observable altitudes.  When $\CHfour$-CO interconversion ceases to become kinetically efficient,
for example, both $\CHfour$ and CO will quench, but the quenching will be most obvious for CO
on planets that have cooler upper atmospheres (e.g., brown dwarfs or giant planets far from their
host stars) and most obvious for $\CHfour$ on planets with warmer upper atmospheres (e.g., highly
irradiated hot Jupiters).  The location of the quench point --- such as whether it is in the
$\CHfour$- or CO-dominated regime --- can also play a role determining which molecule appears to
be the most influenced by quenching.

In our kinetics-transport models that include neutral carbon, nitrogen, and oxygen chemistry, 
we find that species like CO, $\HtwoO$, $\COtwo$, and $\Ntwo$ are relatively unaffected by 
disequilibrium chemistry because of their strong bonds and/or efficient recycling.  The 
vertical profiles of these species are expected to follow equilibrium, except at very low 
pressures ($P$ $\lta$ 1 $\mu$bar), where photochemical processes can play a dominant role 
under some conditions.  In contrast, the mole fractions of $\CHfour$ and $\NHthree$ are 
expected to be significantly enhanced ({\it i.e.}, by several orders of magnitude) over 
equilibrium predictions in the bulk of the ``photosphere'' of these planets because of 
transport-induced quenching.  In the upper stratosphere, however, photochemical processes 
abruptly and effectively remove $\NHthree$ and $\CHfour$ on HD 189733b and HD 209458b in 
favor of photochemical products like HCN, $\CtwoHtwo$, N, and C  
\citep[cf.][]{liang04,zahnle09,zahnle11,line10}.  

As was first discussed by 
\citet{liang03,yelle04,garcia07}, one major consequence of photochemistry on close-in 
transiting planets like HD 209458b and HD 189733b is the production of huge quantities of 
atomic H due to catalytic cycles initiated by $\HtwoO$ photolysis.  In our model, as with 
some models of thermospheric photochemistry \citep[e.g.,][]{yelle04,garcia07}, H replaces 
$\Htwo$ as the most-abundant constituent near the base of the thermosphere, and H will be 
the dominant neutral component throughout the bulk of the thermosphere of these planets.  
Our model differs from those of \citet{liang03,liang04} and \citet{line10} in the 
effectiveness of $\Htwo$ destruction at high altitudes.  Other significant photochemical 
products in our models include HCN, unsaturated hydrocarbons like $\CtwoHtwo$, atomic O, C, 
and N, and certain radicals like OH, $\CHthree$, and $\NHtwo$.

This disequilibrium chemistry will have some observational consequences, as the disequilibrium products 
can affect the photometric and spectral signatures during the primary transit and secondary eclipse of 
transiting planets like HD 189733b and HD 209458b.  Increased opacity from HCN, $\CHfour$, $\NHthree$, 
and under certain conditions $\CtwoHtwo$ could be particularly important, especially for cooler
planets like HD 189733b or for transit spectra that are influenced by the atmospheric transmission 
at the colder terminators.  We recommend that investigators who analyze transit 
and eclipse data should include the possible effects of nitrogen-bearing species like HCN and NH$_3$ in
their analyses.  The additional disequilibrium species will also likely affect the thermal structure
of the planets --- a factor that was not considered in our present analysis --- and future radiative
models should include the effects of potential disequilibrium amounts of species like HCN, CH$_4$,
NH$_3$, and $\CtwoHtwo$.  We note that spectroscopic information on the hot bands of these molecules 
is needed to better predict the effects of these species on the observed spectra, and measurements
of ultraviolet photoabsorption cross sections at high temperatures would improve the model
predictions.

We find that although we produce some benzene, complex nitriles, and other high-molecular-weight 
species in our HD 189733b and HD 209458b models, the production of these and other potential soot 
precursors is not very favorable with our currently adopted neutral carbon, oxygen, and nitrogen 
photochemistry, in part because CO 
rather that $\CHfour$ is the dominant carbon-bearing constituent at high altitudes in our models.  In 
this respect, we disagree with the conclusions based on the cool, isothermal model atmospheres of
\citet{zahnle09,zahnle11}.  On both HD 189733b and HD 209458b, we expect the CO $\leftrightarrow$ $\CHfour$
interconversion quench point to reside within the CO stability field, such that CO rather than $\CHfour$ 
is the dominant carbon-bearing constituent in the cooler upper atmospheres of these planets ---
unless the eddy diffusion coefficient is extremely high ($K_{zz}$ $>$ 10$^{11}$ cm$^2$ $\smone$ on
HD 189733b).  Since cool 
($\lta$ 1500 K) isothermal models are already quenched at the bottom boundary, where $\CHfour$ 
is dominant, they predict unrealistically high $\CHfour$ abundances on planets like HD 189733b 
\citep{zahnle11}.  The likelihood of soot formation on HD 189733b has therefore been overemphasized in the
\citet{zahnle11} models.  On warmer planets like HD 209458b, complex hydrocarbons will have particular 
difficulty surviving, as the high H abundance and efficient thermochemical kinetics will greatly increase 
net destruction rates for these species.  However, some soot formation may be possible on cooler planets 
like HD 189733b, particularly if one considers the possibility of ion chemistry that is initiated from 
extreme ultraviolet radiation or charged particles from the parent star.  On many planets with reduced 
atmospheres within our own solar system, high-altitude hazes are produced from such processes
\citep[e.g.,][]{west81,pryor91,gerard95,wong00,wong03,friedson02,waite07,imanaka07,vuitton07,vuitton08}.
We encourage the consideration of Titan-like ion chemistry in future photochemical models of extrasolar 
giant planets.

Our models of neutral carbon, nitrogen, and oxygen chemistry shed little light on the source of 
the absorber responsible for the stratospheric thermal inversions inferred for HD 209458b and other
exoplanets
\citep[e.g.,][]{burrows07,burrows08,knut08,knut09,fort08a,machalek09,machalek10,madhu09,madhu10inv,odonovan10}.
TiO was originally deemed to be a leading candidate because it would be expected to condense 
on cooler, less-irradiated hot Jupiters while remaining in gas phase on warmer, more highly 
irradiated hot Jupiters \citep{hubeny03,burrows08,fort08a}; however, \citet{knut10} argue that 
the lack of a convincing correlation between the incident flux and the presence of an inferred 
thermal inversion --- as well as an observed apparent correlation between less chromospherically
active stars and exoplanet thermal inversions --- suggest that photochemistry might play a role, 
perhaps through destruction of the responsible absorber.  In our models, we do not see any molecular 
absorbers on HD 209458b that are not also present on HD 189733b.  In fact, the cool stratosphere 
of HD 189733b supports {\it more}\/ disequilibrium molecular constituents than on HD 209458b.  
Atomic species are more prevalent on HD 209458b, and it is possible that some unidentified atom 
could be the culprit absorber, but that suggestion leads to a circular argument, as it is the high 
temperatures that cause atomic species to be more prevalent in the first place.  Chromospherically 
active stars have more flares and produce more cosmic rays, and one might suggest that ion chemistry 
induced by charged particles from the host star could produce a high-altitude haze that could 
obscure the culprit absorber --- despite the fact that photochemical hazes tend to be absorptive 
themselves and could thus heat the atmosphere (see \citealt{pont08} for evidence of a 
Rayleigh-scattering atmosphere at visible wavelengths on HD 189733b such that the Na absorption
features are obscured, suggesting a high-altitude haze, and see \citealt{zahnle11} who champion 
soot formation at high altitudes).  Alternatively, some sulfur photochemical product might 
contribute to absorption \citep{zahnle09}, or there could be a difference in 
dynamical mixing on different exoplanets.  In any case, neutral C-N-O chemistry seems unlikely 
to explain the difference in stratospheric thermal profiles on HD 189733b and HD 209458b because 
of the expected similar photochemical products on the two planets, but ion chemistry might 
contribute to the destruction of gas-phase absorber or to the formation of complex C- and 
N-bearing photochemical products that are ultimately condensible.

The quantitative results of our models are very sensitive to factors such as the thermal structure, 
the eddy diffusion coefficient profile, and the planetary metallicity.  Most of these parameters 
are not well constrained by observations.  The eddy diffusion coefficients, in particular, are 
difficult to obtain either theoretically or observationally.  If eddy diffusion coefficients in 
the lower stratosphere and upper troposphere were as low as are suggested by \citet{youdin10}, then
transport-induced quenching of constituents would be much less important on HD 189733b and HD
209458b.  We note that if the $\CHfour$ abundance and metallicity could be accurately constrained 
by observations, then one could place constraints on the value of $K_{zz}$ at the $\CHfour$ 
$\rightarrow$ CO quench point in the deep atmosphere.  Similarly, one might suggest that the 
observed presence of heavy atomic species like O, C, and metals might provide some constraints on 
the eddy diffusion coefficients in the upper atmosphere, given that low $K_{zz}$ values of $\lta$ 
10$^8$ cm$^2$ $\smone$ result in a very low homopause altitude in our models, making it difficult 
for heavy atomic species to be mixed into the thermosphere.  Note, however, that this conclusion 
is based on hydrostatic models that do not include the hydrodynamic vertical wind, and it is 
likely that the hydrodynamic wind could dominate in cases where $K_{zz}$ is low, such that the 
concept of a homopause level has little meaning for close-in transiting exoplanets; heavy species 
could be dragged into the thermosphere regardless of $K_{zz}$ in these situations, and our low 
$K_{zz}$ models are likely not very realistic.  

Simple time-constant arguments suggest that advection might effectively homogenize the longitudinal 
composition such that the terminator-vs.-dayside differences in the abundance of quenched species 
like $\CHfour$ and $\NHthree$ that are predicted by our simple 1-D models would be washed out at 
pressures greater than a few mbar.   Therefore, observations that indicate a difference in the
column abundance of these species during transit as compared with eclipse observations could provide 
clues to the rate of horizontal transport in the lower stratosphere.  In the upper stratospheres at 
pressures less than $\lta$ 1 mbar, photochemical time constants likely dominate, suggesting that 
the composition could be highly variable with longitude at high altitudes.  One 
important outcome is the potential conversion of HCN and $\CtwoHtwo$ back to $\CHfour$ at high 
altitudes on the nightside hemisphere.  When that high-altitude methane flows back around to the 
dayside, it may strongly absorb stellar radiation, with some interesting radiative consequences, 
including potential emission in the $\nu_3$ band of methane that has some strong lines that 
could potential absorb at high altitudes if $\CHfour$ is abundant enough (see the observations 
of \citealt{swain10}).  The quantitative details need to be worked out to determine if this 
process is viable.

The qualitative details of the photochemistry described here will be similar to other hot 
Jupiters and hot Neptunes.  On very cool planets like GJ 436b, $\CHfour$ will dominate over 
CO, with a resulting increase in the production of photochemical products like hydrocarbons and 
nitriles.  However, $\CHfour$ will not be efficiently removed from the lower atmosphere of 
GJ 436b with our standard neutral C, O, N chemistry, and we cannot explain the low inferred 
$\CHfour$ abundance on this planet \citep{stevenson10}.  Sulfur photochemistry could potentially 
help \citep{zahnle09,zahnle11}, but only if the total sulfur mole fraction approaches the 
$\CHfour$ mole fraction ({\it i.e.}, for S/C ratios $\gg$ solar) or if as-yet-unidentified 
catalytic cycles can permanently convert the methane into other hydrocarbons or C-O bearing 
species.

We eagerly await future results from the upcoming {\it James\ Webb\ Space\ Telescope\/} 
({\it JWST\/}), whose broad wavelength coverage and other spectral capabilities should help 
reduce current uncertainties in the determination of atmospheric composition on extrasolar 
giant planets.



\acknowledgments

We thank A.~Garc\'ia Mu\~noz for sending us his HD 209458b thermospheric model results, 
and Michael Line, Kevin Zahnle, and Roger Yelle for interesting chemistry discussions.  We gratefully 
acknowledge support from the NASA Planetary Atmospheres Program grant numbers NNX10AF65G 
(JM), NNX10AF64G (CV), NNH09AK24I (SK), and now NNX11AD64G.




\clearpage

\clearpage

\end{document}